\documentclass[preprints,article,accept,moreauthors,pdftex]{mdpi} 

\firstpage{1} 
\pubvolume{xx}
\issuenum{1}
\articlenumber{5}
\pubyear{2020}
\copyrightyear{2020}
\history{Received: 16 April 2020; Accepted: 17 June 2020; Published: 23 June 2020} 
\updates{yes} 

\usepackage{soul,upgreek}
\usepackage{ifpdf}

\usepackage{array}
\usepackage{tabularx}

\usepackage[labelformat=simple]{subcaption}

\DeclareCaptionLabelFormat{subcaptionlabel}{\normalfont(\textbf{#2}\normalfont)}
\usepackage{rotating}

\usepackage{tablefootnote}

\usepackage{pgf,pgfplots,pgfplotstable}
\pgfplotsset{compat=newest}
\pgfplotsset{
xlabel near ticks,
ylabel near ticks,
label style={font=\footnotesize},
tick label style={font=\footnotesize},
legend style={font=\scriptsize},
xticklabel style={/pgf/number format/set thousands separator={\,}},
yticklabel style={/pgf/number format/set thousands separator={\,}},
tick scale binop=\times,
try min ticks=6,
legend pos=outer north east
}
\usepgfplotslibrary{external} 
\usepgfplotslibrary{units}

\usepackage{tkz-base,xstring}
\usetikzlibrary{shapes,decorations,shadows}
\usetikzlibrary{decorations.pathreplacing}
\usetikzlibrary{decorations.pathmorphing}
\usetikzlibrary{decorations.shapes}
\usetikzlibrary{decorations.text}
\usetikzlibrary{decorations.footprints}
\usetikzlibrary{decorations.fractals}
\usetikzlibrary{fadings}
\usetikzlibrary{patterns}
\usetikzlibrary{calc}
\usetikzlibrary{shapes.geometric}
\usetikzlibrary{shapes.gates.logic.IEC}
\usetikzlibrary{shapes.gates.logic.US}
\usetikzlibrary{fit,chains}
\usetikzlibrary{positioning}
\usepgflibrary{shapes}
\usetikzlibrary{scopes}
\usetikzlibrary{arrows}
\usetikzlibrary{arrows.meta}
\usetikzlibrary{backgrounds}
\usetikzlibrary{intersections}
\usetikzlibrary{matrix}
\usetikzlibrary{pgfplots.units}
\usetikzlibrary{trees}
\usetikzlibrary{hobby}

\newlength\figureheight
\newlength\figurewidth
\usepackage{stanli}

\usepackage[all]{xy}

\usepackage{amssymb,amsfonts,bm}
\usepackage{textcomp,wasysym,eurosym,stmaryrd}
\usepackage{mathrsfs}
\usepackage{mathtools}
\usepackage{pifont}
\usepackage{upgreek}
\usepackage{centernot}
\usepackage{algorithm,algorithmic}

\usepackage{xspace}
\setlist[itemize]{leftmargin=*,labelsep=5.8mm}
\setlist[enumerate]{leftmargin=*,labelsep=4.9mm}

\setitemize{parsep=6pt,itemsep=0pt,leftmargin=*,labelsep=5.5mm}
\setenumerate{parsep=6pt,itemsep=0pt,leftmargin=*,labelsep=5.5mm}
\setlist[description]{itemsep=0mm}

\newcommand{\ab}{\boldsymbol{a}}
\newcommand{\bb}{\boldsymbol{b}}

\newcommand{\fb}{\boldsymbol{f}}

\newcommand{\hb}{\boldsymbol{h}}

\newcommand{\nb}{\boldsymbol{n}}

\newcommand{\ssb}{\boldsymbol{s}}

\newcommand{\ub}{\boldsymbol{u}}

\newcommand{\xb}{\boldsymbol{x}}

\newcommand{\zerob}{\boldsymbol{0}}

\newcommand{\epsilonb}{\boldsymbol{\varepsilon}}

\newcommand{\sigmab}{\boldsymbol{\sigma}}

\newcommand{\ellb}{\boldsymbol{\ell}}

\newcommand{\Bb}{\boldsymbol{B}}
\newcommand{\Cb}{\boldsymbol{C}}

\newcommand{\Hb}{\boldsymbol{H}}

\newcommand{\Lb}{\boldsymbol{L}}

\newcommand{\Sb}{\boldsymbol{S}}

\newcommand{\Ub}{\boldsymbol{U}}

\newcommand{\varSigmab}{\boldsymbol{\varSigma}}

\newcommand{\Ac}{\mathcal{A}}
\newcommand{\Bc}{\mathcal{B}}

\newcommand{\Ec}{\mathcal{E}}

\newcommand{\Hc}{\mathcal{H}}

\newcommand{\Jc}{\mathcal{J}}

\newcommand{\Lc}{\mathcal{L}}

\newcommand{\Sc}{\mathcal{S}}

\newcommand{\Ecb}{\boldsymbol{\Ec}}

\newcommand{\Jcb}{\boldsymbol{\Jc}}

\newcommand{\Ebb}{\mathbb{E}}

\newcommand{\Rbb}{\mathbb{R}}

\DeclareFontFamily{U}{bbold}{}
\DeclareFontShape{U}{bbold}{m}{n}{<-5.5> bbold5 <5.5-7.5> bbold7 <7.5-> bbold10}{}
\newcommand{\onebb}{{\text{\usefont{U}{bbold}{m}{n}1}}}

\newcommand{\abs}[1]{\lvert#1\rvert}

\newcommand{\norm}[1]{\lVert#1\rVert}

\newcommand{\set}[1]{\{#1\}}

\newcommand{\scalprod}[2]{#1\cdot#2}

\DeclareMathOperator{\divb}{\textbf{div}}
\DeclareMathOperator{\gradd}{grad\kern-.5em{grad}}

\DeclareMathOperator{\nablab}{\boldsymbol{\nabla}}

\DeclareMathOperator*{\argmax}{arg\,max}
\DeclareMathOperator*{\argmin}{arg\,min}

\newcommand{\interval}[4]{\mathopen{#1}#2 \mathclose{}\mathpunct{},#3 \mathclose{#4}}
\newcommand{\intervalcc}[2]{\interval{[}{#1}{#2}{]}}

\newcommand{\intervaloo}[2]{\interval{]}{#1}{#2}{[}}

\renewcommand{\(}{\left(}
\renewcommand{\)}{\right)}

\renewcommand{\leq}{\leqslant}
\let\oldtimes\times
\renewcommand{\times}{\!\oldtimes\!}

\newcommand{\etc}{etc.\xspace}

\Title{{Robust Multiscale Identification of Apparent Elastic Properties at Mesoscale for Random Heterogeneous Materials with Multiscale Field Measurements}} 

\Author{Tianyu Zhang $^{\dagger}$, Florent Pled *$^{,\dagger}$\orcidA{} and Christophe Desceliers *$^{,\dagger}$\orcidB{}}

\AuthorNames{Tianyu Zhang, Florent Pled and Christophe Desceliers}

\address[1]{%
Univ Gustave Eiffel, MSME UMR 8208, F-77454 Marne-la-Vall\'ee, France; tianyu.zhang@univ-eiffel.fr}

\corres{\hangafter=1 \hangindent=1.05em \hspace{-0.82em}Correspondence: florent.pled@univ-eiffel.fr (F.P.); christophe.desceliers@univ-eiffel.fr (C.D.)}

\firstnote{\hangafter=1 \hangindent=1.05em \hspace{-0.82em}These authors contributed equally to this work.}

\abstract{{The aim of this work is to efficiently and robustly solve the statistical inverse problem related to} the identification of the elastic properties at both macroscopic and mesoscopic scales of heterogeneous anisotropic materials with a complex microstructure that usually cannot be properly described in terms of their mechanical constituents at microscale. {Within the context of linear elasticity theory, the apparent elasticity tensor field at a given mesoscale} is modeled by a {prior} non-Gaussian tensor-valued random field. A general methodology using multiscale displacement field measurements simultaneously made at both macroscale and mesoscale has been recently proposed for the identification the hyperparameters of such a {prior} stochastic model by solving a multiscale statistical inverse problem using a {stochastic} computational model and some information from displacement fields at both macroscale and mesoscale. This paper contributes to the improvement of the computational efficiency, accuracy and robustness of such a method by introducing (i) a mesoscopic numerical indicator related to the spatial correlation length(s) of kinematic fields, allowing the time-consuming global optimization algorithm (genetic algorithm) used in a previous work to be replaced with a more efficient algorithm and (ii) an \emph{ad hoc} stochastic representation of the hyperparameters involved in the {prior} stochastic model in order to enhance both the robustness and the precision of the statistical inverse identification method. Finally, the proposed improved method is first validated on \emph{in silico} materials within the framework of 2D plane stress and 3D linear elasticity (using multiscale simulated data obtained through numerical computations) and then exemplified on a real heterogeneous biological material (beef cortical bone) within the framework of 2D plane stress linear elasticity (using multiscale experimental data obtained through mechanical testing monitored by digital 
image correlation).}



\begin{document}


\section{Introduction}
\label{sec:introduction}

Within the framework of linear elasticity theory, the numerical modeling and simulation of heterogeneous materials with hierarchical complex random microstructure give rise to many scientific challenges. Their modeling is a topical issue with numerous applications in diverse material sciences, including for instance sedimentary rocks, natural composites, fiber- or nano-reinforced composites, some concretes and cementitious materials, some porous media, some living biological tissues, among many others \cite{Tor02}. Although such materials are often considered and modeled as deterministic and homogeneous elastic media at macroscale in most practical applications, they are not only random and heterogeneous at microscale but they also usually cannot be explicitly described by any local morphological and mechanical properties of their constituents and easily reconstructed in a computational framework in the presence of multiple interfaces. The modeling and identification of their elastic properties at meso- or microscales have been the subject of many research works in recent decades. Nowadays, with the recent developments achieved around the construction of stochastic models for tensor-valued random elasticity fields and their experimental inverse identification using field imaging techniques, one of the most promising {ways} consists in introducing a {prior} stochastic model of the apparent elasticity tensor field of heterogeneous materials of the considered microstructure at a given mesoscale. Note that this mesoscopic scale allows the introduction of the spatial correlation length(s) of the microstructure, and that for materials with a hierarchical structure, such as cortical bone or tendon, different mesoscopic scales can be defined. {Such a mesoscopic stochastic modeling of random heterogeneous elastic media can further be used to characterize the macroscopic mechanical properties in the context of the stochastic homogenization over a representative volume element (RVE) subdomain. This representative volume element should be, provided that it exists, sufficiently large compared to the microscale and sufficiently small compared to the macroscale.} In the present probabilistic context, a major question concerns the statistical inverse identification of a {prior} stochastic model parameterized by a small or moderate number of hyperparameters using only partial and limited experimental data.

\subsection{Overview of Inverse Methods for the Mechanical Characterization of Micro/Meso-Structural Properties}\label{sec:inverse_methods}

The inverse methods for the experimental identification of elastic properties of homogeneous or heterogeneous materials at macroscale and/or mesoscale have been the subject of numerous research works over the three past decades. The first methods related to the experimental characterization and description of random microstructural morphologies by using image analysis techniques have been introduced and developed by the end of the 1980s \cite{Jeu87,Jeu89,Jeu00, Jeu01b,Jeu12} for the numerical modeling and simulation of random microstructures made up with heterogeneous materials. Since the early 1990s, significant technological advances in the field of optical measuring instruments, such as digital cameras equipped with Charge-Coupled Device (CCD) or Complementary MetalOxideSemiconductor (CMOS) image sensors and microscope objectives, have widely contributed to the emergence of imaging techniques such as two-dimensional (2D) or three-dimensional (3D) digital image correlation (DIC) for identification purposes. DIC techniques \cite{Pan09a, Sut09,Hild12a} are now commonly used in solid mechanics and material sciences for experimental measurements of elastic displacement fields of samples under external loading \cite{Kah94,Ven98a,Ven98b, Hild06,Roux08a,Ret08, Bor10} in order to identify mechanical properties of complex microstructures for heterogeneous materials \cite{Con95,Bax00, Gey02,Gey03,Gra03,Bon05,Hild06,Avr07,Avr08a} with different classes of material symmetries. The recent milestones achieved around data acquisition systems and processing softwares for 3D images obtained for example by X-ray computed microtomography ($\mu$CT) \cite{Fla87,Kak88, Bar00, Stock08,Des10,Maire14}, magnetic resonance imaging (MRI) \cite{vandenEls93, Mai98, Lia00,Hill01}, optical coherence tomography (OCT) \cite{Bea98,Sch99, Fed10,Gam11,Pop11} or any other non-invasive and non-destructive testing technique for the reconstruction of 3D images in high resolution, have~allowed the development of three-dimensional measurements of displacement fields by digital volume correlation (DVC)~\cite{Bay99,Ver04, Bay08,Roux08b,Ret08, Rann10, Hild12a,Mad13, Rob14,Fed14, Hild16, Bou17,Bul18}. Such 3D full-field measurements offer the potential of identifying stochastic models of 3D tensor-valued random elasticity fields at different scales for the mechanical characterization of 3D real microstructures made up of heterogeneous materials.

In the mid 2000s, many research works have been carried out on the statistical inverse identification of stochastic models of the tensor-valued random elasticity field in low or high stochastic dimension at macroscopic and/or mesoscopic scale for complex microstructures modeled by random heterogeneous isotropic or anisotropic linear elastic media \cite{Des06,Gha06,Des07,Mar07,Arn08,Das08,Das09b,Des09,Gui09,Ma09,Mar09a,Arn10,Das10,Ta10,Soi10,Soi11a}. The~proposed methodologies for solving the statistical inverse problem related to the identification of a non-Gaussian tensor-valued random field in high stochastic dimension using available, partial~and limited experimental data are mostly based on (i) the mathematical formulations of functional analysis for stochastic boundary value problems, (ii) the statistical tools derived from probability theory, information theory, mathematical statistics and stochastic optimization, such as the least-squares (LS) method \cite{Law95,Soi17a}, the maximum likelihood estimation (MLE) method \cite{Ser80,Pap02,Spa05a,Soi17a}, the~maximum entropy (MaxEnt) principle \cite{Jay57a,Jay57b,Sob90,Kap92,Jum00,Jay03,Cov06,Soi17a}, the nonparametric statistics \cite{Ser80,Bow97}, the~Bayesian inference method~\mbox{\cite{Beck98,Ber01,Con07,Car09,Stu10,Tan10,Soi17a,Rap20}}, the statistical and computational inverse problems and related stochastic optimization algorithms~\mbox{\cite{Col74,Wal97,Spa05a,Kai05,Tar05,Isa06,Cal18,Ast19}}, (iii)~advanced functional representation techniques and probabilistic methods, such as the Karhunen-Lo\`eve (KL) decomposition \cite{Kar46,Loe77,Loe78} to construct reduced-order stochastic models, the polynomial chaos (PC) expansion \cite{Gha91,Gha99,Xiu02,Soi04,Wan06b} for an adapted high-dimensional stochastic representation of non-Gaussian second-order random fields, (iv) the spectral methods \cite{Gha91,Xiu05,Bab07,Nou08b,LeMai10} and sampling-based approaches \cite{Cal98,Sch01, Rub16} for solving stochastic boundary value problems, and (v) the stochastic homogenization methods \cite{San80, San85,Fra86,Suq87,Huet90,Sab92,Nem93, Jikov94, Suq97,And98, Pra98,For00,Bor01, Jeu01b, Tor02,Zao02, Kan03,Bou04,Sab05, Ost06,Ost07, Soi08a,Xu09bis,Too10, Gui11a,Jeu12,Tef18} to bridge the meso- or microscopic scale and the macroscopic scale. Combining such advanced probabilistic and statistical methods has led to early fundamental works on the statistical inverse identification of non-Gaussian scalar- or tensor-valued random fields in low or high stochastic dimension based on partial and limited experimental data. These works have mainly been devoted to the statistical inverse identification of hyperparameters of {prior} stochastic models in low stochastic dimension, such~as a mean field, a dispersion coefficient and some spatial correlation length(s) or the deterministic coefficients of a polynomial chaos expansion of the random field \cite{Des06,Gha06,Des07,Arn08,Das08,Das09b,Des09,Gui09,Ma09,Mar09a,Arn10,Das10,Ta10,Cot11,Des12,Clo13}. To date, the~latest and more advanced works focus on the inverse identification of {posterior} stochastic models, that are high-dimensional stochastic representations of {prior} stochastic models for non-Gaussian scalar- or tensor-valued random fields \cite{Soi10,Soi11a,Soi12,Per12,Per13,Clo13,Nou14}.

\subsection{Multiscale Statistical Identification Method}\label{sec:identification_method}

In keeping with the aforementioned works, an innovative methodology has been recently proposed in Reference \cite{Ngu15} for the multiscale statistical inverse identification of a {prior} stochastic model of the random apparent elasticity field at mesoscale for a heterogeneous anisotropic elastic microstructure. This multiscale identification procedure has been formulated within the framework of 3D linear elasticity theory under the following assumptions: (i) at macroscale, the elasticity tensor is deterministic and homogeneous and therefore independent of the spatial coordinates; (ii) at a given mesoscale, the~tensor-valued random elasticity field is the restriction to a mesoscopic subdomain of a statistically homogeneous random field indexed by $\Rbb^3$, allowing to be consistent with the assumption for the existence of a representative volume element in the framework of stochastic homogenization \cite{Soi08a,Soi17a}.
 
The proposed method allows for the multiscale inverse identification of (i) the tensor-valued random field that models the apparent elasticity tensor field at a given mesoscale, and (ii) the effective elasticity tensor at macroscale, for a heterogeneous anisotropic elastic material with a random microstructure whose morphological and mechanical properties cannot be properly described and reconstructed in a computational framework from the local topology and mechanical behavior of its constitutive phases. The {prior} stochastic model of the random elasticity field is constructed by using the MaxEnt principle \cite{Jay57a,Jay57b,Sob90,Kap92,Jum00,Jay03,Cov06,Soi17a}, initially derived within the general framework of information theory~\cite{Sha48,Sha01,Bal68}. We then obtain a second-order mean-square continuous non-Gaussian positive-definite symmetric real matrix-valued random field. In addition, an explicit algebraic representation has been established in Reference \cite{Soi06}. Such a {prior} stochastic model of random elasticity field has been used, in~particular, for stochastic boundary value problems, such as static linear elasticity problems~\cite{Soi06,Soi08a,Soi17a}. It is classically parameterized by a small or moderate number of scalar-, vector-~and/or tensor-valued hyperparameters, namely the mean function of the random elasticity field, a dispersion coefficient controlling the level of statistical fluctuations of the random elasticity field around its mean function and spatial correlation lengths characterizing the spatial correlation structure of the random elasticity field. The statistical inverse problem for the identification of this {prior} stochastic model is formulated as a multi-objective optimization problem for which the optimal parameters are the optimal values of the hyperparameters of the stochastic model. However, within the framework of this identification methodology, it can be shown that the mean function of the random elasticity field cannot directly be identified using only the available experimental kinematic field measurements at mesoscale. The~experimental values of the stress fields associated with the kinematic fields observed experimentally at mesoscale should also be known, but these values are not available in practice. Conversely, it can also be shown that the other hyperparameters (dispersion coefficient and spatial correlation lengths) controlling the statistical fluctuations of the random elasticity field cannot directly be identified using only the available experimental kinematic field measurements at macroscale. Consequently, such a statistical inverse identification procedure requires multiscale experimental field measurements that must be made simultaneously at both macroscopic and mesoscopic scales, since by assumption only a single specimen submitted to a given external loading at macroscale is experimentally tested. A stochastic homogenization method is then used to propagate the uncertainties at mesoscale towards the macroscale under the classical assumption of scale separation between macroscale and mesoscale, so that a sufficiently large mesoscopic subdomain can be defined within the macroscopic domain and considered as a representative volume element. However, it should be noted that it is not necessary for this representative volume element to be the same size as the mesoscopic domain(s) of observation on which the experimental measurements are performed. Thus, the multiscale statistical inverse problem is formulated as a multi-objective optimization problem that consists in minimizing a (vector-valued) multi-objective cost function defined by three numerical indicators corresponding to single-objective cost functions \cite{Ngu15}, namely (i) a macroscopic numerical indicator allowing the distance between the measured experimental fields and the computed numerical fields to be quantified at macroscale, (ii) a mesoscopic numerical indicator allowing the distance between the statistical fluctuations exhibited by the measured experimental fields and the ones exhibited by the computed numerical fields to be quantified at mesoscale, and (iii) a multiscale numerical indicator allowing the distance between the elasticity tensor at macroscale and the effective elasticity tensor constructed by computational stochastic homogenization of the random apparent elasticity field in a representative volume element at mesoscale.

\subsection{Drawbacks and Limitations of the Multiscale Identification Method}
\label{sec:drawbacks_method}

The multiscale identification method proposed in Reference \cite{Ngu15} has been first validated by numerical simulations on \emph{in silico} materials and then successfully applied to the experimental characterization of the elastic properties of a biological tissue (beef cortical bone) within the framework of 2D plane stress linear elasticity from multiscale optical measurements of displacement fields performed at both macroscopic and mesoscopic scales on a single cortical bone specimen under static external loading at macroscale \cite{Ngu16}. Nevertheless, the proposed identification method has some drawbacks that limit its use. First, it should be noted that the cost functions introduced for the multi-objective optimization problem are not dedicated to a particular hyperparameter of the {prior} stochastic model of the random field to be identified. Therefore, the only approach considered for solving the multi-objective optimization problem was to use a global optimization algorithm (genetic algorithm) that belongs to the class of random search, genetic and evolutionary algorithms \cite{DaCun67,Cen77,Yu85,Dauer86,Gol89,Deb01a,Mar04,Kon06,Coe06,Coe07,Deb14} to randomly explore the admissible set of hyperparameters. Despite a suitable parameterization (population size at each new generation, random generation of initial population, selection procedure for reproduction including crossover and mutation operators, elite count, stopping criteria, \etc) of the genetic algorithm used in Reference \cite{Ngu15} and the use of parallel processing and computing, the computational cost for solving the multi-objective optimization problem is high. This is due in particular to the large stochastic dimension of the tensor-valued random elasticity field. Secondly, during the validation and implementation of the multiscale identification method proposed in Reference \cite{Ngu15}, it was found that, for different mesoscopic domains of observation within the same macroscopic domain, the resolution of the multi-objective optimization problem led to different optimal values of hyperparameters from one domain to another. Indeed, the~experimental field measurements over each mesoscopic domain of observation can be modeled as different random fields, and therefore the multi-objective cost function on each mesoscopic domain of observation is a deterministic function of these random fields. This explains why the statistics of the multi-objective cost function are different from one mesoscopic domain of observation to another. In~Reference \cite{Ngu15}, the multi-objective cost function has been replaced by the statistical average of the multi-objective cost functions calculated over each of the mesoscopic domains of observation.

\subsection{Improvements of the Multiscale Identification Method and Novelty of the Paper}
\label{sec:improvements_method}

{In order to overcome the issues outlined above, this research work aims to present} two major improvements of the methodology initially proposed in Reference \cite{Ngu15} allowing the statistical inverse identification of the tensor-valued random elasticity field at mesoscale to be performed with a better computational efficiency, higher accuracy and improved robustness. First, we introduce an additional mesoscopic numerical indicator allowing the distance between the spatial correlation length(s) of the measured experimental kinematic fields and the one(s) of the computed numerical kinematic fields to be quantified at mesoscale, so that each hyperparameter of the {prior} stochastic model has its own dedicated single-objective cost function, thus allowing the time-consuming global optimization algorithm (genetic algorithm) used in Reference \cite{Ngu15} to be avoided and replaced with a more efficient algorithm, such as a fixed-point iterative algorithm, for solving the underlying multi-objective optimization problem. Secondly, in the case where experimental field measurements are available on several mesoscopic domains of observation, we propose to not replace ``naively'' the multi-objective cost function by its empirical mean over all the mesoscopic domains of observation, but to consider a multi-objective optimization problem for each mesoscopic domain of observation. Thus, each~mesoscopic domain of observation leads to a possible solution of the values of the hyperparameters. Each of these values is then considered as a realization of a random vector of hyperparameters whose {prior} stochastic model is constructed by using the MaxEnt principle, and~whose hyperparameters can be determined by using the MLE method, in order to improve both the robustness and the accuracy of the inverse identification method of the {prior} stochastic model.

\subsection{Outline of the Paper}
\label{sec:outline}

The paper is organized as follows. Following this introduction, Section~\ref{sec:assumptions} presents the general assumptions for solving the underlying multiscale statistical inverse problem. Then, Section~\ref{sec:exp_config} is dedicated to the description of the multiscale experimental test configuration for obtaining experimental data at both macroscale and mesoscale. Section~\ref{sec:stochastic_model} describes the {prior} stochastic model of the fourth-order tensor-valued random elasticity field and its parameterization. Section~\ref{sec:objectives_strategy} focuses on the objectives of the multiscale statistical inverse problem and the multiscale identification strategy. Next, Section~\ref{sec:indicators} presents the construction of the macroscopic, mesoscopic and multiscale numerical indicators that are used for solving the multiscale statistical inverse problem as a multi-objective optimization problem. In this section, a focus is made on the improvements proposed by this paper in the definition of these numerical indicators with respect to the previous work presented in Reference \cite{Ngu15}. The multi-objective optimization problem is then set in Section~\ref{sec:formulation_optim_problem} and some numerical methods for solving such a multi-objective problem are presented in Section~\ref{sec:solving_optim_problem}. Section~\ref{sec:randomization} discusses an improvement proposed in this paper for a robust identification when some experimental field measurements are available on several mesoscopic domains of observation. Section~\ref{sec:validation} presents a numerical validation of the proposed multiscale identification methodology on \emph{in silico} test specimens within the framework of 3D linear elasticity under 2D plane stress assumption and in the general 3D case, for which the multiscale experimental data have been numerically simulated. Finally, Section~\ref{sec:application} presents an experimental application to a real heterogeneous biological material constituted of beef cortical bone within the framework of linear elasticity under 2D plane stress assumption, for which the multiscale experimental data have been obtained from a single static uniaxial compression test performed on a specimen of beef femoral cortical bone and monitored by 2D digital image correlation at both macroscale and mesoscale. Lastly, Section~\ref{sec:conclusion} gives some conclusions and potential perspectives of this~work.

\section{Assumptions for Solving the Multiscale Statistical Inverse Problem}
\label{sec:assumptions}

In the present work, we address the statistical inverse identification of the elastic properties for a complex microstructure made up of a heterogeneous anisotropic material and considered as a random linear elastic medium. In this section, we first state suitable assumptions for solving this multiscale statistical inverse problem. Within the framework of linear elasticity theory, probability theory and computational stochastic homogenization in micromechanics and multiscale mechanics of heterogeneous materials, the following assumptions related to scale separation, stationarity and ergodicity properties are introduced:
\begin{itemize}
\item there exists a scale separation between macroscale and mesoscale, so that a mesoscopic subdomain can be defined and for which the dimensions are sufficiently large with respect to the size of the heterogeneities and sufficiently small with respect to the size of the macroscopic domain. Such a mesoscopic subdomain can then be considered as a representative volume element;

\item the random apparent elasticity tensor field at mesoscale is the restriction to one or more bounded mesoscopic subdomain(s) of a second-order stationary random field indexed by $\Rbb^3$, and~consequently the mean function of the random elasticity field at mesoscale is independent of the spatial coordinates;

\item the random apparent elasticity tensor field at mesoscale is ergodic in average in the mean-square sense, so that the homogenized elasticity tensor at macroscale calculated by stochastic homogenization of the random apparent elasticity field in a mesoscopic subdomain corresponding to a representative volume element can be considered as almost deterministic, in the sense that (i)~its spatial average reaches an asymptotic convergence with a very high level of probability for a sufficiently large mesoscopic subdomain size, and therefore (ii) its level of statistical fluctuations around its mean function at macroscale can be considered as negligible, thus yielding a deterministic homogenized elasticity tensor at macroscale.
\end{itemize}

{In this work, we focus on the class of heterogeneous materials that can be considered as random elastic media and for which the hypothesis stated on the scale separation between macroscale and mesoscale is verified}. It should be noted that, if such a scale separation assumption was not satisfied, then the multiscale statistical inverse problem under consideration would be an ill-posed problem if only a single experimental field measurement at macroscale was available, because in this case the macroscopic elasticity (or compliance) tensor must be modeled by a random tensor and a single experimental measurement is not sufficient to identify its stochastic model. {The proposed identification methodology is therefore not adapted to this case and would require several experimental field measurements at macroscale as well as modifications of the macroscopic and multiscale indicators introduced in Section~\ref{sec:indicators}, and also the introduction of additional numerical indicators at macroscale.} Hereinafter, since the present identification methodology is developed within the framework of linear elasticity theory, we will use the terminology ``strain field'' to make reference to the ``linearized strain field'' for the sake of conciseness.

\section{Multiscale Experimental Test Configuration}
\label{sec:exp_config}

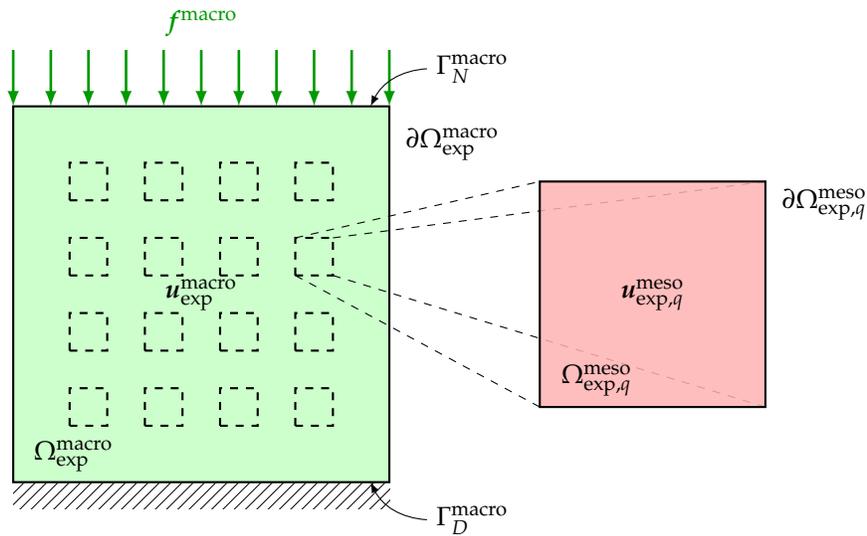
\begin{figure}[H]
\centering
\tikzsetnextfilename{multiscale_experimental_configuration}
\begin{tikzpicture}[every node/.style={minimum size=1cm},on grid]
\def \L {5}
\def \l {0.5}

\coordinate (A) at (0,0);
\coordinate (B) at (\L,\L);
\coordinate (C) at (0,\L);
\coordinate (D) at (\L,0);
\draw[black,thick,fill=green!20
] (A) rectangle (B);
\coordinate (lineloadVarA1) at ($ (C)!0!90:(B) $);
\coordinate (lineloadVarB1) at ($ (B)!0!-90:(C) $);
\coordinate (lineloadVarA2) at ($ (C)!{0.75cm}!90:(B) $);
\coordinate (lineloadVarB2) at ($ (B)!{0.75cm}!-90:(C) $);
\draw[green!60!black,force,->] (lineloadVarA2) -- (lineloadVarA1);
\draw[green!60!black,force,->] (lineloadVarB2) -- (lineloadVarB1);
\pgfmathsetmacro{\lineloadIntervalBegin}{\lineloadInterval/2/\scalingParameter}
\pgfmathsetmacro{\lineloadIntervalStep}{\lineloadInterval/2/\scalingParameter*2}
\pgfmathsetmacro{\lineloadIntervalEnd}{1-\lineloadInterval/2/\scalingParameter}
\foreach \i in {\lineloadIntervalBegin,\lineloadIntervalStep,...,\lineloadIntervalEnd}
\draw[green!60!black,force,->] ($(lineloadVarA2)!\i!(lineloadVarB2)$)-- ($(lineloadVarA1)!\i!(lineloadVarB1)$);
\node[green!60!black,above=0.6] at ($(B)!0.5!(C)$) {$\fb^{\text{macro}}$};
\begin{scope}
	\clip (0,-\supportBasicHeight) rectangle (\L,0);
	\draw[hatching](\L+1,0) -- ++(-\L-1,0);
\end{scope}
\foreach \i in {4,1,2,3}{
	\foreach \j in {1,...,4}{
		\coordinate (E) at (\j*\L/5-\l/2,\i*\L/5-\l/2);
		\coordinate (F) at (\j*\L/5+\l/2,\i*\L/5+\l/2);
		\coordinate (G) at (\j*\L/5-\l/2,\i*\L/5+\l/2);
		\coordinate (H) at (\j*\L/5+\l/2,\i*\L/5-\l/2);
		\draw[black,thick,dashed
		] (E) rectangle (F);
	}
}
\node[right=-0.2] at ($(A)!0.07!(B)$) {$\Omega^{\text{macro}}_{\text{exp}}$};
\coordinate (GN) at (\L+0.5,\L+0.5);
\draw[<-,>=latex] ($(C)!0.95!(B)$) to[bend left] (GN) node[right] {$\Gamma_N^{\text{macro}}$};
\coordinate (GD) at (\L+0.5,-0.5);
\draw[<-,>=latex] ($(A)!0.95!(D)$) to[bend right] (GD) node[right] {$\Gamma_D^{\text{macro}}$};
\node[right=0.1] at ($(B)!0.1!(D)$) {$\partial \Omega^{\text{macro}}_{\text{exp}}$};
\node at ($(A)!0.5!(B)$) {$\ub^{\text{macro}}_{\text{exp}}$};

\begin{scope}[
	xshift=\L cm+1cm,
	]
	\coordinate (I) at (\L/5,\L/5);
	\coordinate (J) at (4*\L/5,4*\L/5);
	\coordinate (K) at (\L/5,4*\L/5);
	\coordinate (L) at (4*\L/5,\L/5);
\end{scope}

\draw[black,dashed,thin] (E) to (I);
\draw[black,dashed,thin] (F) to (J);
\draw[black,dashed,thin] (G) to (K);
\draw[black,dashed,thin] (H) to (L);

\begin{scope}[
	xshift=\L cm+1cm,
	]
	\draw[black,thick,fill=red!30,fill opacity=0.85] (I) rectangle (J);
	\node[right=-0.2] at ($(I)!0.12!(J)$) {$\Omega^{\text{meso}}_{\text{exp},q}$};
	\node[right=0.1] at ($(J)!0.1!(L)$) {$\partial \Omega^{\text{meso}}_{\text{exp},q}$};
	\node at ($(I)!0.5!(J)$) {$\ub^{\text{meso}}_{\text{exp},q}$};
\end{scope}

\end{tikzpicture}
\caption{Multiscale experimental configuration: displacement field $\ub^{\text{macro}}_{\text{exp}}$ measured in the macroscopic domain of observation $\Omega_{\text{exp}}^{\text{macro}}$ and displacement field $\ub^{\text{meso}}_{\text{exp}, q}$ measured in each mesoscopic domain of observation $\Omega^{\text{meso}}_{\text{exp},q}$, for $q=1,\dots,Q$.}
\label{fig:multiscale_experimental_configuration}
\end{figure}

The difficulties related to the acquisition of the experimental measurements for the inverse identification procedure to be carried out are induced not only by the complex nature of the heterogeneous anisotropic elastic microstructure but also by the need to obtain multiscale kinematic field measurements at two different scales (macroscale and mesoscale) for a single test specimen under given static loading conditions through a multiscale DIC performed simultaneously at both macroscale and mesoscale. To overcome such difficulties, a suitable experimental protocol, including~the preparation of the test specimen, the development of a measuring bench, the acquisition system of digital images and the DIC method, has been set up in Reference \cite{Ngu16} for the acquisition of 2D multiscale optical measurements of displacement fields performed at both macroscale and mesoscale on a single beef cortical bone specimen submitted to a static vertical uniaxial compression test. {Such a living biological tissue with a complex hierarchical microstructure is of particular interest in the present context of multiscale modeling and identification for random heterogeneous materials.} The multiscale experimental test configuration is briefly recalled here. A sketch of the multiscale experimental configuration of the specimen at macroscale and mesoscale is represented in Figure~\ref{fig:multiscale_experimental_configuration}.

The test specimen has a cubic shape and is submitted to a simple external load. On the upper side of the specimen, a surface force field is applied, while the opposite side of the specimen is clamped. Then, during the same and unique experimental loading, the displacement fields at both macroscale and mesoscale are simultaneously measured, for instance in using two optical digital cameras equipped with CCD imaging sensors with different spatial resolutions for the simultaneous acquisition of displacement field optical measurements at both macroscopic and mesoscopic scales. The measurements are performed on the domain $\Omega^{\text{macro}}_{\text{exp}}$ at macroscale and on the domain $\Omega^{\text{meso}}_{\text{exp}}$ at mesoscale that are 2D or 3D parts of the specimen at macroscale and mesoscale, respectively. These~domains can be 3D in the case of microtomography techniques for the acquisition of 3D experimental data, or they can be 2D in the case of digital camera techniques for the acquisition of 2D experimental data. Note that in case the dimensions of the mesoscopic domain of observation $\Omega^{\text{meso}}_{\text{exp}}$ are very small with respect to the dimensions of the macroscopic domain of observation $\Omega_{\text {exp}}^{\text{macro}}$, then~more information can be used by collecting additional experimental field measurements at mesoscale on $Q$ non-overlapping mesoscopic domains of observation $\Omega^{\text{meso}}_{\text{exp}, 1},\dots,\Omega^{\text{meso}}_{\text{exp}, Q}$ for which the relative mutual locations into the test specimen are not necessarily recorded. The experimental database is then constituted of the vector-valued experimental displacement fields $\ub^{\text{macro}}_{\text{exp}}$ and $\ub^{\text{meso}}_{\text{exp},1}, \dots, \ub^{\text{meso}}_{\text{exp},Q}$, respectively, at macroscale on $\Omega^{\text{macro}}_{\text{exp}}$ and at mesoscale on $\Omega^{\text{meso}}_{\text{exp}, 1},\dots,\Omega^{\text{meso}}_{\text{exp}, Q}$. The~experimental tensor-valued strain fields $\epsilonb^{\text{macro}}_{\text{exp}}$ and $\epsilonb^{\text{meso}}_{\text{exp}, 1}, \dots, \epsilonb^{\text{meso}}_{\text{exp}, Q}$, respectively associated to the experimental displacement fields $\ub^{\text{macro}}_{\text{exp}}$ and $\ub^{\text{meso}}_{\text{exp}, 1},\dots,\ub^{\text{meso}}_{\text{exp}, Q}$, can be calculated by post-processing through interpolation techniques.

\section{Prior Multiscale Stochastic Model and Its Hyperparameters}
\label{sec:stochastic_model}

At the macroscale, the specimen under test is modeled as a deterministic homogeneous linear elastic medium for which the effective mechanical properties are represented by a deterministic model of the fourth-order elasticity tensor $C^{\text{macro}}(\ab)$ that is independent of spatial position $\xb$ and parameterized by a vector $\ab$ belonging to an admissible set $\Ac^{\text{macro}}$. The vector-valued parameter $\ab$ is constituted of the algebraically independent coefficients spanning the macroscopic elasticity tensor $C^{\text{macro}}(\ab)$ having a given symmetry class induced by linear elastic material symmetries. At~the mesoscale, the specimen under test is modeled as a random heterogeneous linear elastic medium for which the apparent mechanical properties are represented by a {prior} stochastic model of the fourth-order tensor-valued random elasticity field. In Reference \cite{Soi06}, the ensemble SFE$^+$ of non-Gaussian second-order stationary random fields has been introduced and constructed in using the theory of information, the MaxEnt principle and the theory of random matrices. Such a family of tensor-valued random fields is completely parameterized by the values of their mean function, a dispersion coefficient usually denoted as $\delta$, and $d \, n (n+1)/2 = (d^3(d+1)^2 + 2\,d^2(d+1))/8 = 63$ possibly different spatial correlation lengths, with $d=3$ and $n = d(d+1)/2 = 6$ in 3D linear elasticity {(see References \cite{Soi06,Soi08a} for a definition of the spatial correlation lengths of a random field)}. All these parameters are independent of the spatial position $\xb$ since every tensor-valued random field in SFE$^+$ {is} second-order stationary on $\Rbb^3$ by construction. In addition, the dispersion coefficient $\delta$ introduced in Reference \cite{Soi06} is such that
\begin{equation}\label{dispersionparametercondition}
0 \leq \delta < \delta_{\text{sup}}, \quad \text{with} \quad \delta_{\text{sup}} = \sqrt{(n+1)/(n+5)} = \sqrt{7/11} \approx 0.7977 <1,
\end{equation}
where $n = d(d+1)/2 = 6$ with $d=3$ in 3D linear elasticity. Hence, any tensor-valued random field in SFE$^+$ has no statistical fluctuations when $\delta = 0$ and consequently its values are almost surely (a.s.) equal to its mean function. In addition, the level of statistical fluctuations of any tensor-valued random field in SFE$^+$ increases with the value of $\delta$. Consequently, the highest statistical fluctuations are obtained when $\delta = \delta_{\text{sup}}$. Ensemble SFE$^+$ has been especially constructed in Reference \cite{Soi06} for offering a {prior} stochastic model that can be used for modeling the tensor-valued apparent elasticity (or compliance) fields at mesoscale. Consequently, in this paper, we will use the same approach and the {prior} stochastic model of the elasticity tensor field $\Cb^{\text{meso}}$ (resp. the compliance tensor field $\Sb^{\text{meso}}$) will be defined as the restriction to a given bounded subdomain in $\Rbb^3$ of a random tensor field belonging to SFE$^+$ and indexed by $\Rbb^3$. The {prior} stochastic model of $\Cb^{\text{meso}}$ or $\Sb^{\text{meso}}$ can then be deduced from each other by inverse of each other. In this work, we will only consider the special case for which the spatial correlation structure of $\Cb^{\text{meso}}$ (resp. $\Sb^{\text{meso}}$) is defined by only $3$ (instead of $63$) different values $\ell_1,\ell_2,\ell_3$ for the spatial correlation lengths and consequently some of the $63$ spatial correlation lengths are mutually equal to each other. Furthermore, the mean function of $\Cb^{\text{meso}}$ (resp. $\Sb^{\text{meso}}$) can be represented by a set of $n_{\text{sym}}\leq n(n+1)/2$ parameters $\underline{h}_1,\dots,\underline{h}_{n_{\text{sym}}}$ that might have or not physical meaning in mechanical engineering such as Young's moduli, Poisson's ratios, bulk and shear moduli, and so forth (see~for instance Section~\ref{sec:validation}). Finally, the hyperparameters of the {prior} stochastic model of $\Cb^{\text{meso}}$ (resp. $\Sb^{\text{meso}}$) are $\delta$, $\ell_1,\ell_2,\ell_3$ and $\underline{h}_1,\dots,\underline{h}_{n_{\text{sym}}}$ that can be gathered into the vector-valued hyperparameter $\bb = (\delta,\ellb,\underline{\hb})$ in which $\ellb = (\ell_1,\ell_2,\ell_3)$ and $\underline{\hb} = (\underline{h}_1,\dots,\underline{h}_{n_{\text{sym}}})$. Hereinafter, the set of all the admissible values of vector $\underline{\hb}$ is denoted by $\Hc^{\text{meso}}$ and the admissible set of vector $\bb$ is denoted by $\Bc^{\text{meso}}$.

\section{Objectives and Strategy for Solving the Multiscale Statistical Inverse Problem} 
\label{sec:objectives_strategy}

\subsection{Objectives of the Multiscale Statistical Inverse Problem}
\label{sec:objectives}

The deterministic model of $C^{\text{macro}}(\ab)$ at macroscale and the {prior} stochastic model of $\Cb^{\text{meso}}(\bb)$ at mesoscale have to be identified by calculating the optimal values $\ab^{\text{macro}}$ and $\bb^{\text{meso}}$ of the vector-valued parameter $\ab \in \Ac^{\text{macro}}$ and the vector-valued hyperparameter $\bb \in \Bc^{\text{meso}}$, respectively, according to the experimental kinematic field measurements available at both macroscale and mesoscale. While the vector-valued parameter $\ab$ can completely be identified by solving a usual deterministic inverse problem using only the available experimental field measurements at macroscale, the~vector-valued hyperparameter $\bb=(\delta,\ellb,\underline{\hb})$ cannot directly be identified by solving a statistical inverse problem using only the available experimental field measurements at mesoscale. More precisely, the~dispersion parameter $\delta$ and the vector of spatial correlation lengths $\ellb$ require only experimental field measurements at mesoscale to be identified, whereas the vector $\underline{\hb}$ requires additional experimental field measurements at macroscale to be identified. {Indeed, the hyperparameters $\delta$ and $\ellb$ controlling respectively the level of statistical fluctuations and the spatial correlation structure of the random elasticity field require experimental field measurements with a sufficiently fine spatial resolution to be identified, while the hyperparameters $\underline{\hb}$ representing the mean elasticity field would require the experimental values of the stress fields associated with the kinematic (displacement or strain) fields observed experimentally at mesoscale to be identified, but these values are not available in practice.} The complete statistical information on random field $\Cb^{\text{meso}}(\bb)$ must then be transferred to the macroscale in order to identify its mean function $\underline{C}^{\text{meso}}$ using the available experimental field measurements at macroscale. A natural choice for such a transfer of information consists in computing the effective elasticity tensor $\Cb^{\text{eff}}(\bb)$ by a computational stochastic homogenization method and in comparing it with the previously identified elasticity tensor $C^{\text{macro}}(\ab)$. Thus, unlike the vector-valued parameter $\ab$, the vector-valued hyperparameter $\bb$ requires multiscale experimental field measurements (at macroscale and mesoscale) to be completely identified, thus leading to a challenging multiscale statistical inverse problem to be solved. Since by assumption only a single specimen is experimentally tested under a given static external loading applied at macroscale, the experimental field measurements must be performed simultaneously at both macroscale and mesoscale on the single test specimen, but~they do not need to be performed on the whole domain of the specimen.

\subsection{Strategy for Solving the Multiscale Statistical Inverse Problem} 
\label{sec:strategy}

Due to the major difficulties stated above and induced by the complexity of the challenging multiscale statistical inverse problem to be solved, a first complete methodology concerning such a multiscale identification has been recently proposed in Reference \cite{Ngu15}, in which a multiscale statistical inverse identification strategy is introduced and developed for an elastic microstructure with heterogeneous anisotropic statistical fluctuations within the framework of 3D linear elasticity theory. The proposed strategy allows for the identification of (i) the optimal value $\ab^{\text{macro}}$ of vector-valued parameter $\ab$, and (ii) the optimal value $\bb^{\text{meso}}$ of vector-valued hyperparameter $\bb$, by using the experimental displacement field measurements at both macroscale and mesoscale. The multiscale experimental identification methodology originally developed in Reference \cite{Ngu15} consists in introducing and constructing three different numerical indicators allowing the multiscale statistical inverse problem to be formulated as a multi-objective optimization problem. In the present work, we develop an improved multiscale experimental identification methodology involving four numerical indicators that are sensitive to the variation of the {parameters and} hyperparameters to be identified, which are:
\begin{enumerate}[leftmargin=*,labelsep=5mm]
  \item A macroscopic numerical indicator $\Jc^{\text{macro}}(\ab)$, dedicated to the identification of parameter $\ab$, that~allows for quantifying the distance between the experimental strain field $\epsilonb^{\text{macro}}_{\text{exp}}$ associated to the experimental displacement field $\ub^{\text{macro}}_{\text{exp}}$ measured at macroscale in the macroscopic domain $\Omega_{\text{exp}}^{\text{macro}}$ and the strain field $\epsilonb^{\text{macro}}(\ab)$ associated to the displacement field $\ub^{\text{macro}}(\ab)$ computed from a deterministic homogeneous linear elasticity boundary value problem (with both Dirichlet and Neumann boundary conditions) that models the experimental test configuration at macroscale and involves the unknown deterministic elasticity tensor $C^{\text{macro}}(\ab)$;
  
  \item A mesoscopic numerical indicator $\Jc^{\text{meso}}_{\delta}(\bb)$, dedicated to the identification of hyperparameter $\delta$, that allows for quantifying the distance between a pseudo-dispersion coefficient $\delta^{\epsilonb}_{\text{exp}}$ modeling the level of spatial fluctuations of the experimental strain field $\epsilonb^{\text{meso}}_{\text{exp}}$ associated to the experimental displacement field $\ub^{\text{meso}}_{\text{exp}}$ measured at mesoscale in a mesoscopic domain of observation $\Omega_{\text{exp}}^{\text{meso}}$, and a random pseudo-dispersion coefficient $D^{\Ecb}(\bb)$ representing the level of statistical fluctuations of the random strain field $\Ecb^{\text{meso}}(\bb)$ associated to the random displacement field $\Ub^{\text{meso}}(\bb)$ computed from a stochastic heterogeneous linear elasticity boundary value problem (with only Dirichlet boundary conditions) that models the experimental test configuration at mesoscale and involves the random elasticity tensor field $\Cb^{\text{meso}}(\bb)$ with an unknown level of statistical fluctuations $\delta$ that must be identified;
    
  \item Another mesoscopic numerical indicator $\Jc^{\text{meso}}_{\ellb}(\bb)$, dedicated to the identification of hyperparameter $\ellb=(\ell_1, \ell_2, \ell_3)$, that allows for quantifying the distance between the $3$ different pseudo-spatial correlation lengths $\ell^{\epsilonb}_{\text{exp},1},\ell^{\epsilonb}_{\text{exp},2},\ell^{\epsilonb}_{\text{exp},3}$ of the experimental strain field $\epsilonb^{\text{meso}}_{\text{exp}}$ in each spatial direction, measured at mesoscale in a mesoscopic domain of observation $\Omega_{\text{exp}}^{\text{meso}}$, and the $3$ pseudo-spatial correlation lengths $L^{\Ecb}_1(\bb),L^{\Ecb}_2(\bb),L^{\Ecb}_3(\bb)$ of the random strain field $\Ecb^{\text{meso}}(\bb)$ in each spatial direction, computed from the same mesoscopic stochastic boundary value problem as for $\Jc^{\text{meso}}_{\delta}(\bb)$ for which the random elasticity tensor field $\Cb^{\text{meso}}(\bb)$ has a spatial correlation structure induced and characterized by an unknown vector of spatial correlation lengths $\ellb=(\ell_1, \ell_2, \ell_3)$ that must be identified;
    
  \item A multiscale numerical indicator $\Jc^{\text{multi}}_{\underline{\hb}}(\ab,\bb)$, dedicated to the identification of hyperparameter $\underline{\hb}$, that allows for quantifying the distance between the homogeneous deterministic elasticity tensor $C^{\text{macro}}(\ab)$ at macroscale and the effective elasticity tensor $\Cb^{\text{eff}}(\bb)$ resulting from a computational stochastic homogenization in a representative volume element $\Omega^{\text{RVE}}$ at mesoscale of the random elasticity tensor field $\Cb^{\text{meso}}(\bb)$ whose mean function $\underline{C}^{\text{meso}}$ is unknown and must be identify.
\end{enumerate}

The multiscale statistical inverse problem then consists in identifying the optimal values $\ab^{\text{macro}}$ and $\bb^{\text{meso}}$ of the parameters $\ab$ in $\Ac^{\text{macro}}$ and hyperparameters $\bb$ in $\Bc^{\text{meso}}$, respectively, by solving a multi-objective optimization problem that consists in minimizing the (vector-valued) multi-objective cost function $\Jcb(\ab,\bb) = \(\Jc^{\text{macro}}(\ab),\Jc^{\text{meso}}_{\delta}(\bb),\Jc^{\text{meso}}_{\ellb}(\bb),\Jc^{\text{multi}}_{\underline{\hb}}(\ab,\bb)\)$ involving the four aforementioned numerical indicators. However, for further computational savings, the multi-objective optimization problem can be decomposed into (i) a single-objective optimization problem that consists in minimizing $\Jc^{\text{macro}}(\ab)$ for identifying the optimal vector-valued parameter $\ab^{\text{macro}}$ using only the experimental field measurements at macroscale, and (ii) a multi-objective optimization problem that consists in minimizing $\Jcb^{\text{meso}}(\bb) = \(\Jc^{\text{meso}}_{\delta}(\bb),\Jc^{\text{meso}}_{\ellb}(\bb),\Jc^{\text{multi}}_{\underline{\hb}}(\ab^{\text{macro}},\bb)\)$ for identifying the optimal vector-valued hyperparameter $\bb^{\text{meso}}$ using the experimental field measurements at mesoscale and exploiting the optimal vector-valued parameter $\ab^{\text{macro}}$ previously identified at step (i).

\section{Construction of the Numerical Indicators for Solving the Multiscale Statistical Inverse Problem}
\label{sec:indicators}

In this section, the construction of the macroscopic, mesoscopic and multiscale numerical indicators for solving the multiscale statistical inverse problem is presented.

\subsection{Deterministic Macroscopic Boundary Value Problem for the Macroscopic Indicator}
\label{sec:deterministic_BVP}

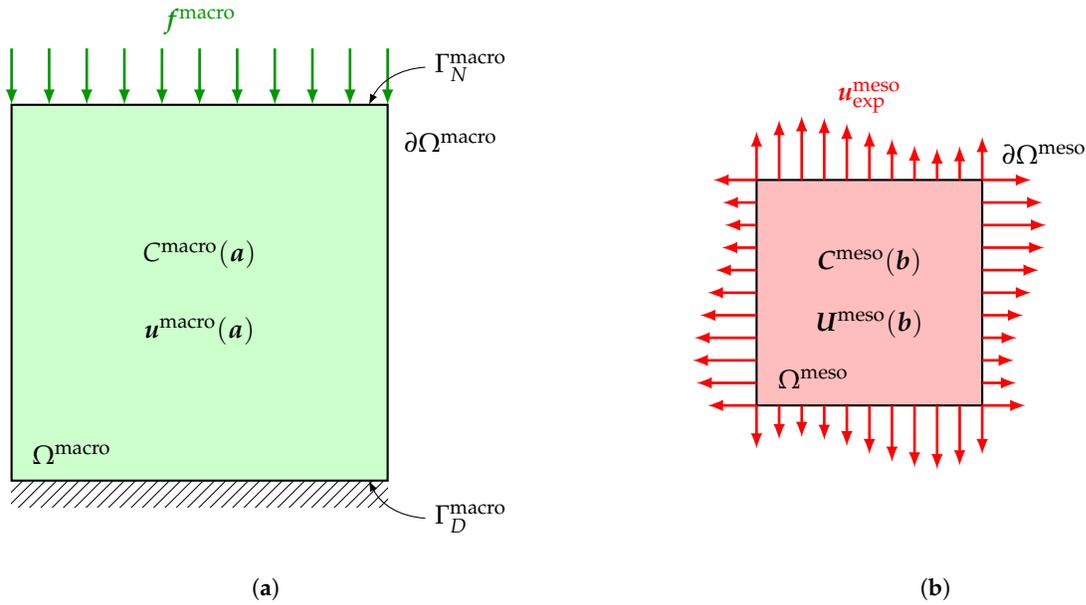
\begin{figure}[H]
\centering
\begin{subfigure}{0.45\textwidth}
	\centering
	\tikzsetnextfilename{boundary_value_problem_macro}
	\begin{tikzpicture}[every node/.style={minimum size=1cm},on grid]
\def \L {5}
\def \l {0.5}

\coordinate (A) at (0,0);
\coordinate (B) at (\L,\L);
\coordinate (C) at (0,\L);
\coordinate (D) at (\L,0);
\draw[black,thick,fill=green!20
] (A) rectangle (B);
\coordinate (lineloadVarA1) at ($ (C)!0!90:(B) $);
\coordinate (lineloadVarB1) at ($ (B)!0!-90:(C) $);
\coordinate (lineloadVarA2) at ($ (C)!{0.75cm}!90:(B) $);
\coordinate (lineloadVarB2) at ($ (B)!{0.75cm}!-90:(C) $);
\draw[green!60!black,force,->] (lineloadVarA2) -- (lineloadVarA1);
\draw[green!60!black,force,->] (lineloadVarB2) -- (lineloadVarB1);
\pgfmathsetmacro{\lineloadIntervalBegin}{\lineloadInterval/2/\scalingParameter}
\pgfmathsetmacro{\lineloadIntervalStep}{\lineloadInterval/2/\scalingParameter*2}
\pgfmathsetmacro{\lineloadIntervalEnd}{1-\lineloadInterval/2/\scalingParameter}
\foreach \i in {\lineloadIntervalBegin,\lineloadIntervalStep,...,\lineloadIntervalEnd}
\draw[green!60!black,force,->] ($(lineloadVarA2)!\i!(lineloadVarB2)$)-- ($(lineloadVarA1)!\i!(lineloadVarB1)$);
\node[green!60!black,above=0.6] at ($(B)!0.5!(C)$) {$\fb^{\text{macro}}$};
\begin{scope}
	\clip (0,-\supportBasicHeight) rectangle (\L,0);
	\draw[hatching](\L+1,0) -- ++(-\L-1,0);
\end{scope}
\node[right=-0.2] at ($(A)!0.07!(B)$) {$\Omega^{\text{macro}}$};
\coordinate (GN) at (\L+0.5,\L+0.5);
\draw[<-,>=latex] ($(C)!0.95!(B)$) to[bend left] (GN) node[right] {$\Gamma_N^{\text{macro}}$};
\coordinate (GD) at (\L+0.5,-0.5);
\draw[<-,>=latex] ($(A)!0.95!(D)$) to[bend right] (GD) node[right] {$\Gamma_D^{\text{macro}}$};
\node[right=0.1] at ($(B)!0.1!(D)$) {$\partial \Omega^{\text{macro}}$};
\node[above] at ($(A)!0.5!(B)$) {$C^{\text{macro}}(\ab)$};
\node[below] at ($(A)!0.5!(B)$) {$\ub^{\text{macro}}(\ab)$};

\end{tikzpicture}
	\caption{}
	\label{fig:boundary_value_problem_macro}
\end{subfigure}\hfill
\begin{subfigure}{0.45\textwidth}
	\centering
	\tikzsetnextfilename{boundary_value_problem_meso}
	\begin{tikzpicture}[every node/.style={minimum size=1cm},on grid]
\def \L {5}
\def \l {0.5}
\def \initforcevalue {0.65}
\def \endforcevalue {0.5}

\begin{scope}[white]
\coordinate (A) at (0,0);
\coordinate (B) at (\L,\L);
\coordinate (C) at (0,\L);
\coordinate (D) at (\L,0);
\draw[white,thick,fill=white
] (A) rectangle (B);
\coordinate (lineloadVarA1) at ($ (C)!0!90:(B) $);
\coordinate (lineloadVarB1) at ($ (B)!0!-90:(C) $);
\coordinate (lineloadVarA2) at ($ (C)!{0.75cm}!90:(B) $);
\coordinate (lineloadVarB2) at ($ (B)!{0.75cm}!-90:(C) $);
\draw[white,force,->] (lineloadVarA2) -- (lineloadVarA1);
\draw[white,force,->] (lineloadVarB2) -- (lineloadVarB1);
\pgfmathsetmacro{\lineloadIntervalBegin}{\lineloadInterval/2/\scalingParameter}
\pgfmathsetmacro{\lineloadIntervalStep}{\lineloadInterval/2/\scalingParameter*2}
\pgfmathsetmacro{\lineloadIntervalEnd}{1-\lineloadInterval/2/\scalingParameter}
\foreach \i in {\lineloadIntervalBegin,\lineloadIntervalStep,...,\lineloadIntervalEnd}
\draw[white,force,->] ($(lineloadVarA2)!\i!(lineloadVarB2)$)-- ($(lineloadVarA1)!\i!(lineloadVarB1)$);
\node[white,above=0.6] at ($(B)!0.5!(C)$) {$\fb^{\text{macro}}$};
\begin{scope}
	\clip (0,-\supportBasicHeight) rectangle (\L,0);
	\draw[hatching](\L+1,0) -- ++(-\L-1,0);
\end{scope}
\node[right=-0.2] at ($(A)!0.07!(B)$) {$\Omega^{\text{macro}}$};
\coordinate (GN) at (\L+0.5,\L+0.5);
\draw[<-,>=latex] ($(C)!0.95!(B)$) to[bend left] (GN) node[right] {$\Gamma_N^{\text{macro}}$};
\coordinate (GD) at (\L+0.5,-0.5);
\draw[<-,>=latex] ($(A)!0.95!(D)$) to[bend right] (GD) node[right] {$\Gamma_D^{\text{macro}}$};
\node[right=0.1] at ($(B)!0.1!(D)$) {$\partial \Omega^{\text{macro}}$};
\node[above] at ($(A)!0.5!(B)$) {$C^{\text{macro}}(\ab)$};
\node[below] at ($(A)!0.5!(B)$) {$\ub^{\text{macro}}(\ab)$};
\end{scope}

\coordinate (I) at (\L/5,\L/5);
\coordinate (J) at (4*\L/5,4*\L/5);
\coordinate (K) at (\L/5,4*\L/5);
\coordinate (L) at (4*\L/5,\L/5);
\draw[black,thick,fill=red!30,fill opacity=0.85] (I) rectangle (J);
\coordinate (lineloadVarA1) at ($ (K)!0!90:(J) $);
\coordinate (lineloadVarB1) at ($ (J)!0!-90:(K) $);
\coordinate (lineloadVarA2) at ($ (K)!{\initforcevalue cm}!90:(J) $);
\coordinate (lineloadVarB2) at ($ (J)!{\endforcevalue cm}!-90:(K) $);
\pgfmathsetmacro{\lineloadIntervalBegin}{\lineloadInterval/2/\scalingParameter}
\pgfmathsetmacro{\lineloadIntervalStep}{\lineloadInterval/2/\scalingParameter*2}
\pgfmathsetmacro{\lineloadIntervalEnd}{1-\lineloadInterval/2/\scalingParameter}
\foreach \i in {0,\lineloadIntervalBegin,\lineloadIntervalStep,...,\lineloadIntervalEnd,1} {
\coordinate (lineloadVarC1) at ($(lineloadVarA1)!\i!(lineloadVarB1)$);
\coordinate (lineloadVarC2) at ($(lineloadVarA1)!\i!(lineloadVarB1)+(0,{(\i-0.5)*(\i-0)*(4*(\i-1)+(\endforcevalue-\initforcevalue)/(1-0))+\initforcevalue})$);
\draw[red,force,<-] (lineloadVarC2)-- (lineloadVarC1);
}

\coordinate (lineloadVarA1) at ($ (J)!0!90:(L) $);
\coordinate (lineloadVarB1) at ($ (L)!0!-90:(J) $);
\coordinate (lineloadVarA2) at ($ (J)!{0.75cm}!90:(L) $);
\coordinate (lineloadVarB2) at ($ (L)!{0.5cm}!-90:(J) $);
\pgfmathsetmacro{\lineloadIntervalBegin}{\lineloadInterval/2/\scalingParameter}
\pgfmathsetmacro{\lineloadIntervalStep}{\lineloadInterval/2/\scalingParameter*2}
\pgfmathsetmacro{\lineloadIntervalEnd}{1-\lineloadInterval/2/\scalingParameter}
\foreach \i in {0,\lineloadIntervalBegin,\lineloadIntervalStep,...,\lineloadIntervalEnd,1} {
\coordinate (lineloadVarC1) at ($(lineloadVarA1)!\i!(lineloadVarB1)$);
\coordinate (lineloadVarC2) at ($(lineloadVarA1)!\i!(lineloadVarB1)+({(\i-0.5)*(\i-0)*(4*(\i-1)+(\endforcevalue-\initforcevalue)/(1-0))+\initforcevalue},0)$);
\draw[red,force,<-] (lineloadVarC2)-- (lineloadVarC1);
}

\coordinate (lineloadVarA1) at ($ (L)!0!90:(I) $);
\coordinate (lineloadVarB1) at ($ (I)!0!-90:(L) $);
\coordinate (lineloadVarA2) at ($ (L)!{0.75cm}!90:(I) $);
\coordinate (lineloadVarB2) at ($ (I)!{0.5cm}!-90:(L) $);
\pgfmathsetmacro{\lineloadIntervalBegin}{\lineloadInterval/2/\scalingParameter}
\pgfmathsetmacro{\lineloadIntervalStep}{\lineloadInterval/2/\scalingParameter*2}
\pgfmathsetmacro{\lineloadIntervalEnd}{1-\lineloadInterval/2/\scalingParameter}
\foreach \i in {0,\lineloadIntervalBegin,\lineloadIntervalStep,...,\lineloadIntervalEnd,1} {
\coordinate (lineloadVarC1) at ($(lineloadVarA1)!\i!(lineloadVarB1)$);
\coordinate (lineloadVarC2) at ($(lineloadVarA1)!\i!(lineloadVarB1)-(0,{(\i-0.5)*(\i-0)*(4*(\i-1)+(\endforcevalue-\initforcevalue)/(1-0))+\initforcevalue})$);
\draw[red,force,<-] (lineloadVarC2)-- (lineloadVarC1);
}

\coordinate (lineloadVarA1) at ($ (I)!0!90:(K) $);
\coordinate (lineloadVarB1) at ($ (K)!0!-90:(I) $);
\coordinate (lineloadVarA2) at ($ (I)!{0.75cm}!90:(K) $);
\coordinate (lineloadVarB2) at ($ (K)!{0.5cm}!-90:(I) $);
\pgfmathsetmacro{\lineloadIntervalBegin}{\lineloadInterval/2/\scalingParameter}
\pgfmathsetmacro{\lineloadIntervalStep}{\lineloadInterval/2/\scalingParameter*2}
\pgfmathsetmacro{\lineloadIntervalEnd}{1-\lineloadInterval/2/\scalingParameter}
\foreach \i in {0,\lineloadIntervalBegin,\lineloadIntervalStep,...,\lineloadIntervalEnd,1} {
\coordinate (lineloadVarC1) at ($(lineloadVarA1)!\i!(lineloadVarB1)$);
\coordinate (lineloadVarC2) at ($(lineloadVarA1)!\i!(lineloadVarB1)-({(\i-0.5)*(\i-0)*(4*(\i-1)+(\endforcevalue-\initforcevalue)/(1-0))+\initforcevalue},0)$);
\draw[red,force,<-] (lineloadVarC2)-- (lineloadVarC1);
}

\node[red,above=0.6] at ($(J)!0.5!(K)$) {$\ub^{\text{meso}}_{\text{exp}}$};
\node[right=-0.2] at ($(I)!0.12!(J)$) {$\Omega^{\text{meso}}$};
\node[above right=0.2] at ($(J)!0.1!(L)$) {$\partial \Omega^{\text{meso}}$};
\node[above=-0.1] at ($(I)!0.5!(J)$) {$\Cb^{\text{meso}}(\bb)$};
\node[below=-0.1] at ($(I)!0.5!(J)$) {$\Ub^{\text{meso}}(\bb)$};

\end{tikzpicture}
	\caption{}
	\label{fig:boundary_value_problem_meso}
\end{subfigure}
\caption{Boundary value problems at (\textbf{a}) macroscale and (\textbf{b}) mesoscale. (\textbf{a}) Deterministic boundary value problem characterized by deterministic elasticity tensor $C^{\text{macro}}(\ab)$ at macroscale: deterministic displacement field $\ub^{\text{macro}}(\ab)$ computed at macroscale in $\Omega^{\text{macro}}$; (\textbf{b}) Stochastic boundary value problem characterized by random elasticity tensor field $\Cb^{\text{meso}}(\bb)$ at mesoscale: random displacement field $\Ub^{\text{meso}}(\bb)$ computed at mesoscale in $\Omega^{\text{meso}}$.}\label{fig:boundary_value_problems}
\end{figure}

At macroscale, the deterministic boundary value problem modeling the experimental test configuration described in Section~\ref{sec:exp_config} is written over an open bounded domain $\Omega^{\text{macro}} \subset \Rbb^3$ with macroscopic dimensions of the specimen. The experimental domain of observation $\Omega_{\text{exp}}^{\text{macro}}$ is simulated as one given 2D or 3D part $\Omega_{\text{obs}}^{\text{macro}}$ of $\Omega^{\text{macro}}$. The boundary $\partial \Omega^{\text{macro}}$ of $\Omega^{\text{macro}}$ consists of two disjoint and complementary parts $\Gamma_N^{\text{macro}}$, on which Neumann boundary conditions are applied, and~$\Gamma_D^{\text{macro}}$, on which Dirichlet boundary conditions are applied, such that $\partial \Omega^{\text{macro}} = \overline{\Gamma_N^{\text{macro}} \cup \Gamma_D^{\text{macro}}}$ and $\Gamma_N^{\text{macro}} \cap \Gamma_D^{\text{macro}} = \emptyset$, with $\abs{\Gamma_D^{\text{macro}}} \neq 0$, where $\abs{\Gamma_D^{\text{macro}}}$ denotes the 2D measure of $\Gamma_D^{\text{macro}}$. A given deterministic surface force field $\fb^{\text{macro}}$ is applied on $\Gamma_N^{\text{macro}}$, while homogeneous Dirichlet conditions are applied on $\Gamma_D^{\text{macro}}$, so that there is no rigid body motion during the test. Within the context of linear elasticity theory, the deterministic boundary value problem at macroscale consists in finding the vector-valued displacement field $\ub^{\text{macro}}$ and the associated tensor-valued Cauchy stress field $\sigmab^{\text{macro}}$ satisfying the following equilibrium equations, stress-strain constitutive equation and Neumann and Dirichlet boundary conditions
\begin{alignat}{2}
-\divb(\sigmab^{\text{macro}}) &= \zerob &&\quad \text{in } \Omega^{\text{macro}}, \label{equilibriumequationmacro}\\ 
\sigmab^{\text{macro}} &= C^{\text{macro}}(\ab) : \epsilonb^{\text{macro}} &&\quad \text{in } \Omega^{\text{macro}}, \label{consititutiveequationmacro}\\
\scalprod{\sigmab^{\text{macro}}}{\nb^{\text{macro}}} &= \fb^{\text{macro}} &&\quad \text{on } \Gamma_N^{\text{macro}}, \label{Neumannbcmacro}\\ 
\ub^{\text{macro}} &= \zerob &&\quad \text{on } \Gamma_D^{\text{macro}}, \label{Dirichletbcmacro}
\end{alignat}
in which $\divb$ denotes the divergence operator of a second-order tensor-valued field with respect to $\xb${, the colon symbol $:$ denotes the classical twice contracted tensor product}, $\nb^{\text{macro}}$ is the unit normal vector to $\partial \Omega^{\text{macro}}$ pointing outward $\Omega^{\text{macro}}$ and $\epsilonb^{\text{macro}}$ is the classical tensor-valued strain field associated to displacement field $\ub^{\text{macro}}$ and defined by
\begin{equation}\label{kinematicrelationmacro}
\epsilonb^{\text{macro}} = \epsilonb(\ub^{\text{macro}}) = \frac{1}{2}\(\nablab \ub^{\text{macro}} + (\nablab \ub^{\text{macro}})^T\),
\end{equation}
in which $\epsilonb$ denotes the deterministic linear operator mapping the displacement field to the corresponding linearized strain field{, the superscript ${}^T$ denotes the transpose operator} and $\nablab$ denotes the gradient operator of a vector-valued field with respect to $\xb$. Recall that, as the material is assumed to be deterministic and homogeneous at macroscale, the unknown fourth-order deterministic elasticity tensor $C^{\text{macro}}(\ab)$ involved in constitutive Equation \eqref{consititutiveequationmacro} is independent of $\xb$ and parameterized by a parameter $\ab$ belonging to an admissible set $\Ac^{\text{macro}}$ depending on the considered material symmetry class. A sketch of the deterministic boundary value problem at macroscale is represented in Figure~\ref{fig:boundary_value_problems}a.

\subsection{Stochastic Mesoscopic Boundary Value Problem for the Mesoscopic Indicators}
\label{sec:stochastic_BVP}

At mesoscale, the stochastic boundary value problem modeling the experimental test configuration described in Section~\ref{sec:exp_config} is written over an open bounded domain $\Omega^{\text{meso}} \subset \Rbb^3$ with mesoscopic dimensions. A given domain of observation $\Omega_{\text{exp}}^{\text{meso}}$ corresponds to one given 2D or 3D part $\Omega_{\text{obs}}^{\text{meso}}$ of $\Omega^{\text{meso}}$. Within the context of linear elasticity theory, the stochastic boundary value problem at mesoscale consists in finding the vector-valued random displacement field $\Ub^{\text{meso}}$ and the associated tensor-valued random Cauchy stress field $\varSigmab^{\text{meso}}$ satisfying the following equilibrium equations, stress-strain constitutive equation and Dirichlet boundary conditions
\begin{alignat}{2}
-\divb(\varSigmab^{\text{meso}}) &= \zerob &&\quad \text{in } \Omega^{\text{meso}}, \label{equilibriumequationmeso}\\ 
\varSigmab^{\text{meso}} &= \Cb^{\text{meso}}(\bb) : \Ecb^{\text{meso}} &&\quad \text{in } \Omega^{\text{meso}}, \label{consititutiveequationmeso}\\
\Ub^{\text{meso}} &= \ub^{\text{meso}}_{\text{exp}} &&\quad \text{on } \partial \Omega^{\text{meso}}, \label{Dirichletbcmeso}
\end{alignat}
where $\Ecb^{\text{meso}}$ is the tensor-valued random strain field associated to random displacement field $\Ub^{\text{meso}}$ and defined by
\begin{equation}\label{kinematicrelationmeso}
\Ecb^{\text{meso}} = \epsilonb(\Ub^{\text{meso}}) = \frac{1}{2}\(\nablab \Ub^{\text{meso}} + (\nablab \Ub^{\text{meso}})^T\).
\end{equation}

Note that non-homogeneous Dirichlet boundary conditions \eqref{Dirichletbcmeso} are prescribed on the whole boundary $\partial \Omega^{\text{meso}}$ of $\Omega^{\text{meso}}$, which correspond to the displacement field $\ub^{\text{meso}}_{\text{exp}}$ that is experimentally measured over a given domain of observation $\Omega_{\text{exp}}^{\text{meso}}$ on the test specimen at mesoscale. Note also that~\eqref{consititutiveequationmeso} can equivalently be rewritten as
\begin{equation}\label{consititutiveequationmeso_compliance}
\varSigmab^{\text{meso}} = (\Sb^{\text{meso}}(\bb))^{-1} : \Ecb^{\text{meso}} \quad \text{in } \Omega^{\text{meso}},
\end{equation}
where $\Sb^{\text{meso}}(\bb) = (\Cb^{\text{meso}}(\bb))^{-1}$ is the random compliance tensor field of the considered material at mesoscale. For some linear elasticity problems, such as with 2D plane stress assumption, constitutive~Equation \eqref{consititutiveequationmeso_compliance} is more appropriate than \eqref{consititutiveequationmeso}.
A sketch of the stochastic boundary value problem at mesoscale is represented in Figure~\ref{fig:boundary_value_problems}b.

\subsection{Macroscopic Numerical Indicator}
\label{sec:macro_indicator}

Within the context of inverse identification, the optimal identified value $\ab^{\text{macro}}$ of parameter $\ab$ can be determined by exploiting the sensitivity of the model strain field $\epsilonb^{\text{macro}}$ with respect to $\ab$ and using the experimental strain field $\epsilonb^{\text{macro}}_{\text{exp}}$, which is obtained in $\Omega_{\text{exp}}^{\text{macro}}$ but can be rewritten in $\Omega^{\text{macro}}_{\text{obs}}$, through~the introduction of a macroscopic numerical indicator $\Jc^{\text{macro}}(\ab)$ defined for any vector $\ab\in \Ac^{\text{macro}}$ by
\begin{equation}\label{indicatormacro}
\Jc^{\text{macro}}(\ab) 
= \dfrac{1}{\abs{\Omega^{\text{macro}}_{\text{obs}}}} \int_{\Omega^{\text{macro}}_{\text{obs}}} \norm{\epsilonb^{\text{macro}}(\xb;\ab) - \epsilonb^{\text{macro}}_{\text{exp}}(\xb)}_F^2 \, d\xb,
\end{equation}
where $\abs{\Omega^{\text{macro}}_{\text{obs}}}$ denotes the measure of domain $\Omega^{\text{macro}}_{\text{obs}}$ and $
\Vert \cdot \Vert_F$ denotes the Frobenius (or~Hilbert-Schmidt) norm. Macroscopic numerical indicator $\Jc^{\text{macro}}(\ab)$ allows for quantifying the spatial average over the macroscopic domain $\Omega_{\text{obs}}^{\text{macro}}$ of the distance between the model strain field $\epsilonb^{\text{macro}}(\ab)$ and the experimental strain field $\epsilonb^{\text{macro}}_{\text{exp}}$ at macroscale. The optimal vector-valued parameter $\ab^{\text{macro}}$ can then be identified by minimizing $\Jc^{\text{macro}}(\ab)$ over all vector-valued parameter $\ab$ in $\Ac^{\text{macro}}$, provided~that the model strain field $\epsilonb^{\text{macro}}(\ab)$ computed by solving the deterministic boundary value problem~\eqref{equilibriumequationmacro}-\eqref{kinematicrelationmacro} is sufficiently sensitive to parameter $\ab$.

\subsection{Mesoscopic and Multiscale Numerical Indicators}
\label{sec:meso_indicators}

Within the context of statistical inverse identification, the optimal identified values $\bb^{\text{meso}} = (\delta^{\text{meso}},\ellb^{\text{meso}},\underline{\hb}^{\text{meso}})$ of $\bb=(\delta, \ellb, \underline{\hb})$ can be determined by exploiting the sensitivity of some quantities of interest of the stochastic boundary value problem~\eqref{equilibriumequationmeso}-\eqref{kinematicrelationmeso} with respect to $\delta$, $\ellb=(\ell_1,\ell_2,\ell_3)$ and $\underline{\hb}$, respectively, and using their counterparts coming from the experimental measurements through the introduction of two mesoscopic numerical indicators $\Jc^{\text{meso}}_{\delta}(\bb)$ and $\Jc^{\text{meso}}_{\ellb}(\bb)$ and one multiscale numerical indicator $\Jc^{\text{multi}}_{\underline{\hb}}(\ab,\bb)$.

\subsubsection{Mesoscopic Numerical Indicator Associated to the Dispersion Parameter}
\label{sec:meso_indicator_dispersion}

A first mesoscopic numerical indicator $\Jc^{\text{meso}}_{\delta}(\bb)$ is introduced to identify the dispersion parameter $\delta$ controlling the level of statistical fluctuations of random elasticity field $\Cb^{\text{meso}}(\bb)$ at mesoscale and defined for any vector $\bb\in \Bc^{\text{meso}}$ by
\begin{equation}\label{indicatormesodelta}
\Jc^{\text{meso}}_{\delta}(\bb) = \(\dfrac{\Ebb\{D^{\Ecb}(\bb)\} - \delta^{\epsilonb}_{\text{exp}}}{\delta^{\epsilonb}_{\text{exp}}}\)^2,
\end{equation}
where $\Ebb$ denotes the mathematical expectation, $D^{\Ecb}(\bb)$ is a positive-valued random variable that models the random level of spatial fluctuations of the random solution obtained by solving the stochastic boundary value problem~\eqref{equilibriumequationmeso}-\eqref{kinematicrelationmeso} at mesoscale and where $\delta^{\epsilonb}_{\text{exp}}$ is its counterpart for the experimental test specimen at mesoscale, such that
\begin{equation}\label{dispersionstrainfield}
D^{\Ecb}(\bb) = \dfrac{\sqrt{V^{\Ecb}(\bb)}}{\norm{\underline{\Ecb}^{\text{meso}}(\bb)}_F} \quad \text{and} \quad \delta^{\epsilonb}_{\text{exp}} = \dfrac{\sqrt{V^{\epsilonb}_{\text{exp}}}}{\norm{\underline{\epsilonb}^{\text{meso}}_{\text{exp}}}_F},
\end{equation}
where $\underline{\Ecb}^{\text{meso}}(\bb)$ and $\underline{\epsilonb}^{\text{meso}}_{\text{exp}}$ are the spatial averages of random strain field $\Ecb^{\text{meso}}(\bb)$ and experimental strain field $\epsilonb^{\text{meso}}_{\text{exp}}$, respectively, and where
\begin{align}
V^{\Ecb}(\bb) &= \dfrac{1}{\abs{\Omega^{\text{meso}}_{\text{obs}}}} \int_{\Omega^{\text{meso}}_{\text{obs}}} \norm{\Ecb^{\text{meso}}(\xb;\bb) - \underline{\Ecb}^{\text{meso}}(\bb)}_F^2 \, d\xb,\label{variancestrainfield_model}\\
V^{\epsilonb}_{\text{exp}} &= \dfrac{1}{\abs{\Omega^{\text{meso}}_{\text{obs}}}} \int_{\Omega^{\text{meso}}_{\text{obs}}} \norm{\epsilonb^{\text{meso}}_{\text{exp}}(\xb) - \underline{\epsilonb}^{\text{meso}}_{\text{exp}}}_F^2 \, d\xb,\label{variancestrainfield_exp}
\end{align}
where $\abs{\Omega^{\text{meso}}_{\text{obs}}}$ denotes the measure of domain $\Omega^{\text{meso}}_{\text{obs}}$. Note that it can easily be shown that $\underline{\Ecb}^{\text{meso}}(\bb) = \underline{\epsilonb}^{\text{meso}}_{\text{exp}}$ for all $\bb \in \Bc^{\text{meso}}$ a.s. and consequently $\underline{\Ecb}^{\text{meso}}(\bb)$ is a deterministic tensor. Also, since random strain field $\Ecb^{\text{meso}}(\bb)$ is \emph{a priori} nor statistically homogeneous neither ergodic in average, $\underline{\Ecb}^{\text{meso}}(\bb)$ does not correspond to the statistical mean function of $\Ecb^{\text{meso}}(\bb)$ and therefore $V^{\Ecb}(\bb)$ (resp. $D^{\Ecb}(\bb)$) does not correspond to the variance (resp. dispersion coefficient) of $\Ecb^{\text{meso}}(\bb)$.
The mesoscopic numerical indicator $\Jc^{\text{meso}}_{\delta}(\bb)$ defined by \eqref{indicatormesodelta} allows for quantifying the relative distance between the statistical mean value of $D^{\Ecb}(\bb)$ and its experimental observation $\delta^{\epsilonb}_{\text{exp}}$. It should also be noted that a mesoscopic numerical indicator similar to this one was introduced in Reference \cite{Ngu15}, but with different expressions than that of \eqref{indicatormesodelta}, \eqref{variancestrainfield_model} and \eqref{variancestrainfield_exp} for the definitions of $\Jc^{\text{meso}}_{\delta}(\bb)$ and $V^{\Ecb}(\bb)$, respectively.

\subsubsection{Mesoscopic Numerical Indicator Associated to the Spatial Correlation Lengths}
\label{sec:meso_indicator_corr}

A second mesoscopic numerical indicator $\Jc^{\text{meso}}_{\ellb}(\bb)$ is introduced to identify the vector of spatial correlation lengths $\ellb = (\ell_1,\ell_2,\ell_3)$ characterizing the spatial correlation structure of random elasticity field $\Cb^{\text{meso}}(\bb)$ (or random compliance field $\Sb^{\text{meso}}(\bb)$) and defined for any vector $\bb\in \Bc^{\text{meso}}$ by
\begin{equation}\label{indicatormesocorr}
\Jc^{\text{meso}}_{\ellb}(\bb) = \sum_{\alpha=1}^3 \(\dfrac{\Ebb\{L^{\Ecb}_{\alpha}(\bb)\} - \ell^{\epsilonb}_{\text{exp},\alpha}}{\ell^{\epsilonb}_{\text{exp},\alpha}}\)^2,
\end{equation}
where $L^{\Ecb}_{\alpha}(\bb)$ is a positive-valued random variable that models the spatial correlation length along the $\alpha$-th spatial direction (relative to the spatial coordinate $x_{\alpha}$) characterizing the spatial correlation structure of the statistical fluctuations of random strain field $\Ecb^{\text{meso}}(\bb)$ and where $\ell^{\epsilonb}_{\text{exp},\alpha}$ is its observation for the experimental test specimen at mesoscale. Usual signal processing methods (such as the periodogram method) are used for estimating $L^{\Ecb}_{\alpha}(\bb)$ and $\ell^{\epsilonb}_{\text{exp},\alpha}$ by considering the approximation that they are independent of $\xb$ which is not the case since $\Ecb^{\text{meso}}(\bb)$ and $\epsilonb^{\text{meso}}_{\text{exp}}$ are usually not statistically homogeneous because of the non-homogeneous Dirichlet boundary conditions \eqref{Dirichletbcmeso} involving the experimental displacement field $\ub^{\text{meso}}_{\text{exp}}$ on $\partial \Omega^{\text{meso}}$. 
The mesoscopic numerical indicator $\Jc^{\text{meso}}_{\ellb}(\bb)$ defined by \eqref{indicatormesocorr} allows for quantifying the relative distance between the statistical mean values of 
$L^{\Ecb}_1(\bb),L^{\Ecb}_2(\bb),L^{\Ecb}_3(\bb)$ 
and their experimental observations $\ell^{\epsilonb}_{\text{exp},1},\ell^{\epsilonb}_{\text{exp},2},\ell^{\epsilonb}_{\text{exp},3}$. 

\subsubsection{Multiscale Numerical Indicator Associated to Computational Stochastic Homogenization}
\label{sec:macro_meso_indicator_homog}

A multiscale numerical indicator $\Jc^{\text{multi}}_{\underline{\hb}}(\ab,\bb)$ is introduced to identify the mean function $\underline{C}^{\text{meso}}(\bb)$ of the random elasticity field $\Cb^{\text{meso}}(\bb)$ at mesoscale and defined for any vector $\ab \in \Ac^{\text{macro}}$ and any vector $\bb \in \Bc^{\text{meso}}$ by
\begin{equation}\label{indicatormulti}
\Jc^{\text{multi}}_{\underline{\hb}}(\ab,\bb) = \(\dfrac{\norm{\Ebb\{\Cb^{\text{eff}}(\bb)\} - C^{\text{macro}}(\ab)}_F}{\norm{C^{\text{macro}}(\ab)}_F}\)^2,
\end{equation}
where $\Cb^{\text{eff}}(\bb)$ is the effective elasticity tensor constructed by computational stochastic homogenization of $\Cb^{\text{meso}}(\bb)$ in an open bounded mesoscopic domain $\Omega^{\text{RVE}}$, which is assumed to be a representative volume element. It should be noted that, under scale separation assumption, $\Cb^{\text{eff}}(\bb)$ is actually a random tensor for which the level of statistical fluctuations tends to zero when the size of domain $\Omega^{\text{RVE}}$ tends to infinity \cite{Soi08a,Gui11a,Soi17a}. This is the reason why the statistical mean value $\Ebb\{\Cb^{\text{eff}}(\bb)\}$ has been considered in the definition \eqref{indicatormulti} of $\Jc^{\text{multi}}_{\underline{\hb}}(\ab,\bb)$ instead of the effective elasticity tensor $\Cb^{\text{eff}}(\bb)$ itself. The multiscale indicator $\Jc^{\text{multi}}_{\underline{\hb}}(\ab,\bb)$ defined by \eqref{indicatormulti} allows for quantifying the relative distance between (i) the macroscopic elasticity tensor $C^{\text{macro}}(\ab)$ involved in the deterministic boundary value problem~\eqref{equilibriumequationmacro}-\eqref{kinematicrelationmacro} at macroscale, and (ii) the statistical mean value of the effective elasticity tensor $\Cb^{\text{eff}}(\bb)$ calculated by a computational stochastic homogenization method in the mesoscopic subdomain $\Omega^{\text{RVE}}$ of the random elasticity field $\Cb^{\text{meso}}(\bb)$ involved in the stochastic boundary value problem~\eqref{equilibriumequationmeso}-\eqref{kinematicrelationmeso} at~mesoscale.

\subsection{Comments}
\label{sec:improvements}

It should be noted that in the original formulation initially proposed \cite{Ngu15}, the numerical indicator $\Jc^{\text{meso}}_{\ellb}(\bb)$ was not introduced. The improved formulation proposed in the present work is more advanced than the original formulation initially proposed in Reference \cite{Ngu15} to the extent that it involves an additional mesoscopic numerical indicator, namely $\Jc^{\text{meso}}_{\ellb}(\bb)$, so that the parameter $\ab$ and the three components $\delta$, $\ellb$ and $\underline{\hb}$ of the hyperparameter $\bb$ each have their own dedicated numerical indicator. Thus, the number of single-objective cost functions being equal to the number of parameters to optimize, it is possible to substitute the computationally expensive global search algorithm used in~Reference \cite{Ngu15}, which belongs to the class of random search, genetic and evolutionary algorithms \cite{DaCun67,Cen77,Yu85,Dauer86,Gol89,Deb01a,Mar04,Kon06,Coe06,Coe07,Deb14}, with~a more computationally efficient optimization algorithm, such as the fixed-point iterative algorithm considered in the present work {(see Section~\ref{sec:solving_optim_problem})}. Indeed, even using parallel processing and computing tools, the computational cost incurred by the global optimization algorithm (genetic algorithm) used in Reference \cite{Ngu15} remains high due to the large stochastic dimension of the tensor-valued random elasticity field $\Cb^{\text{meso}}(\bb)$, so that the multi-objective optimization problem can be numerically intractable, with the current available computer resources, in very high stochastic dimension for large-scale (non-)linear computational models of three-dimensional random microstructures. {The computational cost of the genetic algorithm is compared to the one of the fixed-point iterative algorithm in terms of the number of evaluations of the stochastic computational model in the 2D validation example presented in Section~\ref{sec:validation_2D}. It provides a measure of the computational efficiency that is independent of the computer hardware used to perform the numerical simulations.} Lastly, it should be noted that an alternative mesoscopic numerical indicator $\Jc^{\text{meso}}_{\delta}(\bb)$ is used compared to the previous work in Reference \cite{Ngu15} without degrading the performance in terms of accuracy.

\section{Multiscale Statistical Inverse Problem Formulated as a Multi-Objective Optimization Problem}
\label{sec:formulation_optim_problem}

The multiscale statistical inverse identification of parameter $\ab$ and hyperparameter $\bb$ can be performed simultaneously by formulating the multiscale statistical inverse problem as a multi-objective optimization problem, that is
\begin{equation}\label{optimizationpbmin}
(\ab^{\text{macro}},\bb^{\text{meso}}) = \argmin_{\ab \in \Ac^{\text{macro}}, \bb \in \Bc^{\text{meso}}} \Jcb(\ab,\bb),
\end{equation}
where $\Jcb(\ab,\bb)$ is the (vector-valued) multi-objective cost function consisting of the four aforementioned numerical indicators as single-objective cost functions and defined for any vector $\ab \in \Ac^{\text{macro}}$ and any vector $\bb \in \Bc^{\text{meso}}$ by
\begin{equation}\label{multiobjectivecostfunction}
\Jcb(\ab,\bb) = \(\Jc^{\text{macro}}(\ab),\Jc^{\text{meso}}_{\delta}(\bb),\Jc^{\text{meso}}_{\ellb}(\bb),\Jc^{\text{multi}}_{\underline{\hb}}(\ab,\bb)\).
\end{equation}

In accordance with the strategy for solving the multiscale statistical inverse problem (see~Section~\ref{sec:strategy}), for a better computational efficiency, the multiscale statistical inverse identification of $\ab$ and $\bb$ is performed sequentially by splitting the multi-objective optimization problem into two subproblems solved one after the other:
\begin{enumerate}[leftmargin=*,labelsep=5mm]
\item a macroscale inverse problem formulated as a single-objective optimization problem that consists in calculating the optimal value $\ab^{\text{macro}}$ of parameter $\ab$ in $\Ac^{\text{macro}}$ that minimizes the macroscopic numerical indicator $\Jc^{\text{macro}}(\ab)$, that is
\begin{equation}\label{optimizationpbmacro}
\ab^{\text{macro}} = \argmin_{\ab\in \Ac^{\text{macro}}} \Jc^{\text{macro}}(\ab);
\end{equation}

\item a mesoscale statistical inverse problem formulated as a multi-objective optimization problem that consists in calculating the optimal value $\bb^{\text{meso}} $ of hyperparameter $\bb$ in $\Bc^{\text{meso}}$ that minimizes the two mesoscopic numerical indicators $\Jc^{\text{meso}}_{\delta}(\bb)$ and $\Jc^{\text{meso}}_{\ellb}(\bb)$ as well as the multiscale numerical indicator $\Jc^{\text{multi}}_{\underline{\hb}}(\ab^{\text{macro}},\bb)$ simultaneously, that is
\begin{equation}\label{optimizationpbmesomin}
\bb^{\text{meso}} = \argmin_{\bb\in \Bc^{\text{meso}}} \Jcb^{\text{meso}}(\bb),
\end{equation}
where $\Jcb^{\text{meso}}(\bb)$ is the (vector-valued) multi-objective cost function defined for any vector $\bb \in \Bc^{\text{meso}}$ by
\begin{equation}\label{multiobjectivecostfunctionmeso}
\Jcb^{\text{meso}}(\bb) = \(\Jc^{\text{meso}}_{\delta}(\bb),\Jc^{\text{meso}}_{\ellb}(\bb),\Jc^{\text{multi}}_{\underline{\hb}}(\ab^{\text{macro}},\bb)\).
\end{equation}
\end{enumerate}

\section{Numerical Methods for Solving the Multi-Objective Optimization Problem}
\label{sec:solving_optim_problem}

The deterministic boundary value problem~\eqref{equilibriumequationmacro}-\eqref{kinematicrelationmacro} defined on domain $\Omega^{\text{macro}}$ at macroscale and the stochastic boundary value problem~\eqref{equilibriumequationmeso}-\eqref{kinematicrelationmeso} defined on a subdomain $\Omega^{\text{meso}} \subset \Omega^{\text{macro}}$ at mesoscale are both discretized using a classical displacement-based finite element method (FEM) \cite{Hug87, Zie05}. The mathematical expectations of the quantities of interest of the stochastic boundary value problem~\eqref{equilibriumequationmeso}-\eqref{kinematicrelationmeso} involved in the three numerical indicators $\Jc^{\text{meso}}_{\delta}(\bb)$, $\Jc^{\text{meso}}_{\ellb}(\bb)$ and $\Jc^{\text{multi}}_{\underline{\hb}}(\ab^{\text{macro}},\bb)$ are estimated using the Monte Carlo numerical simulation method \cite{Kal86,Fis96,Cal98,Sch01, Rub16} with $N_s$ independent realizations $\set{\Cb^{\text{meso}}(\theta_r)}_{1\leq r \leq N_s}$ of $\Cb^{\text{meso}}$. For the computation of the optimal value $\ab^{\text{macro}}$, the classical single-objective optimization problem~\eqref{optimizationpbmacro} is solved using the Nelder-Mead simplex algorithm~\cite{Nel65, Walt91, Lag98, McKin98, Kol03}. For the computation of the optimal value $\bb^{\text{meso}}$, the non-trivial multi-objective optimization problem~\eqref{optimizationpbmesomin} does not admit a single global optimal solution, but~inherently gives rise to a set of optimal solutions (called Pareto optima) resulting from a trade-off among the three components $\Jc^{\text{meso}}_{\delta}(\bb)$, $\Jc^{\text{meso}}_{\ellb}(\bb)$ and $\Jc^{\text{multi}}_{\underline{\hb}}(\ab^{\text{macro}},\bb)$ of the multi-objective cost function $\Jcb^{\text{meso}}(\bb)$ which are competing and \emph{a priori} conflicting. Based on the concept of noninferiority~\cite{Zad63} (also called Pareto optimality) for characterizing the components of a multi-objective function, a~noninferior (or Pareto optimal) solution is such that an improvement in any objective function requires a degradation of some of the other objective functions, whereas an inferior solution is such that an improvement can be attained in all the objective functions. The set of all the noninferior solutions in the parameter space is called the Pareto optimal set and the corresponding objective function values in the multidimensional objective function space is called the Pareto optimal front. The interested reader can refer to~\mbox{References \cite{Deb01a,Mar04,Kon06,Coe06,Coe07,Deb14,Sen17}} and the references therein for an overview of nonlinear multi-objective optimization methods including the fundamental principles, some Pareto (near-)optimality conditions and a number of traditional and evolutionary optimization algorithms. In Reference \cite{Ngu15}, the multi-objective optimization problem under consideration has been successfully solved by using the genetic algorithm~\cite{Deb01a,Deb14} that allows for constructing and finding a set of local Pareto optimal solutions that should be sufficiently representative of the whole Pareto optimal set and as many and diverse as possible for further selection \cite{Kon06,Sen17}. The best compromise optimal solution is selected among all the potential Pareto optimal solutions as the one that minimizes the distance to a utopian solution that is constituted by the individual optimal solutions of the conflicting components of the multi-objective function, which corresponds to the origin of the Pareto front.

In the present work, a dedicated numerical indicator has been set up specifically for each component of hyperparameter $\bb=(\delta,\ellb,\underline{\hb})$, allowing for the use of a simpler and more efficient multi-objective optimization algorithm, namely a fixed-point iterative algorithm. Starting from an \emph{ad hoc} initial guess, it consists in sequentially minimizing $\Jc^{\text{meso}}_{\delta}(\bb)$, $\Jc^{\text{meso}}_{\ellb}(\bb)$ and $\Jc^{\text{multi}}_{\underline{\hb}}(\ab^{\text{macro}},\bb)$ respectively with respect to $\delta$, $\ellb$ and $\underline{\hb}$ in their sets of admissible values that are such that $\bb=(\delta,\ellb,\underline{\hb})$ belongs to $\Bc^{\text{meso}}$. The iterative process is stopped when the residual norm between two iterates becomes lower than a user-specified prescribed tolerance for each of the three single-objective optimization problems. Numerical results have shown that, for the problem under consideration, such~a fixed-point iterative algorithm can achieve the same precision as the genetic algorithm in terms of convergence but with a lower overall computational cost (see the numerical examples in Sections~\ref{sec:validation} and \ref{sec:application}). The main drawback of such a numerical optimization algorithm lies in the choice of the initial values used to start the algorithm that may be critical for the localization of the final global convergence region. {Besides, note that the fixed-point iterative method introduced in this work could \emph{a priori} be applied to the original formulation proposed in Reference \cite{Ngu15}, but it would lead to minimize the objective function $\Jc^{\text{meso}}_{\delta}(\bb)$ with respect to $\delta$ and $\ellb$ simultaneously given the other hyperparameters $\underline{\hb}$. Although it is possible, the problem is that $\Jc^{\text{meso}}_{\delta}(\bb)$ is very sensitive to $\delta$ but less sensitive with respect to $\ellb$, since it has been tailored to perform the identification of the optimal value $\delta^{\text{meso}}$ of $\delta$ and not the one $\ellb^{\text{meso}}$ of $\ellb$. Consequently, using such a fixed-point iterative strategy would yield uncertainties on the identified value $\ellb^{\text{meso}}$ of $\ellb$. It is the reason why the additional objective function $\Jc^{\text{meso}}_{\ellb}(\bb)$ has been introduced and for which the sensitivity is of first order with respect to $\ellb$ and of second order with respect to $\delta$.}

\section{Probabilistic Model for a Robust Identification of the Hyperparameters}
\label{sec:randomization}

When several non-overlapping mesoscopic domains of observation $\Omega^{\text{meso}}_{\text{exp},1},\dots,\Omega^{\text{meso}}_{\text{exp},Q}$ are available for experimental measurements for the same test specimen instead of a unique observation domain $\Omega^{\text{meso}}_{\text{exp}}$, then the solution of the multi-objective optimization problem presented in Section~\ref{sec:formulation_optim_problem} can yield different optimal values $\bb^{\text{meso}}_1,\dots,\bb^{\text{meso}}_Q$ of hyperparameter $\bb^{\text{meso}}$ when experimental data comes from one mesoscopic domain of observation to another since mesoscopic indicators $\Jc^{\text{meso}}_{\delta}(\bb)$ and $\Jc^{\text{meso}}_{\ellb}(\bb)$ depend on the values of experimental displacement fields $\ub^{\text{meso}}_{\text{exp},1},\dots,\ub^{\text{meso}}_{\text{exp},Q}$ that are measured on each of them. Consequently, the optimal value $\bb^{\text{meso}}$ of hyperparameter $\bb$ should be considered as uncertain and should be modeled as a vector-valued random variable $\Bb = \(D,\Lb,\underline{\Hb}\)$ for which $\bb^{\text{meso}}_1,\dots,\bb^{\text{meso}}_Q$ are assumed to be $Q$ independent realizations. Thus, in Reference \cite{Ngu15}, a robust identification of the optimal value $\bb^{\text{opt}}$ is proposed by averaging the identified values $\bb^{\text{meso}}_1,\dots,\bb^{\text{meso}}_Q$. Nevertheless, in the present work, an improved strategy is proposed that consists in constructing a {prior} stochastic model of the vector-valued hyperparameter $\Bb$ by using the MaxEnt principle \cite{Jay57a,Jay57b,Jay03,Soi17a} and the available information allowing for the explicit construction and parametric representation of the probability density function $p_{\Bb} \colon \bb \mapsto p_{\Bb}(\bb)$ of random vector $\Bb$. A robust identified value $\bb^{\text{opt}}$ is finally obtained using the MLE method \cite{Ser80,Pap02,Spa05a,Soi17a} with the independent realizations $\bb^{\text{meso}}_1,\dots,\bb^{\text{meso}}_Q$. The available information for constructing the {prior} stochastic model of $\Bb$ is as follows: (i) random variables $D$, $\Lb$ and $\underline{\Hb}$ are mutually statistically independent, (ii) random variable $D$ takes its values a.s. in $\intervaloo{0}{\delta_{\text{sup}}}$ with $\delta_{\text{sup}} = \sqrt{(n+1)/(n+5)} = \sqrt{7/11} \approx 0.7977 < 1$ (with $n=6$ in linear elasticity), (iii)~the random components of random vector $\Lb$ are (statistically independent) positive-valued random variables a.s. for which {the mean value is given in $\intervaloo{0}{+\infty}$ and} the values are unlikely near zero by construction of the mesoscale modeling, otherwise it would mean the current scale of the computational model is not correct and too large, (iv) the random components of random vector $\underline{\Hb}$ take their values a.s. in the admissible set $\Hc^{\text{meso}}$. We then have for all $\bb = (\delta,\ellb,\hb) \in \Bc^{\text{meso}}$,
\begin{equation}\label{pdfB}
p_{\Bb}(\bb) = p_{D}(\delta)\,p_{\Lb}(\ellb)\, p_{\underline{\Hb}}(\hb),
\end{equation}
where
\begin{equation}\label{pdfDelta}
p_{D}(\delta) = \dfrac{1}{\delta_{\text{sup}}} \onebb_{\intervaloo{0}{\delta_{\text{sup}}}}(\delta),
\end{equation}
\begin{equation}\label{pdfLcorr}
p_{\Lb}(\ellb) = \prod_{\alpha=1}^3 p_{L_{\alpha}}(\ell_{\alpha}) \quad \text{with}\quad p_{L_{\alpha}}(\ell_{\alpha}) = \onebb_{\intervaloo{0}{+\infty}}(\ell_{\alpha}) \dfrac{1}{{b_{\alpha}}^{a_{\alpha}} \Gamma(a_{\alpha})} {\ell_{\alpha}}^{a_{\alpha}-1} \exp(-\ell_{\alpha}/b_{\alpha}).
\end{equation}
in which $\onebb_{\intervaloo{0}{\delta_{\text{sup}}}}$ is the indicator function of the interval $\intervaloo{0}{\delta_{\text{sup}}}$ such that $\onebb_{\intervaloo{0}{\delta_{\text{sup}}}}(\delta) = 1$ if $\delta \in \intervaloo{0}{\delta_{\text{sup}}}$ and $\onebb_{\intervaloo{0}{\delta_{\text{sup}}}}(\delta) = 0$ if $\delta \not\in \intervaloo{0}{\delta_{\text{sup}}}$, where $\ssb_1 = (a_1,b_1),\ssb_2 = (a_2,b_2),\ssb_3 = (a_3,b_3)$ are positive parameters to be identified. We refer the reader to Reference \cite{Gui13a} for a detailed construction of the {prior} stochastic model of $\underline{\Hb}$ and a rigorous characterization of the statistical dependence between the components of random elasticity tensors exhibiting a.s. some given material symmetry properties for the six highest levels of linear elastic symmetries. For the special case of isotropic materials, we~have $\Hc^{\text{meso}} = \intervaloo{0}{+\infty} \times \intervaloo{0}{+\infty}$ and the {prior} probability density function $p_{\underline{\Hb}}$ of random vector $\underline{\Hb}$ is written as for all $\hb = (h_1,h_2) \in \Hc^{\text{meso}}$,
\begin{equation}\label{pdfC}
p_{\underline{\Hb}}(\hb) = p_{\underline{H}_1}(h_1) \times p_{\underline{H}_2}(h_2),
\end{equation}
in which
\begin{align}
p_{\underline{H}_1}(h_1) &= \onebb_{\Rbb^+}(h_1) k_1 {h_1}^{-\lambda} \exp\(-\lambda_1 h_1\),\label{pdfC1}\\
p_{\underline{H}_2}(h_2) &= \onebb_{\Rbb^+}(h_2) k_2 {h_2}^{-5\lambda} \exp\(-\lambda_2 h_2\),\label{pdfC2}
\end{align}
where $k_1 = {\lambda_1}^{1-\lambda} / \Gamma(1-\lambda)$ and $k_2 = {\lambda_2}^{1-5\lambda} / \Gamma(1-5\lambda)$ are two positive normalization constants. The probabilistic model of $\underline{\Hb}$ is then parameterized by the vector-valued hyperparameter $\underline{\ssb} = (\lambda,\lambda_1,\lambda_2) \in \intervaloo{-\infty}{1/5} \times \intervaloo{0}{+\infty}^2$. {The mean values of $\underline{H}_1$ and $\underline{H}_2$ are respectively equal to $(1-\lambda)/\lambda_1$ and $(1-5\lambda)/\lambda_2$, and the dispersion coefficients of $\underline{H}_1$ and $\underline{H}_2$ are respectively equal to $1/\sqrt{1-\lambda}$ and $1/\sqrt{1-5\lambda}$. Note that the probability density functions of $\underline{H}_1$ and $\underline{H}_2$ both involve the same hyperparameter $\lambda<1/5$ that controls the level of statistical fluctuations of both $\underline{H}_1$ and $\underline{H}_2$. In addition, $\underline{H}_1$ and $\underline{H}_2$ cannot be deterministic variables, since their dispersion coefficients are non zero whatever the value of $\lambda<1/5$.} Finally, the probabilistic model of $\Bb = \(D,\Lb,\underline{\Hb}\)$ involves the unknown vector-valued hyperparameter $\ssb = (\ssb_1,\ssb_2,\ssb_3,\underline{\ssb}) = (a_1,b_1,a_2,b_2,a_3,b_3,\lambda,\lambda_1,\lambda_2)$ belonging to the admissible set $\Sc = (\intervaloo{0}{+\infty}^2)^3 \times \intervaloo{-\infty}{1/5} \times \intervaloo{0}{+\infty}^2$. The optimal value $\ssb^{\text{opt}}$ of $\ssb$ is determined using the MLE method with the available data that are the $Q$ independent realizations $\bb^{\text{meso}}_1,\dots,\bb^{\text{meso}}_Q$ of random vector $\Bb$. The MLE method consists in computing $\ssb^{\text{opt}}$ by solving the following optimization~problem
\begin{equation}\label{optimalhyperparameter}
\ssb^{\text{opt}} = \argmax_{\ssb\in \Sc} \Lc(\ssb;\bb^{\text{meso}}_1,\dots,\bb^{\text{meso}}_Q),
\end{equation}
where $\ssb \mapsto \Lc({\ssb;\bb^{\text{meso}}_1,\dots,\bb^{\text{meso}}_Q})$ is the log-likelihood function for the $Q$ independent realizations $\bb^{\text{meso}}_1,\dots,\bb^{\text{meso}}_Q$ of $\Bb$ which is defined for all $\ssb\in \Sc$ by
\begin{equation}\label{loglikelihood}
\Lc(\ssb;\bb^{\text{meso}}_1,\dots,\bb^{\text{meso}}_Q) = \sum_{q=1}^{Q} \log(p_{\Bb}(\bb^{\text{meso}}_q;\ssb)).
\end{equation}

The accuracy of the identified optimal value $\ssb^{\text{opt}}$ is then all the higher as the number $Q$ of mesoscopic domains of observation is large but at the expense of a higher computational cost. Lastly, the optimal value $\bb^{\text{opt}}$ of vector-valued hyperparameter $\bb \in \Bc^{\text{meso}}$ is computed by solving the following optimization problem
\begin{equation}\label{optimalhyperparametermeso}
\bb^{\text{opt}} = \argmax_{\bb\in \Bc^{\text{meso}}} p_{\Bb}(\bb;\ssb^{\text{opt}}).
\end{equation}

Hence, optimal value $\bb^{\text{opt}}$ corresponds to the most probable value of random vector $\Bb$ according to the identified probability distribution represented by its probability density function $p_{\Bb}(\cdot;\ssb^{\text{opt}})$ parameterized by $\ssb^{\text{opt}}$. Note that the averaging approach presented in Reference \cite{Ngu15} is a particular case of the MLE method presented in this section if the {prior} stochastic models of $D$, $\Lb$ and $\underline{\Hb}$ are uniform random variables. It is the reason why a better robust identification is expected since the {prior} stochastic model of $\Bb$ has been improved in this work. {In the present work, since $D$ is modeled as a uniform random variable on $\intervaloo{0}{\delta_{\text{sup}}}$, the optimal value $\delta^{\text{opt}}$ of $\delta$ is simply obtained by averaging the $Q$ independent realizations $\delta^{\text{meso}}_1,\dots,\delta^{\text{meso}}_Q$ of $D$. A more advanced prior stochastic model for $D$ could have been considered, for instance by adding as available information that its mean value is given and its values are unlikely near zero, thus leading to a unimodal probability density function $p_D$ with support $\intervaloo{0}{\delta_{\text{sup}}}$ and with a higher parameterization than the simple uniform probability density function considered here.}

\section{Numerical Validation of the Multiscale Identification Method on In Silico Materials in 2D Plane Stress and 3D Linear Elasticity}
\label{sec:validation}

In this section, we present a numerical application of the improved multiscale identification methodology proposed in the present work within the framework of 2D plane stress and 3D linear elasticity theories by using \emph{in silico} materials for which the macroscopic and mesoscopic mechanical properties are known. The required multiscale ``experimental'' kinematic fields have been obtained through numerical simulations using one random realization of the random elasticity field in SFE$^+$ (see Section~\ref{sec:stochastic_model}) not restricted from $\Rbb^3$ to some mesoscopic domain $\Omega^{\text{meso}}$ but restricted to the whole macroscopic domain $\Omega^{\text{macro}}$ for a given experimental value $\bb^{\text{meso}}_{\text{exp}}$ of hyperparameter $\bb \in \Bc^{\text{meso}}$. The~solution of a deterministic boundary value problem over this macroscopic domain $\Omega^{\text{macro}}$ is then computed for a heterogeneous random elasticity field whose spatial correlation lengths correspond to the characteristic sizes of the heterogeneities at microscale. This deterministic boundary value problem is solved using a classical numerical method (FEM) whose computational cost is high and potentially prohibitive in 3D, what can be avoided by computational homogenization methods, but it is required to completely simulate the multiscale ``experimental'' measurements.

\subsection{Validation on an In Silico Specimen in Compression Test in 2D Plane Stress Linear Elasticity}
\label{sec:validation_2D}

For this first numerical validation example, a 2D plane stress assumption is considered. Macroscopic domain of observation $\Omega_{\text{exp}}^{\text{macro}}$ is a 2D square domain and it exactly corresponds to the cross-section of macroscopic domain $\Omega^{\text{macro}}$ and such that $\Omega_{\text{obs}}^{\text{macro}} = \Omega_{\text{exp}}^{\text{macro}}$ since the test specimen is \emph{in silico}. The dimensions of 2D macroscopic domain of observation $\Omega_{\text{exp}}^{\text{macro}}$ are $1\times 1$~cm$^2$ in a fixed Cartesian frame $(O,x_1,x_2)$ of $\Rbb^2$. It is possible to introduce a set of $Q=16$ non-overlapping 2D square mesoscopic domains of observation $\Omega^{\text{meso}}_{\text{exp}, 1},\dots,\Omega^{\text{meso}}_{\text{exp}, Q}\subset\Omega_{\text{exp}}^{\text{macro}}$ for which the mesoscale dimensions are $1\times 1$~mm$^2$ {(see Figure~\ref{fig:multiscale_experimental_configuration} for a schematic representation of domains of observation $\Omega_{\text{exp}}^{\text{macro}}$ and $\Omega^{\text{meso}}_{\text{exp}, 1},\dots,\Omega^{\text{meso}}_{\text{exp}, Q}$)}. Consequently, observation domain $\Omega_{\text{obs}}^{\text{meso}}$, for which the dimensions are also $1\times 1$~mm$^2$, is defined as the 2D square cross-section of mesoscopic domain $\Omega^{\text{meso}}$. Deterministic surface force field $\fb^{\text{macro}}$ is uniformly distributed on the top boundary of macroscopic domain $\Omega^{\text{macro}}$ and applied along the (downward vertical) $-x_2$ direction with an intensity of $5$~kN such that $\norm{\fb^{\text{macro}}} = 5$~kN/cm$^2 = 5\times10^7$~N/m$^2$, while the bottom boundary of macroscopic domain $\Omega^{\text{macro}}$ is clamped.

\subsubsection{Parameterization of the Macroscopic and Mesoscopic Models}
\label{sec:model_validation_2D}

At macroscale, the solution of deterministic boundary value problem~\eqref{equilibriumequationmacro}-\eqref{kinematicrelationmacro} with 2D plane stress assumption depends only on $6$ components $\{S^{\text{macro}}(\ab)\}_{ijkh}$ of deterministic compliance tensor $S^{\text{macro}}(\ab)$ with $i,j,k,h\in\set{1,2}$. Consequently, the solution at macroscale depends only on the components of a 2D fourth-order compliance tensor $S^{\text{macro}}_{\text{2D}}(\ab)$ that is defined by $\{S^{\text{macro}}_{\text{2D}}(\ab)\}_{ijkh} = \{S^{\text{macro}}(\ab)\}_{ijkh}$ for all $i,j,k,h\in\set{1,2}$. Then, a 2D fourth-order elasticity tensor at macroscale can be introduced and defined by $C^{\text{macro}}_{\text{2D}}(\ab) = (S^{\text{macro}}_{\text{2D}}(\ab))^{-1}$. Since within the framework of linear elasticity theory, any isotropic material is completely characterized by a bulk modulus $\kappa$ and a shear modulus $\mu$ at macroscale, then we have the vector-valued parameter $\ab = (\kappa,\mu)$. In particular, we have chosen the experimental value $\ab^{\text{macro}}_{\text{exp}} = (\kappa^{\text{macro}}_{\text{exp}},\mu^{\text{macro}}_{\text{exp}})$ with $\kappa^{\text{macro}}_{\text{exp}} = 13.901$~GPa and $\mu^{\text{macro}}_{\text{exp}} = 3.685$~GPa, corresponding to a Young's modulus $E^{\text{macro}}_{\text{exp}} = 10.158$~GPa and and a Poisson's ratio $\nu^{\text{macro}}_{\text{exp}} = 0.3782$.

At mesoscale, the solution of stochastic boundary value problem~\eqref{equilibriumequationmeso}-\eqref{kinematicrelationmeso} with 2D plane stress assumption depends only on $6$ components $\{\Sb^{\text{meso}}(\bb)\}_{ijkh}$ of random compliance tensor field $\Sb^{\text{meso}}(\bb)$ with $i,j,k,h\in\set{1,2}$ or equivalently on every $21$ components $\{\Cb^{\text{meso}}(\bb)\}_{ijkh}$ of random elasticity tensor field $\Cb^{\text{meso}}(\bb)$ with $i,j,k,h\in\set{1,2,3}$. It is the reason why we have chosen to construct the {prior} stochastic model of the random compliance tensor field $\Sb^{\text{meso}}(\bb)$ as presented in Section~\ref{sec:stochastic_model} and the stochastic boundary value problem~\eqref{equilibriumequationmeso}-\eqref{kinematicrelationmeso} is solved in using \eqref{consititutiveequationmeso_compliance} rather than \eqref{consititutiveequationmeso}. Furthermore, its~mean function $\underline{S}^{\text{meso}}$ is spatially constant and models an isotropic elastic medium that is completely characterized by a mean bulk modulus $\underline{\kappa}$ and a mean shear modulus $\underline{\mu}$ at mesoscale. Consequently, the vector-valued hyperparameter $\bb = (\delta,\ell,\underline{\hb})$ involves only (i) a dispersion parameter $\delta$, (ii) a spatial correlation length $\ell$ that is such that $\ell_1 = \ell_2 = \ell$ in order to be consistent with the effective model at macroscale for which the material is assumed to be isotropic and with $\ell_3 = +\infty$ in order to be consistent with the 2D plane stress assumption, and (iii) a vector-valued hyperparameter $\underline{\hb} = (\underline{\kappa},\underline{\mu})$ gathering the mean bulk modulus $\underline{\kappa}$ and the mean shear modulus $\underline{\mu}$ at mesoscale. In particular, we have chosen the experimental value $\bb_{\text{exp}}^{\text{meso}}=(\delta_{\text{exp}}^{\text{meso}},\ell_{\text{exp}}^{\text{meso}},\underline{\kappa}_{\text{exp}}^{\text{meso}},\underline{\mu}_{\text{exp}}^{\text{meso}})$ with $\delta_{\text{exp}}^{\text{meso}} = 0.40$, $\ell_{\text{exp}}^{\text{meso}} = 125~\upmu$m, $\underline{\kappa}_{\text{exp}}^{\text{meso}} = 13.75$~GPa and $\underline{\mu}_{\text{exp}}^{\text{meso}} = 3.587$~GPa, corresponding to a mean Young's modulus $\underline{E}_{\text{exp}}^{\text{meso}} = 9.900$~GPa and a mean Poisson's ratio $\underline{\nu}_{\text{exp}}^{\text{meso}} = 0.380$~GPa. 
For identification purposes and further computational savings, we consider a reduced admissible set $\Bc^{\text{meso}}_{\text{ad}} \subset \Bc^{\text{meso}}$ for the vector-valued hyperparameter $\bb = (\delta,\ell,\underline{\kappa},\underline{\mu})$ such that $\delta \in \intervalcc{0.25}{0.50}$, $\ell \in \intervalcc{20}{250}~\upmu$m, $\underline{\kappa} \in \intervalcc{8.5}{17}$~GPa, $\underline{\mu} \in \intervalcc{2.15}{4.50}$~GPa, instead of the full admissible set $\Bc^{\text{meso}} = \intervaloo{0}{\delta_{\text{sup}}} \times \intervaloo{0}{+\infty} \times \intervaloo{0}{+\infty}^2$ with $\delta_{\text{sup}} = \sqrt{(n+1)/(n+5)} = \sqrt{7/11} \approx 0.7977 < 1$ (with $n=6$ in linear elasticity). This reduced admissible set $\Bc^{\text{meso}}_{\text{ad}}$ is then discretized into $n_V=10$ equidistant points in each dimension for which the three numerical indicators $\Jc^{\text{meso}}_{\delta}(\bb)$, $\Jc^{\text{meso}}_{\ellb}(\bb)$ and $\Jc^{\text{multi}}_{\underline{\hb}}(\ab^{\text{macro}},\bb)$ defined in Section~\ref{sec:meso_indicators} are evaluated and compared. {The identified values $\bb^{\text{meso}}_1,\dots,\bb^{\text{meso}}_Q$ of hyperparameters $\bb$ for each of the $Q$ mesoscopic domains of observation $\Omega^{\text{meso}}_{\text{exp},1},\dots,\Omega^{\text{meso}}_{\text{exp},Q}$ are then searched on this multidimensional grid of $n_V\times n_V\times n_V\times n_V$ points in the hypercube $\Bc^{\text{meso}}_{\text{ad}}$.}

Within the framework of linear elasticity under 2D plane stress assumption, both the deterministic boundary value problem~\eqref{equilibriumequationmacro}-\eqref{kinematicrelationmacro} and the stochastic boundary value problem~\eqref{equilibriumequationmeso}-\eqref{kinematicrelationmeso} are solved by discretizing the 2D macroscopic and mesoscopic domains of observation $\Omega^{\text{macro}}_{\text{obs}}$ and $\Omega^{\text{meso}}_{\text{obs}}$ in space using the FEM. The finite element meshes of 2D square domains $\Omega_{\text{obs}}^{\text{macro}}$ and $\Omega_{\text{obs}}^{\text{meso}}$ are structured meshes made up with $4$-nodes linear quadrangular elements with Gauss-Legendre quadrature rule. The stochastic boundary value problem~\eqref{equilibriumequationmeso}-\eqref{kinematicrelationmeso} at mesoscale is solved using the Monte Carlo numerical method. Mesh convergence analyses of the numerical solutions of the deterministic boundary value problem~\eqref{equilibriumequationmacro}-\eqref{kinematicrelationmacro} at macroscale and of the stochastic boundary value problem~\eqref{equilibriumequationmeso}-\eqref{kinematicrelationmeso} at mesoscale have been performed in order to define accurate finite element approximations at both macroscopic and mesoscopic scales. The finite element mesh of 2D macroscopic domain $\Omega_{\text{obs}}^{\text{macro}}$ is a regular grid containing $25 \times 25$ quadrangular elements with uniform element size $h^{\text{macro}} = 0.4~\text{mm} = 4\times 10^{-4}$~m in each spatial direction. It thus comprises $676$ nodes and $625$ elements, with $1300$ unknown degrees of freedom (dofs). The finite element mesh of 2D mesoscopic domain $\Omega_{\text{obs}}^{\text{meso}}$ is a regular grid containing $100 \times 100$ quadrangular elements with uniform element size $h^{\text{meso}}$ = $10~\upmu\text{m} = 10^{-5}$~m in each spatial direction. It thus comprises $10,201$ nodes and $10,000$ elements, with $20,000$ unknown dofs. The number of Gauss integration points per spatial correlation length used for numerical quadrature over 2D macroscopic domain of observation $\Omega_{\text{obs}}^{\text{macro}}$ and 2D mesoscopic domain of observation $\Omega_{\text{obs}}^{\text{meso}}$ is $n_G = 4$ in each spatial~direction.

Concerning the computational stochastic homogenization with 2D plane stress assumption, we~consider a 2D square domain $\Omega^{\text{RVE}}$ of side length $B^{\text{RVE}}$ defined in a Cartesian frame $(O,x_1,x_2)$ and we use the homogenization method with static uniform boundary conditions (i.e., with homogeneous stresses) which is appropriate for linear elasticity under 2D plane stress assumption. Note that only the components $\{\Sb^{\text{eff}}(\bb)\}_{ijkh}$ with $i,j,k,h\in\set{1,2}$ can be calculated. We then obtain a 2D fourth-order effective compliance tensor $\Sb^{\text{eff}}_{\text{2D}}(\bb)$ that is such that $\{\Sb^{\text{eff}}_{\text{2D}}(\bb)\}_{ijkh} = \{\Sb^{\text{eff}}(\bb)\}_{ijkh}$ for all $i,j,k,h\in\set{1,2}$. Then, a 2D fourth-order effective elasticity tensor can be defined as $\Cb^{\text{eff}}_{\text{2D}}(\bb) = (\Sb^{\text{eff}}_{\text{2D}}(\bb) )^{-1}$. A convergence analysis of the statistical estimator of its statistical fluctuations with respect to the representative volume element size $B^{\text{RVE}}$ has been performed. A representative volume element size $B^{\text{RVE}} = 20 \times \ell = 400~\upmu\text{m} = 4\times10^{-4}$~m has been found to be sufficient to reach negligible statistical fluctuations for the construction of the multiscale numerical indicator $\Jc^{\text{multi}}_{\underline{\hb}}(\ab^{\text{macro}},\bb)$ that is calculated by replacing $C^{\text{macro}}(\ab)$ and $\Cb^{\text{eff}}(\bb)$ with $C^{\text{macro}}_{\text{2D}}(\ab)$ and $\Cb^{\text{eff}}_{\text{2D}}(\bb)$, respectively, in \eqref{indicatormulti}.

As the mathematical expectations involved in each of the numerical indicators $\Jc^{\text{meso}}_{\delta}(\bb)$, $\Jc^{\text{meso}}_{\ellb}(\bb)$ and $\Jc^{\text{multi}}_{\underline{\hb}}(\ab^{\text{macro}},\bb)$ are evaluated using the Monte Carlo numerical method, statistical convergence analyses of their statistical estimators with respect to the number of independent realizations $N_s$ have been carried out and a convergence has been reached for $N_s=500$. {Sensitivity analyses of each of the three numerical indicators have been performed with respect to each of the hyperparameters $\delta$, $\ell$, $\underline{\hb}=(\underline{\kappa},\underline{\mu})$, respectively, in the reduced admissible set $\Bc^{\text{meso}}_{\text{ad}} = \intervalcc{0.25}{0.50} \times \intervalcc{20}{250}~\upmu\text{m} \times \intervalcc{8.5}{17}~\text{GPa} \times \intervalcc{2.15}{4.50}~\text{GPa}$. Hence, it can be shown that each numerical indicator is sufficiently sensitive to the variation of its dedicated hyperparameter and that the multi-objective optimization problem~\eqref{optimizationpbmesomin} to be solved is well-posed.}


Recall the multiscale statistical inverse problem has been formulated into two decoupled optimization problems in $\ab$ and $\bb$, respectively, to be solved sequentially (see Section~\ref{sec:formulation_optim_problem}), namely (i) a macroscale single-objective optimization problem~\eqref{optimizationpbmacro} for the inverse identification of the optimal value $\ab^{\text{macro}}$ of parameter $\ab$ in its admissible set $\Ac^{\text{macro}}$, and (ii) a mesoscale multi-objective optimization problem~\eqref{optimizationpbmesomin} for the statistical inverse identification of the global optimal value $\bb^{\text{opt}}$ of hyperparameter $\bb$ in its reduced admissible set $\Bc^{\text{meso}}_{\text{ad}}$.

\subsubsection{Resolution of the Single-Objective Optimization Problem at Macroscale}
\label{sec:results_validation_2D_macro}

In this paragraph, we present the results of the first single-objective optimization problem~\eqref{optimizationpbmacro} at macroscale which consists in minimizing the macroscopic numerical indicator $\Jc^{\text{macro}}(\ab)$ constructed in the macroscopic domain of observation $\Omega^{\text{macro}}_{\text{exp}}$ for identifying the optimal value $\ab^{\text{macro}}$ of $\ab$ at macroscale. The single-objective optimization problem~\eqref{optimizationpbmacro} at macroscale has been solved using the Nelder-Mead simplex algorithm. The identification results are reported in Table~\ref{tab:Res2DMacro} and show that the relative error between the identified optimal value $\ab^{\text{macro}} = (13.901, 3.685)$ in [GPa] and the reference experimental value $\ab^{\text{macro}}_{\text{exp}} = (14.328, 3.670)$ in [GPa] used for the construction of the numerically simulated ``experimental'' database remains small (less than $3\%$ and $0.5\%$ for $\kappa^{\text{macro}}$ and $\mu^{\text{macro}}$, respectively), allowing to validate the proposed identification methodology in 2D plane stress linear elasticity for the resolution of the single-objective optimization problem~\eqref{optimizationpbmacro} at macroscale.

\begin{table}[H]
\caption{Comparison between the identified optimal value $\ab^{\text{macro}}$ and the reference experimental value~$\ab^{\text{macro}}_{\text{exp}}$.}
\label{tab:Res2DMacro}
\centering
\begin{tabular}{ccc}
\toprule
& \boldmath{$\kappa$ \textbf{[GPa]}} & \boldmath{$\mu$ \textbf{[GPa]}} \\
\midrule
$\ab^{\text{macro}}$ & $13.901$ & $3.685$ \\
$\ab^{\text{macro}}_{\text{exp}}$ & $14.328$ & $3.670$ \\
Relative error [\%] & $2.980$ & $0.4009$ \\
\bottomrule
\end{tabular}
\end{table}

\subsubsection{Resolution of the Multi-Objective Optimization Problem at Mesoscale}
\label{sec:results_validation_2D_meso}

In this paragraph, we present the results of the second multi-objective optimization problem~\eqref{optimizationpbmesomin} at mesoscale which consists in simultaneously minimizing the three numerical indicators $\Jc^{\text{meso}}_{\delta}(\bb)$, $\Jc^{\text{meso}}_{\ellb}(\bb)$ and $\Jc^{\text{multi}}_{\underline{\hb}}(\ab^{\text{macro}},\bb)$ constructed in each of the $Q=16$ mesoscopic domains of observation $\Omega^{\text{meso}}_{\text{exp},1},\dots,\Omega^{\text{meso}}_{\text{exp},Q}$ using the optimal parameter $\ab^{\text{macro}}=(13.901,3.685)$ in [GPa] previously identified at macroscale (see the previous paragraph) for identifying the global optimal value $\bb^{\text{opt}}$ of $\bb$ at mesoscale. The multi-objective optimization problem~\eqref{optimizationpbmesomin} at mesoscale has been solved using the fixed-point iterative algorithm on the one hand and the genetic algorithm on the other hand for comparison purposes. In order to analyze the numerical efficiency of these two resolution approaches, instead of evaluating the computing time which strongly depends on the computer hardware used, we choose in this work to compare the number of evaluations of the random solution of the stochastic boundary value problem~\eqref{equilibriumequationmeso}-\eqref{kinematicrelationmeso} at mesoscale (i.e., the number of calls to the deterministic numerical model at mesoscale) required by each algorithm to achieve the desired convergence.

The identification results obtained with the fixed-point iterative algorithm are summarized in Table~\ref{tab:Res2DMesoFP} for the set of $Q=16$ mesoscopic domains of observation $\Omega^{\text{meso}}_{\text{exp},1},\dots,\Omega^{\text{meso}}_{\text{exp},Q}$, namely the set of $Q$ identified values $\bb^{\text{meso}}_1,\dots,\bb^{\text{meso}}_{Q}$ and numbers of iterations $n_1,\dots,n_{Q}$ required to reach the desired convergence, with a convergence criterion on the residual norm between two iterations that must be less than a prescribed tolerance set to $10^{-9}$, and the global optimal value $\bb^{\text{opt}}$ computed by using the MLE method. On the one hand, there are greater variations between the identified values $\ell^{\text{meso}}_1,\dots,\ell^{\text{meso}}_{Q}$ and $\delta^{\text{meso}}_1,\dots,\delta^{\text{meso}}_{Q}$, reflecting the fact that the two associated mesoscopic numerical indicators $\Jc^{\text{meso}}_{\delta}(\bb)$ and $\Jc^{\text{meso}}_{\ellb}(\bb)$ depend directly on the experimental field measurements on each mesoscopic domain of observation. On the other hand, the lower variability between the identified values $\underline{\kappa}^{\text{meso}}_1,\dots,\underline{\kappa}^{\text{meso}}_{Q}$ and $\underline{\mu}^{\text{meso}}_1,\dots,\underline{\mu}^{\text{meso}}_{Q}$ can be explained by the fact that the associated multiscale numerical indicator $\Jc^{\text{multi}}_{\underline{\hb}}(\ab^{\text{macro}},\bb)$ does not depend directly on the experimental field measurements on each mesoscopic domain of observation but is rather conditioned by the identified values $\ell^{\text{meso}}_1,\dots,\ell^{\text{meso}}_{Q}$ and $\delta^{\text{meso}}_1,\dots,\delta^{\text{meso}}_{Q}$. Thus, the relative errors calculated on these two hyperparameters are essentially due to the quality of the discretization of the reduced admissible set $\Bc^{\text{meso}}_{\text{ad}}$. {In particular, the fixed-point iterative algorithm has selected the same identified value $\underline{\mu}^{\text{meso}}_1 = \dots = \underline{\mu}^{\text{meso}}_Q = 3.717$~GPa (among the $n_V=10$ test points in $\intervalcc{2.15}{4.50}$~GPa) for the $Q=16$ mesoscopic domains of observation $\Omega^{\text{meso}}_{\text{exp},1},\dots,\Omega^{\text{meso}}_{\text{exp},Q}$. Clearly, a finer grid (with $n_V > 10$) might yield different values for the identified hyperparameter $\underline{\mu}^{\text{meso}}$ selected by the optimization algorithm. It is the reason why a prior probabilistic model for the identified hyperparameters has been introduced.} The number of evaluations of the stochastic computational model needed by the fixed-point iterative algorithm is given by $n^{\text{FP}}_{\text{tot}}=3 \, n_V \, N_s \sum_{q=1}^{16} n_{q}$, where the superscript ${}^{\text{FP}}$ refers to ``Fixed-Point'' and $n_V$ is the number of evaluations of a numerical indicator to search for the minimum with respect to the associated hyperparameter. Figure~\ref{fig:pdfDLKMFP2D} shows the probability density functions $p_{D}$, $p_{L}$, $p_{\underline{K}}$ and $p_{\underline{M}}$ of random variables $D$, $L$, $\underline{K}$ and $\underline{M}$, respectively, which are defined in Section~\ref{sec:randomization} with the two components $\underline{H}_1 = \underline{K}$ and $\underline{H}_2 = \underline{M}$ of random vector $\underline{\Hb}=(\underline{K},\underline{M})$. {As suggested by the identification results shown in Table~\ref{tab:Res2DMesoFP} and as already mentioned in Section~\ref{sec:randomization}, a more advanced prior stochastic model for $D$ would have been preferable to obtain a unimodal probability density function $p_D$ with support $\intervaloo{0}{\delta_{\text{sup}}}$ and which would be concentrated around the reference experimental value $\delta_{\text{exp}}^{\text{meso}} = 0.4$. Besides, although all the independent realizations $\underline{\mu}^{\text{meso}}_1,\dots,\underline{\mu}^{\text{meso}}_{Q}$ of $\underline{M}$ given in Table~\ref{tab:Res2DMesoFP} are equal to the same identified value $3.717$~GPa, the probability density function $p_{\underline{M}}$ does not correspond to the Dirac measure on $\Rbb$ at point $3.717$~GPa but to a gamma distribution with a very small dispersion around this value, since for the prior probabilistic model of $\underline{\Hb}=(\underline{K},\underline{M})$ considered here, $\underline{K}$ and $\underline{M}$ cannot be deterministic variables (see Section~\ref{sec:randomization}).} We finally obtain the global optimal value $\bb^{\text{opt}}=(0.391,135.328,12.273,3.717)$ in $([-],[\upmu\text{m}],[\text{GPa}],[\text{GPa}])$ with relative errors less than $3\%$, $9\%$, $11\%$ and $4\%$ for $\delta^{\text{opt}}$, $\ell^{\text{opt}}$, $\underline{\kappa}^{\text{opt}}$ and $\underline{\mu}^{\text{opt}}$, respectively, with respect to the reference experimental value $\bb_{\text{exp}}^{\text{meso}}=(0.40,125,13.75,3.587)$ in $([-],[\upmu\text{m}],[\text{GPa}],[\text{GPa}])$ used to construct the numerically simulated ``experimental'' database, allowing to validate the proposed identification methodology in 2D plane stress linear elasticity for the resolution of the multi-objective optimization problem~\eqref{optimizationpbmesomin} at~mesoscale.

\begin{table}[H]
\caption{Fixed-point iterative algorithm: comparison between the global optimal value $\bb^{\text{opt}}$ obtained from the $Q=16$ identified values $\bb^{\text{meso}}_1,\dots,\bb^{\text{meso}}_{Q}$ for each of the $Q$ mesoscopic domains of observation $\Omega^{\text{meso}}_{\text{exp},1},\dots,\Omega^{\text{meso}}_{\text{exp},Q}$, and the reference experimental value $\bb^{\text{meso}}_{\text{exp}}$.}
\label{tab:Res2DMesoFP}
\centering
\begin{tabular}{cccccc}
\toprule
& \boldmath{$\delta$} & \boldmath{$\ell$ \textbf{[$\upmu$m]}} & \boldmath{$\underline{\kappa}$\textbf{ [GPa]}} & \boldmath{$\underline{\mu}$ \textbf{[GPa]}} & \boldmath{$n_{q}$} \\
\midrule
$\bb^{\text{meso}}_{1}$ & $0.306$ & $147.778$ & $13.222$ & $3.717$ & 3 \\
$\bb^{\text{meso}}_{2}$ & $0.500$ & $224.444$ & $11.333$ & $3.717$ & 4 \\
$\bb^{\text{meso}}_{3}$ & $0.417$ & $122.222$ & $12.278$ & $3.717$ & 3 \\
$\bb^{\text{meso}}_{4}$ & $0.417$ & $122.222$ & $12.278$ & $3.717$ & 3 \\
$\bb^{\text{meso}}_{5}$ & $0.444$ & $147.778$ & $12.278$ & $3.717$ & 3 \\
$\bb^{\text{meso}}_{6}$ & $0.417$ & $122.222$ & $12.278$ & $3.717$ & 4 \\
$\bb^{\text{meso}}_{7}$ & $0.361$ & $147.778$ & $12.278$ & $3.717$ & 4 \\
$\bb^{\text{meso}}_{8}$ & $0.361$ & $147.778$ & $12.278$ & $3.717$ & 4 \\
$\bb^{\text{meso}}_{9}$ & $0.444$ & $147.778$ & $12.278$ & $3.717$ & 3 \\
$\bb^{\text{meso}}_{10}$ & $0.333$ & $147.778$ & $12.278$ & $3.717$ & 4 \\
$\bb^{\text{meso}}_{11}$ & $0.333$ & $122.222$ & $12.278$ & $3.717$ & 4 \\
$\bb^{\text{meso}}_{12}$ & $0.389$ & $96.667$ & $12.278$ & $3.717$ & 3 \\
$\bb^{\text{meso}}_{13}$ & $0.389$ & $147.778$ & $12.278$ & $3.717$ & 4 \\
$\bb^{\text{meso}}_{14}$ & $0.389$ & $122.222$ & $12.278$ & $3.717$ & 3 \\
$\bb^{\text{meso}}_{15}$ & $0.389$ & $147.778$ & $12.278$ & $3.717$ & 4 \\
$\bb^{\text{meso}}_{16}$ & $0.361$ & $122.222$ & $12.278$ & $3.717$ & 4 \\
\midrule
$\bb^{\text{opt}}$ & $0.391$ & $135.328$ & $12.273$ & $3.717$ & - \\
$\bb^{\text{meso}}_{\text{exp}}$ & $0.400$ & $125.000$ & $13.750$ & $3.587$ & - \\
Relative error [\%] & $2.344$ & $8.262$ & $10.740$ & $3.611$ & - \\
\midrule
$n^{\text{FP}}_{\text{tot}}$ & \multicolumn{5}{c}{$855,000$} \\
\bottomrule
\end{tabular}
\end{table}
\setlength\figureheight{0.08\textheight}
\begin{figure}[H]
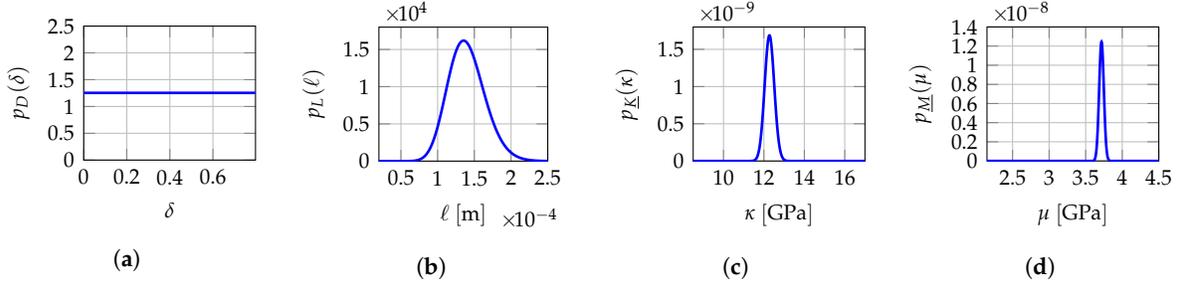

	\centering
	\begin{subfigure}{0.25\linewidth}
		\centering
		\tikzsetnextfilename{pdfDelta_FPMLG500}
%
\begin{tikzpicture}

\begin{axis}[%
width=1.281\figureheight,
height=\figureheight,
at={(0\figureheight,0\figureheight)},
scale only axis,
xmin=0,
xmax=0.797724035217466,
xlabel={$\delta$},
ymin=0,
ymax=2.5,
ylabel={$p_{D}(\delta)$},
axis background/.style={fill=white},
xmajorgrids,
ymajorgrids
]
\addplot [color=blue, line width=1.0pt, forget plot]
  table[row sep=crcr]{%
0	1.25356634105602\\
0.797724035217466	1.25356634105602\\
};
\end{axis}
\end{tikzpicture}%
		\caption{}
		\label{fig:pdfDelta_FPMLG500}
	\end{subfigure}\hfill
	\begin{subfigure}{.25\linewidth}
		\centering
		\tikzsetnextfilename{pdfLcorr_FPMLG500_mdpi}
		\input{pdfLcorr_FPMLG500_mdpi}
		\caption{}
		\label{fig:pdfLcorr_FPMLG500}
	\end{subfigure}\hfill
	\begin{subfigure}{.25\linewidth}
		\centering
		\tikzsetnextfilename{pdfKappa_FPMLG500}
		\input{pdfKappa_FPMLG500}
		\caption{}
		\label{fig:pdfKappa_FPMLG500}
	\end{subfigure}\hfill
	\begin{subfigure}{.25\linewidth}
		\centering
		\tikzsetnextfilename{pdfMu_FPMLG500}
		\input{pdfMu_FPMLG500}
		\caption{}
		\label{fig:pdfMu_FPMLG500}
	\end{subfigure}
	\caption{Fixed-point iterative algorithm: probability density functions $p_{D}$, $p_{L}$, $p_{\underline{K}}$ and $p_{\underline{M}}$ of random variables $D$, $L$, $\underline{K}$ and $\underline{M}$, respectively. (\textbf{a}) $p_{D}(\delta)$; (\textbf{b}) $p_{L}(\ell)$; (\textbf{c}) $p_{\underline{K}}(\kappa)$; (\textbf{d}) $p_{\underline{M}}(\mu)$.}
	\label{fig:pdfDLKMFP2D}
\end{figure}

\begin{figure}[H]
	\centering
	\begin{subfigure}{.25\linewidth}
		\centering
		\tikzsetnextfilename{Q_deltaOpt_FPMLG500}
%
\begin{tikzpicture}

\begin{axis}[%
width=1.281\figureheight,
height=\figureheight,
at={(0\figureheight,0\figureheight)},
scale only axis,
xmin=1,
xmax=16,
xlabel={$Q$},
ymin=0.25,
ymax=0.50,
ylabel={$\delta^{\mathrm{opt}}$},
axis background/.style={fill=white},
xmajorgrids,
ymajorgrids
]
\addplot [color=blue, line width=1.0pt, mark=+, mark options={solid, blue}, forget plot]
  table[row sep=crcr]{%
1	0.305555555555556\\
2	0.402777777777778\\
3	0.407407407407407\\
4	0.409722222222222\\
5	0.416666666666667\\
6	0.416666666666667\\
7	0.408730158730159\\
8	0.402777777777778\\
9	0.407407407407407\\
10	0.4\\
11	0.393939393939394\\
12	0.393518518518519\\
13	0.393162393162393\\
14	0.392857142857143\\
15	0.392592592592593\\
16	0.390625\\
};
\end{axis}
\end{tikzpicture}%
		\caption{}
		\label{fig:Q_deltaOpt_FPMLG500}
	\end{subfigure}\hfill
	\begin{subfigure}{.25\linewidth}
		\centering
		\tikzsetnextfilename{Q_lcorrOpt_FPMLG500_mdpi}
%
\begin{tikzpicture}

\begin{axis}[%
width=1.281\figureheight,
height=\figureheight,
at={(0\figureheight,0\figureheight)},
scale only axis,
xmin=1,
xmax=16,
xlabel={$Q$},
ymin=0.00010,
ymax=0.00020,
ylabel={$\ell^{\mathrm{opt}}$},
axis background/.style={fill=white},
xmajorgrids,
ymajorgrids,
change y base,
y unit=m,
]
\addplot [color=blue, line width=1.0pt, mark=+, mark options={solid, blue}, forget plot]
  table[row sep=crcr]{%
1	0.000147777777777778\\
2	0.000178100669857001\\
3	0.000154023027895193\\
4	0.00014416773712207\\
5	0.000144896493974093\\
6	0.000140367665874062\\
7	0.000141418782223459\\
8	0.000142208732830899\\
9	0.000142824100421231\\
10	0.000143317002961276\\
11	0.000141053585730951\\
12	0.00013587167370285\\
13	0.000136747612595839\\
14	0.000135558841155892\\
15	0.000136333155596568\\
16	0.000135327548198486\\
};
\end{axis}
\end{tikzpicture}%
		\caption{}
		\label{fig:Q_lcorrOpt_FPMLG500}
	\end{subfigure}\hfill
	\begin{subfigure}{.25\linewidth}
		\centering
		\tikzsetnextfilename{Q_kappaOpt_FPMLG500}
%
\begin{tikzpicture}

\begin{axis}[%
width=1.281\figureheight,
height=\figureheight,
at={(0\figureheight,0\figureheight)},
scale only axis,
xmin=1,
xmax=16,
xlabel={$Q$},
ymin=11000000000,
ymax=14000000000,
ylabel={$\underline{\kappa}^{\mathrm{opt}}$},
axis background/.style={fill=white},
xmajorgrids,
ymajorgrids,
change y base,
y unit=Pa,
y SI prefix=giga
]
\addplot [color=blue, line width=1.0pt, mark=+, mark options={solid, blue}, forget plot]
  table[row sep=crcr]{%
1	13222222222.2222\\
2	12241398868.561\\
3	12253513034.2727\\
4	12259573955.1095\\
5	12263213734.0502\\
6	12265640021.2449\\
7	12267372402.1633\\
8	12268673485.145\\
9	12269684820.5707\\
10	12270493249.7362\\
11	12271154956.028\\
12	12271707482.5307\\
13	12272173978.6261\\
14	12272572102.1587\\
15	12272921838.2274\\
16	12273225181.1824\\
};
\end{axis}
\end{tikzpicture}%
		\caption{}
		\label{fig:Q_kappaOpt_FPMLG500}
	\end{subfigure}\hfill
	\begin{subfigure}{.25\linewidth}
		\centering
		\tikzsetnextfilename{Q_muOpt_FPMLG500}
%
\begin{tikzpicture}

\begin{axis}[%
width=1.26\figureheight,
height=\figureheight,
at={(0\figureheight,0\figureheight)},
scale only axis,
xmin=1,
xmax=16,
xlabel={$Q$},
ymin=3712000000,
ymax=3720000000,
ylabel={$\underline{\mu}^{\mathrm{opt}}$},
axis background/.style={fill=white},
xmajorgrids,
ymajorgrids,
change y base,
y unit=Pa,
y SI prefix=giga
]
\addplot [color=blue, line width=1.0pt, mark=+, mark options={solid, blue}, forget plot]
  table[row sep=crcr]{%
1	3716666666.66667\\
2	3714459153.81834\\
3	3715195152.21455\\
4	3715563273.89662\\
5	3715784071.12741\\
6	3715931083.26432\\
7	3716036448.08008\\
8	3716115177.05053\\
9	3716176284.07606\\
10	3716225442.16266\\
11	3716265495.02991\\
12	3716299105.7868\\
13	3716327185.14509\\
14	3716351316.18711\\
15	3716372416.05024\\
16	3716390773.79362\\
};
\end{axis}
\end{tikzpicture}%
		\caption{}
		\label{fig:Q_muOpt_FPMLG500}
	\end{subfigure}
	\caption{Fixed-point iterative algorithm: evolutions of the identified global optimal values $\delta^{\text{opt}}$, $\ell^{\text{opt}}$, $\underline{\kappa}^{\text{opt}}$ and $\underline{\mu}^{\text{opt}}$ with respect to the number $Q$ of mesoscopic domains of observation considered. (\textbf{a})~$\delta^{\text{opt}}(Q)$; (\textbf{b}) $\ell^{\text{opt}}(Q)$; (\textbf{c}) $\underline{\kappa}^{\text{opt}}(Q)$; (\textbf{d}) $\underline{\mu}^{\text{opt}}(Q)$.}
	\label{fig:MDLKM}
\end{figure}
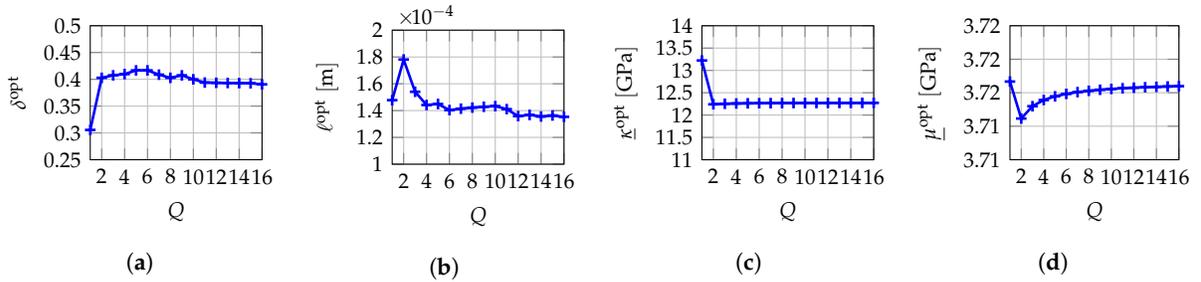

Figure~\ref{fig:MDLKM} shows the evolution of the global optimal values $\delta^{\text{opt}}$, $\ell^{\text{opt}}$, $\underline{\kappa}^{\text{opt}}$ $\underline{\mu}^{\text{opt}}$ estimated by the MLE method as a function of the number $Q$ of independent realizations $\bb^{\text{meso}}_1,\dots,\bb^{\text{meso}}_Q$ of random vector $\Bb = \(D,L,\underline{K},\underline{M}\)$. Although the number $Q$ remains low (less than or equal to $16$), we observe that each of the global optimal values tends to converge towards an objective value when $Q$ increases, which demonstrates that the use of the MLE method with the {prior} probabilistic model of $\Bb$ proposed in this work allows a robust identification of the vector-valued hyperparameter $\bb=(\delta,\ell,\underline{\kappa},\underline{\mu})$.

In terms of computational efficiency, we can see in Table~\ref{tab:Res2DMesoFP} that the numbers of iterations $n_1,\dots,n_{Q}$ required to achieve the desired convergence are relatively low (less than or equal to $4$) for each of the $Q=16$ mesoscopic domains of observation $\Omega^{\text{meso}}_{\text{exp},1},\dots,\Omega^{\text{meso}}_{\text{exp},Q}$, leading to a number of calls to the deterministic numerical model at mesoscale of $855,000$. Table~\ref{tab:Res2DMesoFP550500} contains the global optimal values $\bb^{\text{opt}}$ and the corresponding relative errors (with respect to the reference experimental value $\bb^{\text{meso}}_{\text{exp}}$) obtained for different values $N_s\in \set{5,50,500}$ of the number of independent realizations generated for the statistical estimation of the mathematical expectations involved in the different numerical indicators. It can be seen that a strong decrease in the value of $N_s$ allows a considerable gain in computing time while maintaining similar results for the identified global optimal values, which can be explained by the use of the MLE method which makes the resolution of the statistical inverse identification problem more robust with respect to the convergence of the statistical estimators used in the numerical indicators of the multi-objective optimization problem~\eqref{optimizationpbmesomin}. 
\begin{table}[H]
\caption{Fixed-point iterative algorithm: comparison between the global optimal value $\bb^{\text{opt}}$ and the reference experimental value $\bb^{\text{meso}}_{\text{exp}}$ for different values of the number $N_s$ of independent realizations generated for the statistical estimation of the mathematical expectations involved in the different numerical indicators.}
\label{tab:Res2DMesoFP550500}
\centering
\begin{tabular}{cccccc}
\toprule
& \boldmath{$\delta$} & \boldmath{$\ell$ \textbf{[$\upmu$m]}} & \boldmath{$\underline{\kappa}$ \textbf{[GPa]}} & \boldmath{$\underline{\mu}$ \textbf{[GPa]}} & \boldmath{$n^{\text{FP}}_{\text{tot}}$} \\
\midrule
$\bb^{\text{meso}}_{\text{exp}}$ & $0.400$ & $125.000$ & $13.750$ & $3.587$ & - \\
\midrule
$\bb^{\text{opt}}$ ($N_s=500$) & $0.391$ & $135.328$ & $12.273$ & $3.717$ & $855,000$ \\
Relative error [\%] & $2.344$ & $8.262$ & $10.740$ & $3.611$ & - \\
\midrule
$\bb^{\text{opt}}$ ($N_s=50$) & $0.387$ & $134.859$ & $12.217$ & $3.717$ & $87,000$ \\
Relative error [\%] & $3.212$ & $7.887$ & $11.153$ & $3.611$ & - \\
\midrule
$\bb^{\text{opt}}$ ($N_s=5$) & $0.396$ & $140.220$ & $12.335$ & $3.717$ & $9,000$ \\
Relative error [\%] & $1.042$ & $12.176$ & $10.293$ & $3.611$ & - \\
\bottomrule
\end{tabular}
\end{table}

The identification results obtained with the genetic algorithm are summarized in Table~\ref{tab:Res2DMesoGA} for the set of $Q=16$ mesoscopic domains of observation $\Omega^{\text{meso}}_{\text{exp},1},\dots,\Omega^{\text{meso}}_{\text{exp},Q}$, namely the set of $Q$ identified values $\bb^{\text{meso}}_1,\dots,\bb^{\text{meso}}_{Q}$ and numbers of generations $n_1,\dots,n_{Q}$ required to reach the desired convergence, and the global optimal value $\bb^{\text{opt}}$ computed by using the MLE method. The initial population used to start the genetic algorithm contains $n_I = 40$ individuals. Figure~\ref{fig:ParetoFront} shows an example of different 2D cross-sections of the Pareto front for the first mesoscopic domain of observation $\Omega^{\text{meso}}_{\text{exp},1}$. The best comprise optimal solution corresponds to the point marked with a green circle on the different 2D cross-sections of the Pareto front, because according to the explanations given in Section~\ref{sec:solving_optim_problem}, it is chosen among all the noninferior (Pareto optimal) solutions generated and selected in the optimal Pareto set (represented by red stars in Figure~\ref{fig:ParetoFront}) as the one that minimizes the distance at the origin of the Pareto front in the multidimensional space (of dimension $3$) of the multi-objective cost function $\Jcb^{\text{meso}}(\bb)$. For reasons of limitation in terms of calculation cost, the number $N_s$ of independent realizations used for the statistical estimation of the mathematical expectations involved in the numerical indicators $\Jc^{\text{meso}}_{\delta}(\bb)$, $\Jc^{\text{meso}}_{\ellb}(\bb)$ and $\Jc^{\text{multi}}_{\underline{\hb}}(\ab^{\text{macro}},\bb)$ is reduced to $N_s=50$. Although the statistical convergence of the three numerical indicators is not achieved, the results of Table~\ref{tab:Res2DMesoFP550500} show that, thanks to the probabilistic modeling of the hyperparameters and the maximum likelihood estimation, the results of the statistical inverse identification method are not significantly affected by a decrease in the value of $N_s$ and are therefore robust with respect to the statistical fluctuations of the different numerical indicators. The number of evaluations of the stochastic computational model needed by the genetic algorithm is given by $n^{\text{GA}}_{\text{tot}}=3 \,n_I \, N_s\sum_{q=1}^{16} n_{q}$, where the superscript ${}^{\text{GA}}$ refers to ``Genetic Algorithm''. Figure~\ref{fig:pdfDLKMGA2D} shows the probability density functions $p_{D}$, $p_{L}$, $p_{\underline{K}}$ and $p_{\underline{M}}$ of random variables $D$, $L$, $\underline{K}$ and $\underline{M}$, respectively. We finally deduce the global optimal value $\bb^{\text{opt}}=(0.372,128.401,11.656,3.306)$ in $([-],[\upmu\text{m}],[\text{GPa}],[\text{GPa}])$ with relative errors less than $8\%$, $3\%$, $16\%$ and $8\%$ for $\delta^{\text{opt}}$, $\ell^{\text{opt}}$, $\underline{\kappa}^{\text{opt}}$ and $\underline{\mu}^{\text{opt}}$, respectively, with respect to the reference experimental value $\bb_{\text{exp}}^{\text{meso}}=(0.40,125,13.75,3.587)$ in $([-],[\upmu\text{m}],[\text{GPa}],[\text{GPa}])$, which are acceptable (reasonably good) and similar to the errors obtained by the fixed-point iterative algorithm. There are still some fluctuations in the values $\underline{\kappa}^{\text{meso}}_1,\dots,\underline{\kappa}^{\text{meso}}_{Q}$ and $\underline{\mu}^{\text{meso}}_1,\dots,\underline{\mu}^{\text{meso}}_{Q}$ identified on each of the $Q=16$ mesoscopic domains of observation $\Omega^{\text{meso}}_{\text{exp},1},\dots,\Omega^{\text{meso}}_{\text{exp},Q}$, which was not the case for the fixed-point iterative algorithm. This underlies the numerical resolution of the Pareto front, which depends on the number $n_V$ of values in each dimension of the parameter search space. In terms of computational efficiency, we can see that the number $n^{\text{GA}}_{\text{tot}}=19,176,000$ of evaluations of the stochastic computational model (resulting from the number of individuals $n_I=40$ in the initial population and the number of population generations $n_1,\dots,n_{Q}$) is much higher than that $n^{\text{FP}}_{\text{tot}} = 87,000$ required by the fixed-point iterative algorithm with $N_s=50$ (see Table~\ref{tab:Res2DMesoFP550500}). Finally, the fixed-point iterative algorithm allows significant computational savings (in terms of computational cost) compared to the genetic algorithm.

\setlength\figureheight{0.12\textheight}
\begin{figure}[H]
	\centering
	\begin{subfigure}{0.33\linewidth}
		\centering
		\tikzsetnextfilename{ParetoFront_I1I2}
%
\begin{tikzpicture}

\begin{axis}[%
width=1.268\figureheight,
height=\figureheight,
at={(0\figureheight,0\figureheight)},
scale only axis,
xmin=-0.03,
xmax=0.45,
xlabel={$\mathcal{J}^{\mathrm{meso}}_{\delta}(\boldsymbol{b})$},
ymin=-0.03,
ymax=0.40,
yticklabel style={/pgf/number format/fixed,/pgf/number format/precision=2},
ylabel={$\mathcal{J}^{\mathrm{meso}}_{\boldsymbol{\ell}}(\boldsymbol{b})$},
axis background/.style={fill=white},
xmajorgrids,
ymajorgrids
]
\addplot [color=red, draw=none, mark=star, mark options={solid, red}, forget plot]
  table[row sep=crcr]{%
0.26112915109734	0.193385339908498\\
1.41702838123472e-08	0.0443380080846728\\
0.0263590248694987	0.218329054882815\\
0.0359455848553355	3.06806282464113e-05\\
0.232536371144855	0.148515843740129\\
0.000240724512792364	0.0753826013028044\\
0.267839251102744	0.198503470399037\\
0.129689024452425	0.0625977126666567\\
5.14672200928442e-05	0.0135925008016883\\
0.390962955011006	0.328786152012316\\
0.260419530353455	0.193381423274344\\
0.397169718800871	0.336414540406498\\
0.0267741220092851	0.218675612817673\\
0.0327186439361831	0.000219330409818162\\
};
\addplot [color=green!60!black, line width=1.0pt, draw=none, mark size=5.0pt, mark=o, mark options={solid, green!60!black}, forget plot]
table[row sep=crcr]{%
	0.0327186439361831	0.000219330409818162\\
};
\end{axis}
\end{tikzpicture}%
		\caption{}
		\label{fig:ParetoFront_I1I2}
	\end{subfigure}\hfill
	\begin{subfigure}{0.33\linewidth}
		\centering
		\tikzsetnextfilename{ParetoFront_I1I3}
%
\begin{tikzpicture}

\begin{axis}[%
width=1.268\figureheight,
height=\figureheight,
at={(0\figureheight,0\figureheight)},
scale only axis,
xmin=-0.03,
xmax=0.45,
xlabel={$\mathcal{J}^{\mathrm{meso}}_{\delta}(\boldsymbol{b})$},
ymin=-0.03,
ymax=0.20,
yticklabel style={/pgf/number format/fixed,/pgf/number format/precision=2},
ylabel={$\mathcal{J}^{\mathrm{multi}}_{\underline{\hb}}(\boldsymbol{a}^{\mathrm{macro}},\boldsymbol{b})$},
axis background/.style={fill=white},
xmajorgrids,
ymajorgrids
]
\addplot [color=red, draw=none, mark=star, mark options={solid, red}, forget plot]
  table[row sep=crcr]{%
0.26112915109734	0.000593902678202308\\
1.41702838123472e-08	0.023082908641791\\
0.0263590248694987	4.80911245631627e-05\\
0.0359455848553355	0.123169992778181\\
0.232536371144855	0.000100352037619752\\
0.000240724512792364	0.000101610598394581\\
0.267839251102744	0.000109083305120985\\
0.129689024452425	0.000112323595888177\\
5.14672200928442e-05	0.164439490268496\\
0.390962955011006	0.000235678332909752\\
0.260419530353455	0.0045678045199167\\
0.397169718800871	0.000305424528105527\\
0.0267741220092851	0.00121186111915441\\
0.0327186439361831	0.0794929735369815\\
};
\addplot [color=green!60!black, line width=1.0pt, draw=none, mark size=5.0pt, mark=o, mark options={solid, green!60!black}, forget plot]
table[row sep=crcr]{%
	0.0327186439361831	0.0794929735369815\\
};
\end{axis}
\end{tikzpicture}%
		\caption{}
		\label{fig:ParetoFront_I1I3}
	\end{subfigure}\hfill
	\begin{subfigure}{0.33\linewidth}
		\centering
		\tikzsetnextfilename{ParetoFront_I2I3}
%
\begin{tikzpicture}

\begin{axis}[%
width=1.268\figureheight,
height=\figureheight,
at={(0\figureheight,0\figureheight)},
scale only axis,
xmin=-0.03,
xmax=0.45,
xlabel={$\mathcal{J}^{\mathrm{meso}}_{\boldsymbol{\ell}}(\boldsymbol{b})$},
ymin=-0.03,
ymax=0.20,
yticklabel style={/pgf/number format/fixed,/pgf/number format/precision=2},
ylabel={$\mathcal{J}^{\mathrm{multi}}_{\underline{\hb}}(\boldsymbol{a}^{\mathrm{macro}},\boldsymbol{b})$},
axis background/.style={fill=white},
xmajorgrids,
ymajorgrids
]
\addplot [color=red, draw=none, mark=star, mark options={solid, red}, forget plot]
  table[row sep=crcr]{%
0.193385339908498	0.000593902678202308\\
0.0443380080846728	0.023082908641791\\
0.218329054882815	4.80911245631627e-05\\
3.06806282464113e-05	0.123169992778181\\
0.148515843740129	0.000100352037619752\\
0.0753826013028044	0.000101610598394581\\
0.198503470399037	0.000109083305120985\\
0.0625977126666567	0.000112323595888177\\
0.0135925008016883	0.164439490268496\\
0.328786152012316	0.000235678332909752\\
0.193381423274344	0.0045678045199167\\
0.336414540406498	0.000305424528105527\\
0.218675612817673	0.00121186111915441\\
0.000219330409818162	0.0794929735369815\\
};
\addplot [color=green!60!black, line width=1.0pt, draw=none, mark size=5.0pt, mark=o, mark options={solid, green!60!black}, forget plot]
table[row sep=crcr]{%
0.000219330409818162	0.0794929735369815\\
};
\end{axis}
\end{tikzpicture}%
		\caption{}
		\label{fig:ParetoFront_I2I3}
	\end{subfigure}
	\caption{Different 2D cross-sections of the Pareto front with the noninferior (Pareto optimal) solutions represented by red stars {$\star$} and the best compromise optimal solution surrounded by a green circle \textcolor{green!60!black}{\textbf{$\Circle$}} for the mesoscopic domain of observation $\Omega^{\text{meso}}_{\text{exp},1}$. (\textbf{a}) cross-section $(\Jc^{\mathrm{meso}}_{\delta}(\bb),\Jc^{\mathrm{meso}}_{\ellb}(\bb))$; (\textbf{b}) cross-section $(\Jc^{\mathrm{meso}}_{\delta}(\bb),\Jc^{\mathrm{multi}}_{\underline{\hb}}(\ab^{\mathrm{macro}},\bb))$; (\textbf{c}) cross-section $(\Jc^{\mathrm{meso}}_{\ellb}(\bb),\Jc^{\mathrm{multi}}_{\underline{\hb}}(\ab^{\mathrm{macro}},\bb))$.}
	\label{fig:ParetoFront}
\end{figure}
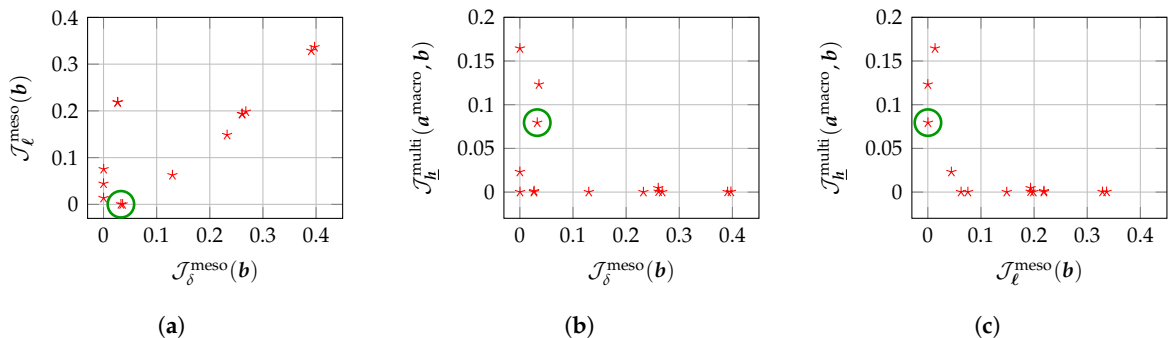

\begin{table}[H]
\caption{Genetic algorithm: comparison between the global optimal value $\bb^{\text{opt}}$ obtained from the $Q=16$ identified values $\bb^{\text{meso}}_1,\dots,\bb^{\text{meso}}_{Q}$ for each of the $Q$ mesoscopic domains of observation $\Omega^{\text{meso}}_{\text{exp},1},\dots,\Omega^{\text{meso}}_{\text{exp},Q}$, and the reference experimental value $\bb^{\text{meso}}_{\text{exp}}$.}
\label{tab:Res2DMesoGA}
\centering
\begin{tabular}{cccccc}
\toprule
& \boldmath{$\delta$} & \boldmath{$\ell$ \textbf{[$\upmu$m]}} & \boldmath{$\underline{\kappa}$ \textbf{[GPa]}} & \boldmath{$\underline{\mu}$ \textbf{[GPa]}} & \boldmath{$n_{q}$} \\
\midrule
$\bb^{\text{meso}}_{1}$ & $0.361$ & $122.222$ & $16.056$ & $2.411$ & 193 \\
$\bb^{\text{meso}}_{2}$ & $0.333$ & $147.778$ & $9.444$ & $2.933$ & 202 \\
$\bb^{\text{meso}}_{3}$ & $0.417$ & $198.889$ & $13.222$ & $3.194$ & 189 \\
$\bb^{\text{meso}}_{4}$ & $0.333$ & $147.778$ & $13.222$ & $3.456$ & 197 \\
$\bb^{\text{meso}}_{5}$ & $0.444$ & $147.778$ & $11.333$ & $4.239$ & 207 \\
$\bb^{\text{meso}}_{6}$ & $0.417$ & $173.333$ & $12.278$ & $2.933$ & 201 \\
$\bb^{\text{meso}}_{7}$ & $0.278$ & $147.778$ & $10.389$ & $3.717$ & 192 \\
$\bb^{\text{meso}}_{8}$ & $0.278$ & $147.778$ & $12.278$ & $3.194$ & 199 \\
$\bb^{\text{meso}}_{9}$ & $0.389$ & $96.667$ & $14.167$ & $3.978$ & 210 \\
$\bb^{\text{meso}}_{10}$ & $0.333$ & $96.667$ & $11.333$ & $2.933$ & 205 \\
$\bb^{\text{meso}}_{11}$ & $0.278$ & $96.667$ & $15.111$ & $2.933$ & 203 \\
$\bb^{\text{meso}}_{12}$ & $0.417$ & $122.222$ & $12.278$ & $4.239$ & 198 \\
$\bb^{\text{meso}}_{13}$ & $0.472$ & $122.222$ & $14.167$ & $3.456$ & 194 \\
$\bb^{\text{meso}}_{14}$ & $0.389$ & $96.667$ & $12.278$ & $2.672$ & 208 \\
$\bb^{\text{meso}}_{15}$ & $0.361$ & $122.222$ & $14.167$ & $3.456$ & 190 \\
$\bb^{\text{meso}}_{16}$ & $0.444$ & $173.333$ & $9.444$ & $3.978$ & 208 \\
\midrule
$\bb^{\text{opt}}$ & $0.372$ & $128.401$ & $11.656$ & $3.306$ & - \\
$\bb^{\text{meso}}_{\text{exp}}$ & $0.400$ & $125.000$ & $13.750$ & $3.587$ & - \\
Relative error [\%] & $7.118$ & $2.721$ & $15.228$ & $7.844$ & - \\
\midrule
$n^{\text{GA}}_{\text{tot}}$ & \multicolumn{5}{c}{$19,176,000$} \\
\bottomrule
\end{tabular}
\end{table}

\setlength\figureheight{0.08\textheight}
\begin{figure}[H]
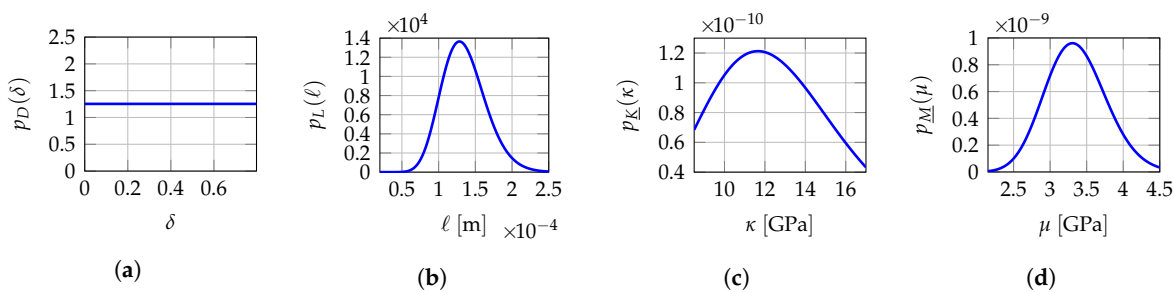

	\centering
	\begin{subfigure}{.25\linewidth}
		\centering
		\tikzsetnextfilename{pdfDelta_GAMLG50}
%
\begin{tikzpicture}

\begin{axis}[%
width=1.281\figureheight,
height=\figureheight,
at={(0\figureheight,0\figureheight)},
scale only axis,
xmin=0,
xmax=0.797724035217466,
xlabel={$\delta$},
ymin=0,
ymax=2.5,
ylabel={$p_{D}(\delta)$},
axis background/.style={fill=white},
xmajorgrids,
ymajorgrids
]
\addplot [color=blue, line width=1.0pt, forget plot]
  table[row sep=crcr]{%
0	1.25356634105602\\
0.797724035217466	1.25356634105602\\
};
\end{axis}
\end{tikzpicture}%
		\caption{}
		\label{fig:pdfDelta_GAMLG50}
	\end{subfigure}\hfill
	\begin{subfigure}{.25\linewidth}
		\centering
		\tikzsetnextfilename{pdfLcorr_GAMLG50_mdpi}
		\input{pdfLcorr_GAMLG50_mdpi}
		\caption{}
		\label{fig:pdfLcorr_GAMLG500}
	\end{subfigure}\hfill
	\begin{subfigure}{.25\linewidth}
		\centering
		\tikzsetnextfilename{pdfKappa_GAMLG50}
		\input{pdfKappa_GAMLG50}
		\caption{}
		\label{fig:pdfKappa_GAMLG50}
	\end{subfigure}\hfill
	\begin{subfigure}{.25\linewidth}
		\centering
		\tikzsetnextfilename{pdfMu_GAMLG50}
		\input{pdfMu_GAMLG50}
		\caption{}
		\label{fig:pdfMu_GAMLG50}
	\end{subfigure}
	\caption{Genetic algorithm: probability density functions $p_{D}$, $p_{L}$, $p_{\underline{K}}$ and $p_{\underline{M}}$ of random variables $D$, $L$, $\underline{K}$ and $\underline{M}$, respectively. (\textbf{a}) $p_{D}(\delta)$; (\textbf{b}) $p_{L}(\ell)$; (\textbf{c}) $p_{\underline{K}}(\kappa)$; (\textbf{d}) $p_{\underline{M}}(\mu)$.}
	\label{fig:pdfDLKMGA2D}
\end{figure}

\subsection{Validation on an In Silico Specimen in Compression Test in 3D Linear Elasticity}
\label{sec:validation_3D}

In this section, we present a second validation example in 3D linear elasticity. We assume there are $Q=3$ test specimens on which are applied exactly the same external loads at macroscale. Recall~that for the validation, the ``experimental'' tests are actually performed \emph{in silico}. Macroscopic domain of observation $\Omega_{\text{exp}}^{\text{macro}}$ is exactly the same 3D cubic domain for each test specimen and corresponds to 3D experimental field measurements on the full volume of each test specimen. As for the previous 2D validation example, since the experimental field measurements are actually performed \emph{in silico}, we~also have $\Omega_{\text{obs}}^{\text{macro}} = \Omega_{\text{exp}}^{\text{macro}}$. The dimensions of each 3D macroscopic domain of observation $\Omega^{\text{macro}}_{\text{exp}}$ are $2\times 2 \times 2$~mm$^3$. For each test specimen, the mesoscale dimensions of 3D mesoscopic domain of observation $\Omega^{\text{meso}}_{\text{exp}}$ are $0.5\times 0.5 \times 0.5$~mm$^3$ (see Figure~\ref{fig:configuration_3D}). Deterministic surface force field $\fb^{\text{macro}}$ is uniformly distributed on the top boundary of macroscopic domain $\Omega^{\text{macro}}_{\text{exp}}$ and applied along the (downward vertical) $-x_3$ direction with an intensity of $2$~kN such that $\norm{\fb^{\text{macro}}} = 50$~kN/$\text{cm}^2 = 5\times10^8$~N/$\text{m}^2$, while the bottom boundary of macroscopic domain $\Omega^{\text{macro}}_{\text{exp}}$ is clamped.

\setlength\figureheight{0.2\textheight}
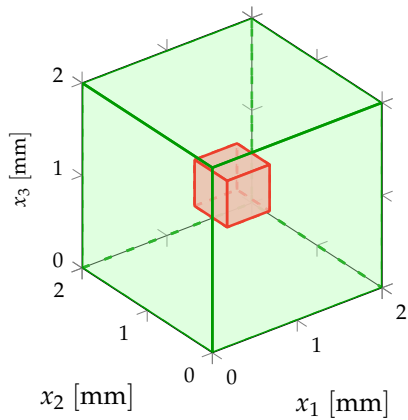
\begin{figure}[H]
	\centering
	\tikzsetnextfilename{configuration_3D}
%
\begin{tikzpicture}

\begin{axis}[%
width=0.895\figureheight,
height=\figureheight,
at={(0\figureheight,0\figureheight)},
scale only axis,
plot box ratio=1 1 1,
xmin=0,
xmax=0.002,
tick align=outside,
xlabel={$x_1$},
ymin=0,
ymax=0.002,
ylabel={$x_2$},
zmin=0,
zmax=0.002,
zlabel={$x_3$},
view={-37.5}{30},
change x base,
change y base,
change z base,
x unit=m,
y unit=m,
z unit=m,
x SI prefix=milli,
y SI prefix=milli,
z SI prefix=milli
]

\addplot3[area legend, loosely dashed, line width=1.0pt, table/row sep=crcr, patch, surf, faceted color=green!60!black, patch type=rectangle, color=green!30, fill opacity=0.20, forget plot, patch table={%
1	3	2	0\\
3	7	6	2\\
1	3	7	5\\
}]
table[row sep=crcr] {%
x	y	z\\
0	0	0\\
0	0.002	0\\
0.002	0	0\\
0.002	0.002	0\\
0	0	0.002\\
0	0.002	0.002\\
0.002	0	0.002\\
0.002	0.002	0.002\\
};

\addplot3[area legend, dashed, line width=1.0pt, table/row sep=crcr, patch, surf, faceted color=red, patch type=rectangle, color=red!30, fill=red!30, fill opacity=0.75, forget plot, patch table={%
1	3	2	0\\
3	7	6	2\\
1	3	7	5\\
}]
table[row sep=crcr] {%
x	y	z\\
0.00075	0.00075	0.00075\\
0.00075	0.00125	0.00075\\
0.00125	0.00075	0.00075\\
0.00125	0.00125	0.00075\\
0.00075	0.00075	0.00125\\
0.00075	0.00125	0.00125\\
0.00125	0.00075	0.00125\\
0.00125	0.00125	0.00125\\
};

\addplot3[area legend, line width=1.0pt, table/row sep=crcr, patch, surf, faceted color=red, patch type=rectangle, color=red!30, fill=red!30, fill opacity=0.75, forget plot, patch table={%
6	7	5	4\\
1	5	4	0\\
0	2	6	4\\
}]
table[row sep=crcr] {%
x	y	z\\
0.00075	0.00075	0.00075\\
0.00075	0.00125	0.00075\\
0.00125	0.00075	0.00075\\
0.00125	0.00125	0.00075\\
0.00075	0.00075	0.00125\\
0.00075	0.00125	0.00125\\
0.00125	0.00075	0.00125\\
0.00125	0.00125	0.00125\\
};

\addplot3[area legend, line width=1.0pt, table/row sep=crcr, patch, surf, faceted color=green!60!black, patch type=rectangle, color=green!30, fill opacity=0.25, forget plot, patch table={%
6	7	5	4\\
1	5	4	0\\
0	2	6	4\\
}]
table[row sep=crcr] {%
x	y	z\\
0	0	0\\
0	0.002	0\\
0.002	0	0\\
0.002	0.002	0\\
0	0	0.002\\
0	0.002	0.002\\
0.002	0	0.002\\
0.002	0.002	0.002\\
};
\end{axis}
\end{tikzpicture}%
	\caption{Illustration of the test specimen occupying the 3D cubic macroscopic domain of observation $\Omega^{\text{macro}}_{\text{exp}} = \Omega^{\text{macro}}$ (in green) which contains a 3D cubic mesoscopic domain of observation $\Omega^{\text{meso}}_{\text{exp}} = \Omega^{\text{meso}}$ (in red) for the numerical validation in 3D linear elasticity.}
	\label{fig:configuration_3D}
\end{figure}

\subsubsection{Parameterization of the Macroscopic and Mesoscopic Models}
\label{sec:model_validation_3D}

Within the framework of linear elasticity theory, any material that is isotropic at macroscale can be completely characterized by a bulk modulus $\kappa$ and a shear modulus $\mu$. Consequently, we have chosen the parameterization $\ab=(\kappa,\mu)$. In particular, we have chosen the experimental value $\ab^{\text{macro}}_{\text{exp}} = (\kappa^{\text{macro}}_{\text{exp}},\mu^{\text{macro}}_{\text{exp}})$ with $\kappa^{\text{macro}}_{\text{exp}} = 138.783$~GPa and $\mu^{\text{macro}}_{\text{exp}} = 64.355$~GPa, corresponding to a Young's modulus $E^{\text{macro}}_{\text{exp}} = 167.218$~GPa and a Poisson's ratio $\nu^{\text{macro}}_{\text{exp}} = 0.2992$.

At mesoscale, we have chosen to construct the {prior} stochastic model of the random elasticity tensor field $\Cb^{\text{meso}}$ as presented in Section~\ref{sec:stochastic_model} and the stochastic boundary value problem~\eqref{equilibriumequationmeso}-\eqref{kinematicrelationmeso} is solved in using \eqref{consititutiveequationmeso} rather than \eqref{consititutiveequationmeso_compliance}. Furthermore, its mean function $\underline{C}^{\text{meso}}$ is spatially constant and models an isotropic elastic medium that is completely characterized by a mean bulk modulus $\underline{\kappa}$ and a mean shear modulus $\underline{\mu}$ at mesoscale. Consequently, the vector-valued hyperparameter $\bb = (\delta,\ell,\underline{\hb})$ involves only (i) a dispersion parameter $\delta$, (ii) a spatial correlation length $\ell$ that is such that $\ell_1 = \ell_2 = \ell_3 = \ell$ in order to be consistent with the effective model at macroscale for which the material is assumed to be isotropic, and (iii) a vector-valued hyperparameter $\underline{\hb} = (\underline{\kappa},\underline{\mu})$ gathering the mean bulk modulus $\underline{\kappa}$ and the mean shear modulus $\underline{\mu}$ at mesoscale. In particular, we have chosen the experimental value $\bb_{\text{exp}}^{\text{meso}}=(\delta_{\text{exp}}^{\text{meso}},\ell_{\text{exp}}^{\text{meso}},\underline{\kappa}_{\text{exp}}^{\text{meso}},\underline{\mu}_{\text{exp}}^{\text{meso}})$ with $\delta_{\text{exp}}^{\text{meso}} = 0.32$, $\ell_{\text{exp}}^{\text{meso}} = 80~\upmu$m, $\underline{\kappa}_{\text{exp}}^{\text{meso}} = 145$~GPa and $\underline{\mu}_{\text{exp}}^{\text{meso}} = 67.3$~GPa, corresponding to a mean Young's modulus $\underline{E}_{\text{exp}}^{\text{meso}} = 174.85$~GPa and a mean Poisson's ratio $\underline{\nu}_{\text{exp}}^{\text{meso}} = 0.2990$~GPa. As already mentioned in Section~\ref{sec:model_validation_2D}, we can restrict the admissible set $\Bc^{\text{meso}} = \intervaloo{0}{\delta_{\text{sup}}} \times \intervaloo{0}{+\infty} \times \intervaloo{0}{+\infty}^2$ (with $\delta_{\text{sup}} = \sqrt{7/11} \approx 0.7977 < 1$) of the vector-valued hyperparameter $\bb = (\delta,\ell,\underline{\kappa},\underline{\mu})$ to a reduced admissible set $\Bc^{\text{meso}}_{\text{ad}} \subset \Bc^{\text{meso}}$ such that $\delta \in \intervalcc{0.20}{0.45}$, $\ell \in \intervalcc{50}{120}$~$\upmu$m, $\underline{\kappa} \in \intervalcc{87.5}{200}$~GPa and $\underline{\mu} \in \intervalcc{40.5}{95.0}$~GPa. This reduced admissible set $\Bc^{\text{meso}}_{\text{ad}}$ is then discretized into $n_V=10$ equidistant points in each dimension for which the three numerical indicators $\Jc^{\text{meso}}_{\delta}(\bb)$, $\Jc^{\text{meso}}_{\ellb}(\bb)$ and $\Jc^{\text{multi}}_{\underline{\hb}}(\ab^{\text{macro}},\bb)$ defined in Section~\ref{sec:meso_indicators} are evaluated and compared. {The identified values $\bb^{\text{meso}}_1,\bb^{\text{meso}}_2,\bb^{\text{meso}}_3$ of hyperparameters $\bb$ for each of the $3$ \emph{in silico} test specimens are then searched on this multidimensional grid of $n_V\times n_V\times n_V\times n_V$ points in the hypercube $\Bc^{\text{meso}}_{\text{ad}}$.}

The classical displacement-based FEM is used for the spatial discretization of (i) the deterministic boundary value problems defined by~\eqref{equilibriumequationmacro}-\eqref{kinematicrelationmacro} in replacing $C^{\text{macro}}$ by $Q=3$ independent realizations of the random apparent elasticity tensor field $\Cb^{\text{meso}}$ on $\Omega^{\text{macro}}$ instead of $\Omega^{\text{meso}}$ to simulate both the ``experimental'' macroscopic displacement field $\ub_{\text{exp}}^{\text{macro}}$ in $\Omega_{\text{exp}}^{\text{macro}} = \Omega^{\text{macro}}$ and the mesoscopic displacement field $\ub_{\text{exp}}^{\text{meso}}$ in $\Omega_{\text{exp}}^{\text{meso}} = \Omega^{\text{meso}}$, (ii) the deterministic boundary value problem defined by~\eqref{equilibriumequationmacro}-\eqref{kinematicrelationmacro} to calculate the macroscopic displacement field $\ub^{\text{macro}}$ in domain $\Omega_{\text{obs}}^{\text{macro}} = \Omega^{\text{macro}}$, and (iii) the stochastic boundary value problems defined by~\eqref{equilibriumequationmeso}-\eqref{kinematicrelationmeso} to calculate the random mesoscopic displacement fields $\Ub^{\text{meso}}$ in using experimental data obtained by solving (i) that are the experimental displacement fields $\ub_{\text{exp}}^{\text{meso}}$ measured on the boundary of domain $\Omega_{\text{exp}}^{\text{meso}} = \Omega_{\text{obs}}^{\text{meso}}$ for each realization of $\Cb^{\text{meso}}$. The~stochastic solver used for solving the stochastic boundary value problem~\eqref{equilibriumequationmeso}-\eqref{kinematicrelationmeso} at mesoscale is the Monte Carlo numerical method. As 3D macroscopic and mesoscopic domains $\Omega^{\text{macro}}$ and $\Omega^{\text{meso}}$ are cubic domains, we consider for each of them a spatial discretization with a structured mesh made up with $8$-nodes linear hexahedral elements with Gauss-Legendre quadrature rule. The finite element mesh of 3D macroscopic domain $\Omega^{\text{macro}}$ is made with the same spatial discretization as the one used for the 2D validation example at macroscale, that is a structured mesh of $25\times25\times25 = 15,625$ hexahedral elements with uniform element size $h^{\text{macro}}=80~\upmu\text{m}=8\times 10^{-5}$~m in each spatial direction. The finite element mesh of 3D mesoscopic domain $\Omega^{\text{meso}}$ is made with the same spatial discretization as the one used for the 2D validation example at mesoscale and whose element size depends on the smallest spatial correlation length considered, that is a structured mesh of $20\times20\times20 = 8000$ hexahedral elements with uniform element size $h^{\text{meso}}=\ell/(n_G/2)=(50~\upmu\text{m})/2=25~\upmu\text{m}=2.5\times 10^{-5}$~m in each spatial direction, with $n_G=4$ Gauss integration points per spatial correlation length.

Concerning the computational stochastic homogenization, as for the 2D validation example, the size $B^{\text{RVE}}$ of representative volume element $\Omega^{\text{RVE}}$ is defined as a function of the spatial correlation length $\ell$ such that $B^{\text{RVE}}=20\times \ell = 20\times 50~\upmu\text{m} = 1~\text{mm} = 10^{-3}$~m.


Recall that the multiscale statistical inverse problem has been formulated into two decoupled optimization problems in $\ab$ and $\bb$, respectively, to be solved sequentially (see Section~\ref{sec:formulation_optim_problem}), namely (i) a macroscale single-objective optimization problem~\eqref{optimizationpbmacro} for the inverse identification of the optimal value $\ab^{\text{macro}}$ of parameter $\ab$ in its admissible set $\Ac^{\text{macro}}$, and (ii) a mesoscale multi-objective optimization problem~\eqref{optimizationpbmesomin} for the statistical inverse identification of the global optimal value $\bb^{\text{opt}}$ of hyperparameter $\bb$ in its reduced admissible set $\Bc^{\text{meso}}_{\text{ad}}$.

\subsubsection{Resolution of the Single-Objective Optimization Problem at Macroscale}
\label{sec:results_validation_3D_macro}

In this paragraph, we present the results of the first single-objective optimization problem~\eqref{optimizationpbmacro} at macroscale which consists in minimizing the macroscopic numerical indicator $\Jc^{\text{macro}}(\ab)$ constructed in each of the $Q=3$ \emph{in silico} test specimens for identifying the optimal value $\ab^{\text{macro}}$ of $\ab$ at macroscale. The single-objective optimization problem~\eqref{optimizationpbmacro} at macroscale has been solved using the Nelder-Mead simplex algorithm. The identification results are reported in Table~\ref{tab:Res3DMacro} and show that the relative error between the identified optimal value $\ab^{\text{macro}} = (138.783,64.355)$ in [GPa] and the reference experimental value $\ab^{\text{macro}}_{\text{exp}} = (138.758,64.377)$ in [GPa] used for the construction of the numerically simulated ``experimental'' database remains very small (less than $0.02\%$ and $0.04\%$ for $\kappa^{\text{macro}}$ and $\mu^{\text{macro}}$, respectively), allowing to validate the proposed identification methodology in 3D linear elasticity for the resolution of the single-objective optimization problem~\eqref{optimizationpbmacro} at macroscale.

\begin{table}[H]
\caption{Comparison between the identified optimal value $\ab^{\text{macro}}$ and the reference experimental value~$\ab^{\text{macro}}_{\text{exp}}$.}
\label{tab:Res3DMacro}
\centering
\begin{tabular}{ccc}
\toprule
& \boldmath{$\kappa$ \textbf{[GPa]}} & \boldmath{$\mu$ \textbf{[GPa]}} \\
\midrule
$\ab^{\text{macro}}$ & $138.783$ & $64.355$ \\
$\ab^{\text{macro}}_{\text{exp}}$ & $138.758$ & $64.377$ \\
Relative error [\%] & $0.018$ & $0.034$ \\
\bottomrule
\end{tabular}
\end{table}

\subsubsection{Resolution of the Multi-Objective Optimization Problem at Mesoscale}
\label{sec:results_validation_3D_meso}

In this paragraph, we present the results of the second multi-objective optimization problem~\eqref{optimizationpbmesomin} at mesoscale which consists in simultaneously minimizing the three numerical indicators $\Jc^{\text{meso}}_{\delta}(\bb)$, $\Jc^{\text{meso}}_{\ellb}(\bb)$ and $\Jc^{\text{multi}}_{\underline{\hb}}(\ab^{\text{macro}},\bb)$ constructed in each of the $Q=3$ \emph{in silico} tests specimens using the optimal parameter $\ab^{\text{macro}}=(138.783,64.355)$ in [GPa] previously identified at macroscale (see the previous paragraph) for identifying the global optimal value $\bb^{\text{opt}}$ of $\bb$ at mesoscale. 

In contrast, unlike the 2D validation example, the multi-objective optimization problem~\eqref{optimizationpbmesomin} has been solved only with the fixed-point iterative algorithm using the same convergence criterion on the residual norm between two iterations that must be less than a tolerance set to $10^{-9}$ and by searching for the solution of the multi-objective optimization problem~\eqref{optimizationpbmesomin} in a multidimensional grid of $n_V\times n_V\times n_V\times n_V$ points in the reduced admissible set $\Bc^{\text{meso}}_{\text{ad}} \subset \Rbb^4$. The genetic algorithm has not been used because the resulting computational cost was too high with the available computational resources. The number of independent realizations for the statistical estimation of the mathematical expectations involved in the different numerical indicators is set to $N_s=500$. The number of evaluations of the stochastic computational model needed by the fixed-point iterative algorithm is given by $n^{\text{FP}}_{\text{tot}}=3 \, n_V \, N_s \sum_{q=1}^{3} n_{q}$. 

Table~\ref{tab:Res3DMesoFP} reports the identification results obtained with the fixed-point iterative algorithm for the set of $Q=3$ \emph{in silico} tests specimens, namely the set of identified values $\bb^{\text{meso}}_1,\bb^{\text{meso}}_2,\bb^{\text{meso}}_3$ and numbers of iterations $n_1,n_2,n_3$ required to reach the desired convergence (with a tolerance set to $10^{-9}$), and the global optimal value $\bb^{\text{opt}}$ computed by using the MLE method. As for the 2D validation example, there are more significant variations between the identified values $\ell^{\text{meso}}_1,\ell^{\text{meso}}_2,\ell^{\text{meso}}_3$ and $\delta^{\text{meso}}_1,\delta^{\text{meso}}_2,\delta^{\text{meso}}_3$, again reflecting the fact that the two associated mesoscopic numerical indicators $\Jc^{\text{meso}}_{\delta}(\bb)$ and $\Jc^{\text{meso}}_{\ellb}(\bb)$ depend directly on the experimental field measurements on each \emph{in silico} test specimen. The identified values $\underline{\kappa}^{\text{meso}}_1,\underline{\kappa}^{\text{meso}}_2,\underline{\kappa}^{\text{meso}}_3$ and $\underline{\mu}^{\text{meso}}_1,\underline{\mu}^{\text{meso}}_2,\underline{\mu}^{\text{meso}}_3$ being almost identical for each \emph{in silico} test specimen, we directly identify the global optimal values $\underline{\kappa}^{\text{opt}}$ and $\underline{\mu}^{\text{opt}}$ without using the MLE method for the random variables $\underline{K}$ and $\underline{M}$. Figure~\ref{fig:pdfDLKM3D} shows the probability density functions $p_{D}$ and $p_{L}$ defined in Section~\ref{sec:randomization} and associated to random variables $D$ and $L$, respectively. We finally obtain the global optimal value $\bb^{\text{opt}}=(0.330,91.236,150.000,64.722)$ in $([-],[\upmu\text{m}],[\text{GPa}],[\text{GPa}])$ with relative errors less than $4\%$ for $\delta^{\text{opt}}$, $\ell^{\text{opt}}$, $\underline{\kappa}^{\text{opt}}$ and $\underline{\mu}^{\text{opt}}$ with respect to the reference experimental value $\bb^{\text{meso}}_{\text{exp}}=(0.32,80,145,67.3)$ in $([-],[\upmu\text{m}],[\text{GPa}],[\text{GPa}])$ used to construct the numerically simulated ``experimental'' database, allowing to validate the proposed identification methodology in 3D linear elasticity for the resolution of the multi-objective optimization problem~\eqref{optimizationpbmesomin} at mesoscale.

\begin{table}[H]
\caption{Fixed-point iterative algorithm: comparison between the global optimal value $\bb^{\text{opt}}$ obtained from the $3$ identified values $\bb^{\text{meso}}_1,\bb^{\text{meso}}_2,\bb^{\text{meso}}_3$ for each of the $3$ \emph{in silico} test specimens and the reference experimental value $\bb^{\text{meso}}_{\text{exp}}$.}
\label{tab:Res3DMesoFP}
\centering
\begin{tabular}{cccccc}
\toprule
& \boldmath{$\delta$} & \boldmath{$\ell$ \textbf{[$\upmu$m]}} & \boldmath{$\underline{\kappa}$ \textbf{[GPa]}} & \boldmath{$\underline{\mu}$ \textbf{[GPa]}} & \boldmath{$n_{q}$} \\
\midrule
$\bb^{\text{meso}}_{1}$ & $0.311$ & $65.556$ & $150.000$ & $64.722$ & 3 \\
$\bb^{\text{meso}}_{2}$ & $0.367$ & $88.889$ & $150.000$ & $64.722$ & 4 \\
$\bb^{\text{meso}}_{3}$ & $0.311$ & $81.111$ & $150.000$ & $64.722$ & 3 \\
\midrule
$\bb^{\text{opt}}$ & $0.330$ & $77.271$ & $150.000$ & $64.722$ & - \\
$\bb^{\text{meso}}_{\text{exp}}$ & $0.320$ & $80.000$ & $145.000$ & $67.300$ & - \\
Relative error [\%] & $3.009$ & $3.411$ & $3.448$ & $3.831$ & - \\
\midrule
$n^{\text{FP}}_{\text{tot}}$ & \multicolumn{5}{c}{$150,000$} \\
\bottomrule
\end{tabular}
\end{table}

\setlength\figureheight{0.2\textheight}
\begin{figure}[H]
	\centering
	\begin{subfigure}{.5\linewidth}
		\centering
		\tikzsetnextfilename{pdfDelta_FPMLG3D50}
%
\begin{tikzpicture}

\begin{axis}[%
width=1.281\figureheight,
height=\figureheight,
at={(0\figureheight,0\figureheight)},
scale only axis,
xmin=0,
xmax=0.797724035217466,
xlabel={$\delta$},
ymin=0,
ymax=2.5,
ylabel={$p_{D}(\delta)$},
axis background/.style={fill=white},
xmajorgrids,
ymajorgrids
]
\addplot [color=blue, line width=1.0pt, forget plot]
  table[row sep=crcr]{%
0	1.25356634105602\\
0.797724035217466	1.25356634105602\\
};
\end{axis}
\end{tikzpicture}%
		\caption{}
		\label{fig:pdfDelta_FPML3D}
	\end{subfigure}\hfill
	\begin{subfigure}{.5\linewidth}
		\centering
		\tikzsetnextfilename{pdfLcorr_FPMLG3D50_mdpi}
		\input{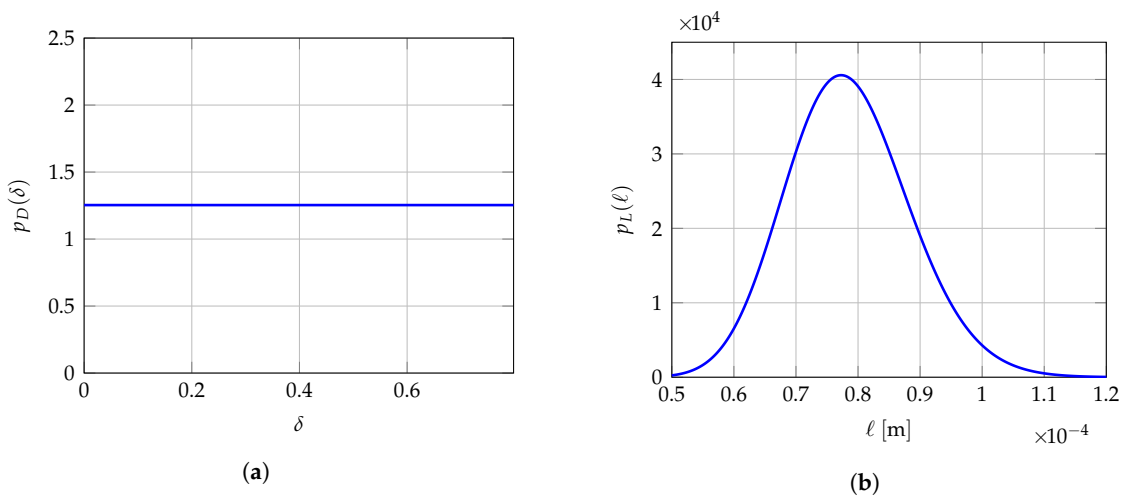}
		\caption{}
		\label{fig:pdfLcorr_FPMLG3D}
	\end{subfigure}
	\caption{Fixed-point iterative algorithm: probability density functions $p_{D}$ and $p_{L}$ of random variables $D$ and $L$, respectively. (\textbf{a}) $p_{D}(\delta)$; (\textbf{b}) $p_{L}(\ell)$.}
	\label{fig:pdfDLKM3D}
\end{figure}

In terms of computational efficiency, we can see in Table~\ref{tab:Res3DMesoFP} that the numbers of iterations $n_1,n_2,n_3$ required to achieve the desired convergence are relatively low (less than or equal to $4$) for each of the $3$ \emph{in silico} test specimens yielding a number of calls to the deterministic numerical model at mesoscale of~$150,000$.

Finally, the results obtained for the identification of the parameters of the deterministic model at macroscale and of the hyperparameters of the {prior} stochastic model at mesoscale for both validation examples in 2D plane stress and 3D linear elasticity, for which the reference experimental values are known \emph{a priori}, demonstrate the efficiency, accuracy and robustness of the improved identification methodology, thereby allowing to apply it in the next section to a real biological material (beef femur cortical bone) with real experimental field measurements. Lastly, let us mention that the fixed-point iterative algorithm introduced in the present work to solve the multi-objective optimization problem allows a considerable gain in terms of computational cost compared to the genetic algorithm used in~Reference \cite{Ngu15}.

\section{Numerical Application of the Multiscale Identification Method to Real Beef Cortical Bone in Plane Stress Linear Elasticity}\label{sec:application}

\begin{figure}[H]
	\centering
	\begin{subfigure}{.25\linewidth}
		\centering
		\includegraphics[width=1.0\linewidth]{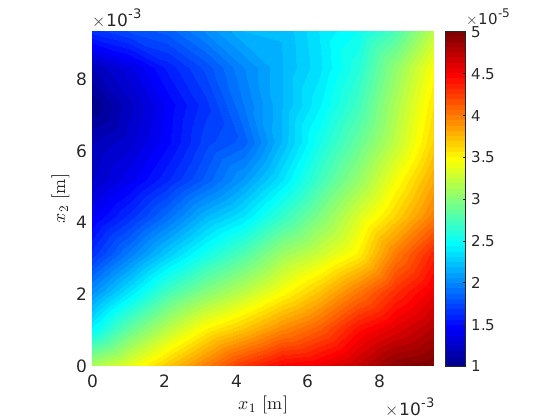}
		\caption{}
		\label{fig:uxMacroExpAOC_Orig}
	\end{subfigure}\hfill
	\begin{subfigure}{.25\linewidth}
		\centering
		\includegraphics[width=1.0\linewidth]{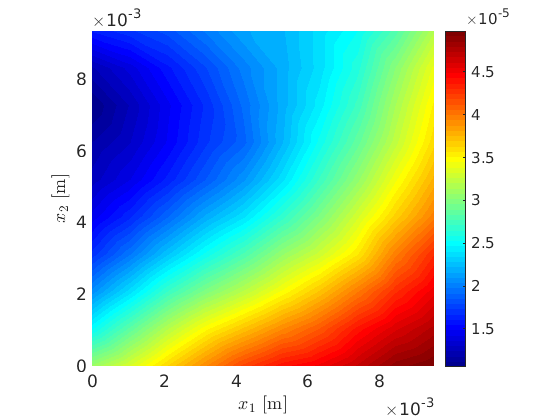}
		\caption{}
		\label{fig:uxMacroExpAOC_Filt}
	\end{subfigure}\hfill
	\begin{subfigure}{.25\linewidth}
		\centering
		\includegraphics[width=1.0\linewidth]{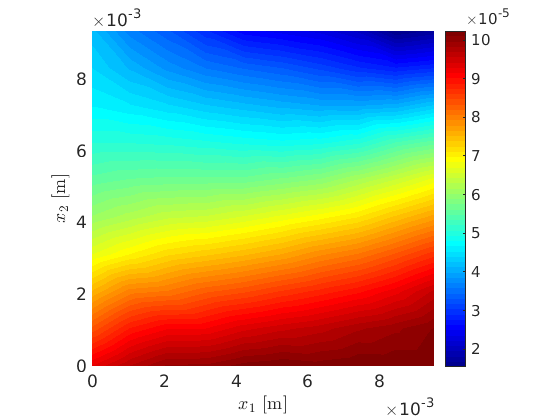}
		\caption{}
		\label{fig:uyMacroExpAOC_Orig}
	\end{subfigure}\hfill
	\begin{subfigure}{.25\linewidth}
		\centering
		\includegraphics[width=1.0\linewidth]{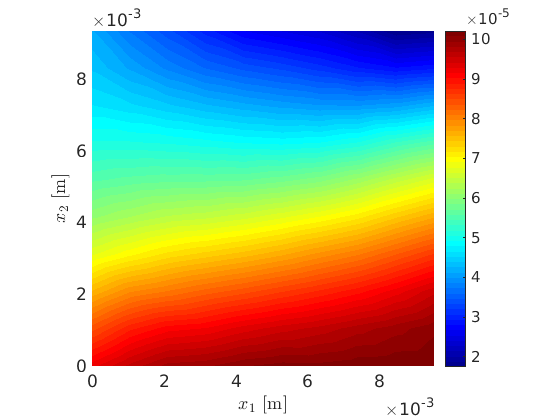}
		\caption{}
		\label{fig:uyMacroExpAOC_Filt}
	\end{subfigure}
	\caption{Components $u^{\text{macro}}_{\text{exp},1}$ and $u^{\text{macro}}_{\text{exp},2}$ of macroscopic experimental displacement field $\ub^{\text{macro}}_{\text{exp}}$ over the 2D macroscopic domain $\Omega^{\text{macro}}_{\text{exp}}$ before and after application of the Gaussian spatial filter. (\textbf{a}) $u^{\text{macro}}_{\text{exp},1}$ unfiltered; (\textbf{b}) $u^{\text{macro}}_{\text{exp},1}$ filtered; (\textbf{c}) $u^{\text{macro}}_{\text{exp},2}$ unfiltered; (\textbf{d}) $u^{\text{macro}}_{\text{exp},2}$ filtered.}
	\label{fig:uMacroExpAOC}
\end{figure}

\begin{figure}[H]
	\centering
	\begin{subfigure}{.25\linewidth}
		\centering
		\includegraphics[width=1.0\linewidth]{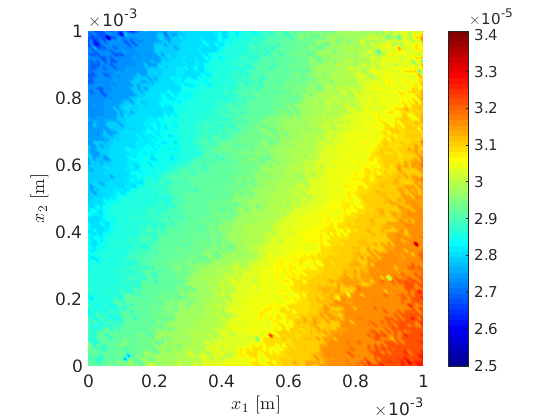}
		\caption{}
		\label{fig:uxMesoExpAOC_Orig}
	\end{subfigure}\hfill
	\begin{subfigure}{.25\linewidth}
		\centering
		\includegraphics[width=1.0\linewidth]{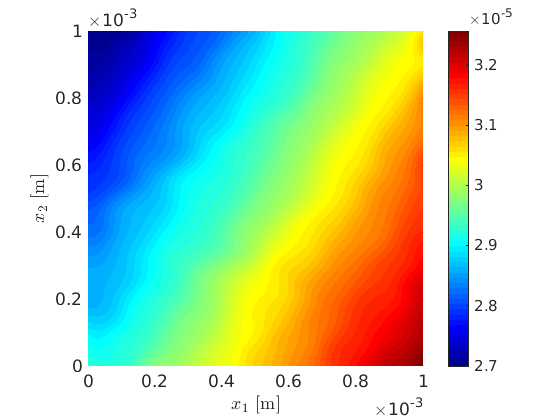}
		\caption{}
		\label{fig:uxMesoExpAOC_Filt}
	\end{subfigure}\hfill
	\begin{subfigure}{.25\linewidth}
		\centering
		\includegraphics[width=1.0\linewidth]{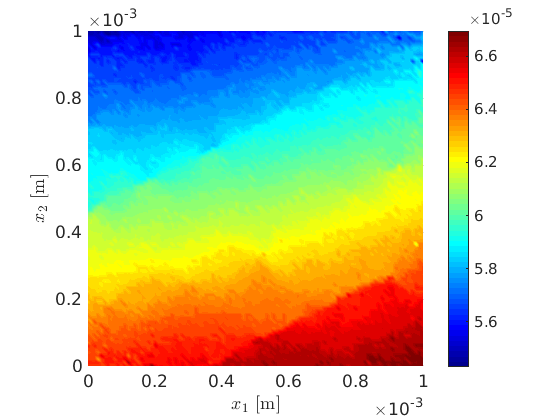}
		\caption{}
		\label{fig:uyMesoExpAOC_Orig}
	\end{subfigure}\hfill
	\begin{subfigure}{.25\linewidth}
		\centering
		\includegraphics[width=1.0\linewidth]{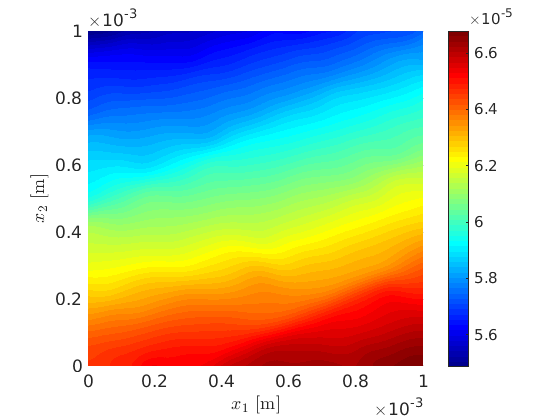}
		\caption{}
		\label{fig:uyMesoExpAOC_Filt}
	\end{subfigure}
	\caption{Components $u^{\text{meso}}_{\text{exp},1}$ and $u^{\text{meso}}_{\text{exp},2}$ of macroscopic experimental displacement field $\ub^{\text{meso}}_{\text{exp}}$ over the 2D mesoscopic domain $\Omega^{\text{meso}}_{\text{exp}}$ before and after application of the Gaussian spatial filter. (\textbf{a}) $u^{\text{meso}}_{\text{exp},1}$ unfiltered; (\textbf{b}) $u^{\text{meso}}_{\text{exp},1}$ filtered; (\textbf{c}) $u^{\text{meso}}_{\text{exp},2}$ unfiltered; (\textbf{d}) $u^{\text{meso}}_{\text{exp},2}$ filtered.}
	\label{fig:uMesoExpAOC}
\end{figure}

In this section, we present a numerical application of the proposed multiscale identification methodology within the framework of 3D linear elasticity with 2D plane stress assumption by using a real experimental database made up of 2D multiscale optical measurements of displacement fields (obtained by DIC method) for only a single test specimen of cortical bone coming from a beef femur. The multiscale experimental test configuration corresponds to the one described in Section~\ref{sec:exp_config} and already considered in the 2D and 3D numerical validation examples presented in Section~\ref{sec:validation}. Technical~details concerning the experimental protocol (specimen preparation, measuring bench, optical image acquisition system and DIC method) for obtaining the multiscale field measurements (performed simultaneously at both macroscale and mesoscale) can be found in Reference \cite{Ngu16}. The unique test specimen at macroscale is a cubic shaped sample with dimensions $1\times 1 \times 1$~cm$^3$ prepared from bovine cortical bone. Even though such a biological tissue is often considered and modeled as a deterministic homogeneous medium with a transversely isotropic linear elastic behavior at macroscale ($\geqslant$10~mm), its microstructure at mesoscale (from $500~\upmu$m to $5$~mm) contains randomly arranged osteons with some resorption cavities (lacuna), that are the principal types of inclusions/inhomogeneities, embedded in a matrix constituted by circumferential interstitial lamella surrounding Haversian canals. As a consequence, it is an anisotropic (heterogeneous) composite material with a complex hierarchical structure, which can be considered and modeled as a random linear elastic medium at mesoscale, and~is therefore well adapted to the experimental application of the multiscale identification methodology developed in the present work. The single specimen is clamped on its lower face and loaded under vertical uniaxial compression onto its upper face with a maximal resultant force of $9$~kN so as to preserve a linear elastic material behavior. In order to reduce the measurement noises (induced by the speckled pattern technique, the lighting of the observed 2D face, the optical image acquisition system, \etc), a Gaussian spatial filter classically used in image processing has been applied to smooth the experimental displacement fields $\ub^{\text{macro}}_{\text{exp}}=(u^{\text{macro}}_{\text{exp},1},u^{\text{macro}}_{\text{exp},2})$ and $\ub^{\text{meso}}_{\text{exp}}=(u^{\text{meso}}_{\text{exp},1},u^{\text{meso}}_{\text{exp},2})$ measured at macroscale and at mesoscale, respectively. {The images of experimental displacement fields at macroscale and at mesoscale have been filtered with a 2D Gaussian smoothing kernel with standard deviation $3.5$. This value has been chosen as a qualitative compromise allowing to regularize/smooth the experimental kinematic fields without removing the spatial fluctuations that are of the same order of magnitude as the lower bound of the search interval for the spatial correlation length $\ell$. Such a spatial filter is also necessary to prevent the optimization algorithms from converging to a zero spatial correlation length.} Figures~\ref{fig:uMacroExpAOC} and \ref{fig:uMesoExpAOC} represent the two components of macroscopic experimental displacement field $\ub^{\text{macro}}_{\text{exp}}$ over the 2D macroscopic domain $\Omega^{\text{macro}}_{\text{exp}}$ and the ones of mesoscopic experimental displacement field $\ub^{\text{meso}}_{\text{exp}}$ over the 2D mesoscopic domain $\Omega_{\text{exp}}^{\text{meso}}$, respectively, before and after application of the Gaussian spatial filter.


\subsection{Parameterization of the Macroscopic and Mesoscopic Models}\label{sec:model_application_2D}

In accordance with the experimental configuration and associated multiscale measurements, 2D plane stresses are assumed and consequently, the deterministic and stochastic computational models at macroscale and mesoscale are the same as those used for the 2D validation example presented in Section~\ref{sec:validation_2D}. Thus, the modeling at macroscale and at mesoscale for the {prior} stochastic model, the hyperparameters and the parameterization are also exactly the same as in Section~\ref{sec:validation_2D}, namely defining $\Sb^{\text{meso}}$ in SFE$^+$ and introducing vector-valued parameter $\ab = (\kappa,\mu)$ and vector-valued hyperparameter $\bb = (\delta,\ell,\underline{\kappa},\underline{\mu})$.
Optimal values of the latter are assumed to be in the reduced admissible set $\Bc^{\text{meso}}_{\text{ad}} \subset \Bc^{\text{meso}}$ constructed from information available in the literature such that $\delta \in \intervalcc{0.30}{0.65}$, $\ell \in \intervalcc{50}{100}$~$\upmu$m, $\underline{\kappa} \in \intervalcc{9.5}{11}$~\text{GPa} and $\underline{\mu} \in \intervalcc{3.5}{5.0}$~GPa, instead of the full admissible space $\Bc^{\text{meso}} = \intervaloo{0}{\delta_{\text{sup}}} \times \intervaloo{0}{+\infty} \times \intervaloo{0}{+\infty}^2$ with $\delta_{\text{sup}} = \sqrt{(n+1)/(n+5)} = \sqrt{7/11} \approx 0.7977 < 1$ (with $n=6$ in linear elasticity). As in the 2D validation example, this reduced admissible set $\Bc^{\text{meso}}_{\text{ad}}$ is discretized into $n_V=10$ points evenly spaced in each dimension for which the three numerical indicators $\Jc^{\text{meso}}_{\delta}(\bb)$, $\Jc^{\text{meso}}_{\ellb}(\bb)$ and $\Jc^{\text{multi}}_{\underline{\hb}}(\ab^{\text{macro}},\bb)$ defined in Section~\ref{sec:meso_indicators} are evaluated and compared.

As for Section~\ref{sec:validation_2D}, both the deterministic boundary value problem~\eqref{equilibriumequationmacro}-\eqref{kinematicrelationmacro} set on the macroscopic domain $\Omega^{\text{macro}}$ and the stochastic boundary value problem~\eqref{equilibriumequationmeso}-\eqref{kinematicrelationmeso} set on the mesoscopic domain $\Omega^{\text{meso}}$ with 2D plane stress assumption are solved by discretizing the 2D domains of observation $\Omega^{\text{macro}}_{\text{obs}}$ and $\Omega^{\text{meso}}_{\text{obs}}$ in space using the FEM. As 2D macroscopic and mesoscopic domains of observation $\Omega^{\text{macro}}_{\text{obs}}$ and $\Omega^{\text{meso}}_{\text{obs}}$ are square domains, we consider for both a spatial discretization with a structured mesh made up with $4$-nodes linear quadrangular elements with Gauss-Legendre quadrature rule, in order to be consistent with the regular grids used for the acquisition and discretization of experimental data. The 2D macroscopic domain $\Omega_{\text{obs}}^{\text{macro}}$ with macroscale dimensions $1\times 1$~cm$^2$ is discretized with a structured mesh of $9\times 9 = 81$ quadrangular elements with uniform element size $h^{\text{macro}}=1.111~\text{mm}=1.111\times10^{-3}$~m in each spatial direction. The 2D mesoscopic domain $\Omega^{\text{meso}}_{\text{obs}}$ with mesoscale dimensions $1\times 1$~mm$^2$ is discretized with a structured mesh of $99 \times 99 = 9801$ quadrangular elements with uniform element size $h^{\text{meso}}=10.10~\upmu\text{m} =1.010\times10^{-5}$~m in each spatial direction. As for the 2D validation example, the size $B^{\text{RVE}}$ of representative volume element $\Omega^{\text{RVE}}$ is defined with respect to the spatial correlation length $\ell$ such that $B^{\text{RVE}}=20\times \ell$. The stochastic boundary value problem~\eqref{equilibriumequationmeso}-\eqref{kinematicrelationmeso} at mesoscale is solved using the Monte Carlo numerical method and statistical convergence analyses have been systematically performed to set the number of independent realizations for the statistical estimation of the mathematical expectations involved in the different numerical indicators to the value $N_s=500$.

\subsection{Numerical Results of the Multiscale Statistical Inverse Identification}\label{sec:results_application_2D}

\subsubsection{Resolution of the Single-Objective Optimization Problem at Macroscale}
\label{sec:results_application_2D_macro}

In this paragraph, we present the results of the first single-objective optimization problem~\eqref{optimizationpbmacro} at macroscale which consists in minimizing the macroscopic numerical indicator $\Jc^{\text{macro}}(\ab)$ constructed in the macroscopic domain of observation $\Omega_{\text{obs}}^{\text{macro}}$ for identifying the optimal value $\ab^{\text{macro}}$ of $\ab$ at macroscale. The single-objective optimization problem~\eqref{optimizationpbmacro} at macroscale has been solved using the Nelder-Mead simplex algorithm. Table~\ref{tab:Res2DMacroAOC} gives the identified optimal value $\ab^{\text{macro}} = (11.335,4.781)$ in [GPa], corresponding to a macroscopic transverse bulk modulus $\kappa^{\text{macro}} = 11.335$~GPa and a macroscopic transverse shear modulus $\mu^{\text{macro}} = 4.781$~GPa, or equivalently to a macroscopic transverse Young's modulus $E^{\text{macro}} = 12.575$~GPa and a macroscopic transverse Poisson's ratio $\nu^{\text{macro}} = 0.3151$, which are in coherence with the values already published and available in the literature for this type of biological material.

\begin{table}[H]
\caption{Identified optimal value $\ab^{\text{macro}}$ of parameter $\ab = (\kappa,\mu)$.}
\label{tab:Res2DMacroAOC}
\centering
\begin{tabular}{ccc}
\toprule
& \boldmath{$\kappa$ \textbf{[GPa]}} & \boldmath{$\mu$ \textbf{[GPa]}}\\
\midrule
$\ab^{\text{macro}}$ & $11.335$ & $4.781$ \\
\bottomrule
\end{tabular}
\end{table}

\subsubsection{Resolution of the Multi-Objective Optimization Problem at Mesoscale}
\label{sec:results_application_2D_meso}

In this paragraph, we present the results of the second multi-objective optimization problem~\eqref{optimizationpbmesomin} at mesoscale which consists in simultaneously minimizing the three numerical indicators $\Jc^{\text{meso}}_{\delta}(\bb)$, $\Jc^{\text{meso}}_{\ellb}(\bb)$ and $\Jc^{\text{multi}}_{\underline{\hb}}(\ab^{\text{macro}},\bb)$ constructed in the mesoscopic domain of observation $\Omega^{\text{meso}}_{\text{obs}}$ using the optimal parameter $\ab^{\text{macro}}=(11.335,4.781)$ in [GPa] previously identified at macroscale (see the last paragraph) for identifying the global optimal value $\bb^{\text{opt}}$ of $\bb$ at mesoscale. The multi-objective optimization problem~\eqref{optimizationpbmesomin} has been solved only by using the fixed-point iterative algorithm (with a convergence criterion on the residual norm between two iterations that must be less than a tolerance set to $10^{-9}$) and by searching for the solution of the multi-objective optimization problem~\eqref{optimizationpbmesomin} in a multidimensional grid of $n_V\times n_V\times n_V\times n_V$ points in the reduced admissible set $\Bc^{\text{meso}}_{\text{ad}} \subset \Rbb^4$. The~number of evaluations of the stochastic computational model needed by the fixed-point iterative algorithm is given by $n^{\text{FP}}_{\text{tot}}=3 \, n_V \, N_s \, n^{\text{FP}}$, where $n^{\text{FP}}$ is the number of iterations required to reach the desired convergence for the considered mesoscopic domain of observation $\Omega^{\text{meso}}_{\text{obs}}$.

Table~\ref{tab:Res2DMesoFPAOC} gives the identified optimal value $\bb^{\text{meso}} = (0.533,61.111,10.500,4.667)$ in $([-],[\upmu\text{m}],[\text{GPa}],[\text{GPa}])$ obtained with the fixed-point iterative algorithm, corresponding to a dispersion parameter $\delta^{\text{meso}} = 0.533$, a spatial correlation length $\ell^{\text{meso}} = 61.111~\upmu$m, a mesoscopic mean transverse bulk modulus $\underline{\kappa}^{\text{meso}} = 10.500$~GPa and a mesoscopic mean transverse shear modulus $\underline{\mu}^{\text{meso}} = 4.667$~GPa, or equivalently to a mesoscopic mean transverse Young's modulus $\underline{E}^{\text{meso}} = 12.194$~GPa and a mesoscopic mean transverse Poisson's ratio $\underline{\nu}^{\text{meso}} = 0.3064$. The number of iterations $n^{\text{FP}}$ required to achieve the desired convergence with the fixed-point iterative algorithm over the mesoscopic subdomain $\Omega^{\text{meso}}$ is $n^{\text{FP}} = 5$, leading to a number of evaluations of the stochastic computational model equal to $n^{\text{FP}}_{\text{tot}} = 7500$. The identification results obtained at mesoscale are also in agreement with the information provided in the literature for this type of biological material. Indeed, from a physical standpoint, the identified spatial correlation length $\ell^{\text{meso}}=61.111$~$\upmu$m turns out to be of the same order of magnitude as the distance between two adjacent lamellae of an osteon in bovine (beef femur) cortical bone. Moreover, such a value of spatial correlation length is in accordance with the assumption of scale separation between macroscale and mesoscale.

\begin{table}[H]
\caption{Fixed-point iterative algorithm: identified optimal value $\bb^{\text{meso}}$ of hyperparameter $\bb=(\delta,\ell,\underline{\kappa},\underline{\mu})$ for the mesoscopic domain of observation $\Omega^{\text{meso}}_{\text{obs}}$.}
\label{tab:Res2DMesoFPAOC}
\centering
\begin{tabular}{cccccc}
\toprule
& \boldmath{$\delta$} & \boldmath{$\ell$ \textbf{[$\upmu$m]}} & \boldmath{$\underline{\kappa}$ \textbf{[GPa]}} & \boldmath{$\underline{\mu}$ \textbf{[GPa]}} & \boldmath{$n^{\text{FP}}$} \\
\midrule
$\bb^{\text{meso}}$ & $0.533$ & $61.111$ & $10.500$ & $4.667$ & 5 \\
\midrule
$n^{\text{FP}}_{\text{tot}}$ & \multicolumn{5}{c}{$7500$} \\
\bottomrule
\end{tabular}
\end{table}

\section{Conclusions}
\label{sec:conclusion}

In the present work, we have revisited the multiscale identification methodology recently proposed in Reference \cite{Ngu15} for the mechanical characterization of the apparent elastic properties of a complex microstructure made up of a heterogeneous anisotropic material 
{that can be considered as a random linear elastic medium}
within the framework of 3D linear elasticity theory. Such a multiscale identification has been performed by solving a challenging multiscale statistical inverse problem (requiring multiscale experimental field measurements) formulated as a multi-objective optimization problem. This latter can be decomposed into a first single-objective optimization problem defined at macroscale and a second multi-objective optimization problem defined at mesoscale, to be solved sequentially and involving cost functions (numerical indicators) sufficiently sensitive to the variation of the parameters and hyperparameters to be identified. These numerical indicators allow for quantifying and minimizing the distance between some relevant quantities of interest resulting from the multiscale experimental field measurements at macroscale and mesoscale on the one hand, and their counterparts obtained through forward numerical simulations of a deterministic computational model at macroscale and of a stochastic computational model at mesoscale corresponding to the experimental configuration on the other hand. We consider an \emph{ad hoc} 
{prior} stochastic model introduced in Reference \cite{Soi06} for the numerical modeling and simulation of the random elasticity field, which is parameterized by a small number of hyperparameters. We also employ a stochastic computational homogenization method for the transfer of statistical information from mesoscale to macroscale. The multiscale identification methodology leads to the identification of the optimal values of (i) the parameters involved in the deterministic model of the effective (deterministic and homogeneous) elasticity tensor at macroscale and (ii) the hyperparameters involved in the {prior} stochastic model of the apparent (random and heterogeneous) elasticity tensor field at mesoscale

In the present paper, we have proposed two main improvements of the multiscale statistical inverse identification methodology of the {prior} stochastic model. First, we have introduced an additional single-objective cost function (numerical indicator) at mesoscale dedicated to the identification of the spatial correlation length(s) involved in the {prior} stochastic model, allowing the newly formulated multi-objective optimization to be solved with a better computational efficiency by using a (computationally cheap) fixed-point iterative algorithm instead the (costly) global optimization algorithm (genetic algorithm) used in Reference \cite{Ngu15}. The identification results obtained with the fixed-point iterative algorithm are promising and comparable to that obtained with the genetic algorithm in terms of accuracy. Second, an \emph{ad hoc} probabilistic modeling of the hyperparameters involved in the {prior} stochastic model and identified on different mesoscopic domains of observation has been proposed in order to improve both the robustness and the precision of the statistical inverse identification method of the {prior} stochastic model. Finally, the improved identification methodology has been first validated on \emph{in silico} materials within the framework of 2D plane stress and 3D linear elasticity with numerically simulated multiscale experimental data, and then {successfully} applied to real heterogeneous biological material within the framework of 2D plane stress linear elasticity with real multiscale experimental measurements of 2D displacement fields obtained from a static uniaxial compression test performed on a single specimen made of bovine cortical bone and monitored by 2D digital image correlation at both {macroscale and mesoscale}. 
In line with this work, several perspectives could be addressed: (i) {the multi-objective optimization problem could be solved by using} machine learning based on artificial neural networks with a numerical database generated from the stochastic computational model to train an artificial neural network in an (offline) preliminary phase and to use the trained neural network to perform the statistical inverse identification in a computationally cheap (online) computing phase for further reducing the computational cost; (ii) {the proposed methodology could be applied} to real multiscale experimental measurements of full 3D displacement fields obtained for example by X-ray computed microtomography and digital volume correlation, and also to other types of random heterogeneous materials; (iii) {the proposed methodology could be improved by identifying} a {posterior} stochastic model of the non-Gaussian random elasticity (or compliance) field in high stochastic dimension at the mesoscale of an anisotropic heterogeneous linear elastic microstructure using the identified {prior} stochastic model.



\vspace{6pt} 



\authorcontributions{Conceptualization, C.D.; methodology, C.D.; software, T.Z., C.D. and F.P.; validation, F.P. and C.D.; formal analysis, T.Z., F.P. and C.D.; investigation, T.Z.; resources, C.D. and F.P.; data curation, F.P. and C.D.; writing---original draft preparation, F.P., {C.D.} and T.Z.; writing---review and editing, F.P. and C.D.; visualization, F.P.; supervision, C.D. and F.P.; project administration, C.D.; funding acquisition, C.D. and F.P. All~authors have read and agreed to the published version of the manuscript.}

\funding{This research received no external funding.}

\acknowledgments{The authors gratefully acknowledge Christian Soize, Professor at Universit\'e Gustave Eiffel, Laboratoire MSME, for helpful discussions and valuable suggestions.}

\conflictsofinterest{The authors declare no conflict of interest.
} 

\abbreviations{The following abbreviations are used in this manuscript:\\


\noindent 
\begin{tabular}{@{}ll}
a.s. & almost surely\\
RVE & Representative Volume Element\\
CCD & Charge-Coupled Device\\
CMOS & Complementary Metal-Oxide-Semiconductor\\
DIC & Digital Image Correlation\\
DVC & Digital Volume Correlation\\
$\mu$CT & micro-Computed Tomography\\
MRI &  Magnetic Resonance Imaging\\
OCT & Optical Coherence Tomography\\
LS & Least Squares\\
MLE & Maximum Likelihood Estimation\\
MaxEnt & Maximum Entropy\\
KL & Karhunen-Lo\`eve\\
PC & Polynomial Chaos\\
FP & Fixed-Point\\
GA & Genetic Algorithm\\
FEM & Finite Element Method
\end{tabular}}

%

\reftitle{References}

\end{document}